\documentclass[useAMS, usenatbib]{mn2e}
\usepackage{graphicx}
\usepackage{longtable}
\usepackage{multicol}
\usepackage{lscape}
\usepackage{natbib}
\usepackage{threeparttable}
\usepackage{gensymb}
\usepackage{subfig}
\usepackage{wrapfig}
\usepackage{hhline}

\title[AGN-driven outflows in local ULIRGs]{Quantifying the AGN-driven outflows in ULIRGs (QUADROS) I: VLT/Xshooter observations of 9 nearby objects}
\author[M. Rose, C.  Tadhunter, C. Ramos Almeida, J. Rodr\'iguez Zaur\'in, F. Santoro, R. Spence]{Marvin Rose$^{1}$\thanks{E-mail:m.rose@sheffield.ac.uk (MR)}, Clive Tadhunter$^{1}$, Cristina Ramos Almeida$^{2,3}$, \newauthor Javier Rodr\'iguez Zaur\'in$^1$, Francesco Santoro$^{4,5}$, Robert Spence$^{1}$\\
$^{1}$Department of Physics and Astronomy, University of Sheffield, Sheffield S3 7RH, UK\\
$^{2}$Instituto de Astrof\'{i}sica, Universidad de La Laguna, E-38205, La Laguna, Tenerife, Spain\\
$^{3}$Departamento de Astrof\'{i}sica, Universidad de La Laguna, E-38206, La Laguna, Tenerife, Spain\\
$^{4}$ASTRON, the Netherlands Institute for Radio Astronomy, PO 2, 79900 AA, Dwingeloo, The Netherlands\\
$^{5}$Kapetyn Astronomical Institute, University of Groningen, PO 800, 9700 AV, Groningen, The Netherlands\\} 
\begin{document}

\date{}

\pagerange{\pageref{firstpage}--\pageref{lastpage}} \pubyear{2017}

\maketitle

\label{firstpage}

\begin{abstract}

\noindent Although now routinely incorporated into hydrodynamic simulations of galaxy evolution, the true importance of the feedback effect of the outflows driven by active galactic nuclei (AGN) remains uncertain from an observational perspective. This is due to a lack of accurate information on the densities, radial scales and level of dust extinction of the outflow regions. Here we use the unique capabilities of VLT/Xshooter to investigate the warm outflows in a representative sample of 9 local (0.06 $<$ z $<$ 0.15) ULIRGs with AGN nuclei and, for the first time, accurately quantify the key outflow properties. We find that the outflows are compact (0.05 $<$ R$_{[OIII]}$ $<$ 1.2 kpc), significantly reddened  (median E(B-V)$\sim$0.5 magnitudes), and have relatively high electron densities (3.4 $<$ log$_{10}$ n$_e$ (cm$^{-3}$) $<$ 4.8). It is notable that the latter densities -- obtained using trans-auroral [SII] and [OII] emission-line ratios -- exceed those typically
assumed for the warm, emission-line outflows in active galaxies, but are similar to
those estimated for broad and narrow absorption line outflow systems detected in
some type 1 AGN. Even if we make the most optimistic assumptions about the true (deprojected) outflow velocities, we find relatively modest mass outflow rates ($0.07 < \dot{M} < 11$ M$_{\sun}$ yr$^{-1}$) and kinetic powers measured as a fraction of the AGN bolometric luminosities  ($4\times10^{-4} < \dot{E}/L_{BOL} <1$\%). Therefore, although warm, AGN-driven outflows have the potential to strongly affect the star formation histories in the inner bulge regions ($r \sim$ 1kpc) of nearby ULIRGs, we lack evidence that they have a significant impact on the evolution of these rapidly evolving systems on larger scales.
\end{abstract}

\begin{keywords}
 galaxies:active --- {\it (galaxies:)}quasars:emission-lines --- \\ {\it (galaxies:)}quasars:general --- galaxies: interactions 
\end{keywords}

\section{Introduction}

Energetic, galactic-scale outflows powered by central active galactic nuclei (AGN) are a potentially important process in the evolution of galaxies through major galaxy mergers. For example, they can regulate the correlations between the black hole masses and the host galaxy properties (\citealt{silk98}; \citealt{fabian99}; \citealt{king15}). Towards the final stages of mergers, when the merging nuclei coalesce, the AGN-induced outflows are predicted to be extremely powerful, disrupting any surrounding molecular clouds and therefore halting star formation in the host galaxy bulges (\citealt{dimatteo05}; \citealt{springel05}; \citealt{hopkins10}). Indeed, outflows have been observed in all gas phases in merging galaxies: ionized (e.g. [OIII]; \citealt{wilman99}; \citealt{lipari03}; \citealt{holt03}; \citealt{holt06}; \citealt{spoon09a}; \citealt{spoon09b}; \citealt{rupke13}; \citealt{javier13}), neutral (e.g. NaID; \citealt{rupke05},\citealt{martin05}; \citealt{teng13}; \citealt{rupke13}) and molecular (e.g. CO, OH; \citealt{cicone12}; \citealt{veilleux13}; \citealt{cicone14}). 

For the AGN-induced outflows to explain the correlations between black hole mass and the properties of the bulges of the host galaxies, models often require the outflows to carry a relatively large fraction (5-10\%; \citealt{fabian99}; \citealt{dimatteo05}; \citealt{springel05}) of the total AGN power. However, the power requirement can be reduced if the outflows are \lq multi-staged' ($\sim$0.5\% $L_{BOL}$, \citealt{hopkins10}). Clearly, it is now important to use observations to determine whether, in reality,  the outflows are as powerful as the models require.

In order to quantify the impact of AGN-induced outflows in the course of galaxy mergers, it is important to identify samples of merging galaxies in which the central supermassive black holes (SMBH) and galaxy bulges are in a phase of rapid growth. Ultraluminous Infrared Galaxies (ULIRGs) have spectral energy distributions (SEDs) that are dominated by high infrared luminosities (L$_{IR}$ $>$ 10$^{12}$ L$_{\sun}$), which represent the dust-reprocessed light of nuclear  starburst and/or  AGN activity \citep{sanders96}. The optical morphologies of the overwhelming majority of local ULIRGs are consistent with them representing major galaxy mergers  (e.g. tidal tails, double nuclei; \citealt{scoville00}; \citealt{veilleux02}). This suggests that low redshift (z$<$0.2) ULIRGs are local analogies of rapidly evolving galaxies at higher redshifts, thus making them laboratories to study the impact of AGN-induced outflows on host galaxies. Indeed, ULIRGs represent just the situation modelled in many of the most recent hydrodynamic simulations of gas-rich mergers.

In this context, it is notable that warm outflows 
have been detected at  both optical and mid-IR wavelengths in a high proportion of nearby ULIRGs with AGN nuclei in the form of broad (Full Width at Half Maximum (FWHM) $>$ 500~km s$^{-1}$), blueshifted (v$_{out}$ $>$ 150 km s$^{-1}$), high ionization emission-lines (e.g. [OIII], [NeIII], [NeV]: \citealt{holt03}; \citealt{spoon09a}; \citealt{spoon09b}; \citealt{holt11}; \citealt{javier13}). Such outflows are specifically associated with the subset of ULIRGs that show optical AGN nuclei. In this subset the emission-line kinematics are often more extreme than those observed in the the general population of nearby Seyfert galaxies and quasars \citep{javier13}. However,  while the emission-line observations clearly demonstrate the presence of AGN-driven warm outflows in the near-nuclear regions of ULIRGs,  many of the key properties of these outflows including the dust extinction, emission-line luminosities (e.g. $L_{H\beta}$), electron densities (n$_{e}$), radial extents ($r$) and kinematics (e.g. outflow velocities, $v_{out}$), have yet to be determined accurately \citep[see discussion in][]{javier13}. Therefore the true mass outflow rates ($\dot{M} \propto L_{H\beta} v_{out} / r n_e$) and kinetic powers ($\dot{E} \propto L_{H\beta} v_{out}^3/ r n_e$) of the warm outflows remain uncertain.

To overcome these issues, we are undertaking a programme of high resolution Hubble Space Telescope (HST) imaging and wide-spectral-coverage spectroscopy observations of a sample of 15 local ULIRGs: the Quantifying Ulirg Agn-DRiven OutflowS (QUADROS) project. HST imaging observations of a subset of 8 objects are presented in Tadhunter et al. (2017; QUADROS paper III), while spectroscopic observations  of 8 (mainly northern) objects from a sample taken with WHT/ISIS are reported in Spence et al. (2017; QUADROS paper II). Here we present deep VLT/Xshooter observations of 7 objects in the southern part of our sample, along with two additional low-redshift ULIRGs.  VLT/Xshooter \citep{vernet11} has particular advantages for this project, because (a) its wide spectral coverage (0.32 -- 2.4$\mu$m) allows access to alternative diagnostic emission-lines e.g. Paschen series for the reddening estimates, and trans-auroral [OII] and [SII] emission-lines for the density estimates \citep{holt11}, and (b) its spectral resolution is sufficiently high ($R\sim5000$) to allow us to separate individual kinematic sub-components in the emission-line blends.  

The paper is organised as follows. In section 2 we describe the observations, data reduction and emission-line fitting procedure we use throughout this paper. Section 3  describes how the key parameters required to quantify the outflow properties are calculated, and presents the results on the reddening, densities and kinematics of the outflows. In section 4 these results are used to determine the mass outflows rates and kinetic powers of the outflows, which are discussed in the context of previous 
studies of AGN-driven outflows. Finally, in section 5 we present our conclusions. Throughout the paper we adopt the cosmological parameters  H$_{0}$ = 73.0 km s$^{-1}$, $\Omega_{\rm m}$ = 0.27 and $\Omega_{\Lambda}$ = 0.73. 

\section{Observations and Data Reduction}

\subsection{Sample}

The QUADROS study concentrates on local ($z < 0.175$)  ULIRGs with optical AGN nuclei. The redshift limit is set to ensure that the targets are sufficiently bright and well resolved that we can study their emission-line regions in detail. This is particularly important because we intend to accurately measure the spatial extents and kinematics of the outflowing gas in ULIRGs. In addition, the redshift limit ensures that key diagnostic emission-lines are observable within the wavelength range of our observations. 

The full sample for the QUADROS project is based on the \citet{kim98} 1 Jy sample of ULIRGs. It comprises all ULIRGs from \citet{kim98} that are classified as having Seyfert-like nuclear spectra by \citet{yuan10} on the basis of the \citet{kewley06} diagnostic diagrams, with redshifts  $z < 0.175$, right ascensions 12 $<$ RA $<$ 02 hr, and declinations $\delta$ $>$ -25 degrees --- 23 objects in total \citep[see][]{javier13}. Of the full sample, three objects -- F12265+0219 (3C273)\footnote{Note that it is debatable whether the quasar 3C273 should be included
as a ULIRG, because its far-IR emission is dominated by non-thermal radiation from its jets, rather than thermal emission from
dust as is the case for most ULIRGs}, F12540+5708, and F21219+1757 -- are type 1 AGN for which the relatively strong, broad Balmer and FeII lines make measurement of the forbidden emission-line kinematics difficult, and for a further three objects -- F01888-0856, F12112+0305, and F23327+2913 -- key emission-lines such as [OIII]$\lambda$5007 have equivalent widths that are too low relative to the stellar continuum to allow the measurement of accurate emission-line kinematics \citep{javier13}. This leaves 17 objects, of which we have made deep spectroscopic observations of 15: 8 with the WHT/ISIS (see Spence et al. 2017) and 7 with the VLT/Xshooter (this paper). Therefore, we have observed 15/17 of the objects that met the original selection criteria of \citet{javier13}, and for which detailed measurements of the optical/near-IR emission-line properties are feasible. 

In this paper we present observations for the 7 ULIRGs that meet our primary selection criteria and have been observed with VLT/Xshooter. By itself, this sample represents 7/9 of all the objects in \citet{kim98} with Seyfert-like optical AGN nuclei, $z < 0.175$, $12 < RA < 17$ hr --- $-25 < \delta < 20$ degrees, and measurable warm gas outflow properties. In addition, we present VLT/Xshooter observations of a further two ULIRGs -- F14378-3651 and F19254-7245 -- that meet our spectral and redshift criteria, but fall outside the declination range of the original QUADROS sample; these were observed to fill
in gaps in our VLT observing schedule when none of other objects in the full sample were observable due wind-related pointing restrictions at the VLT. Overall, we believe that our  VLT/Xshooter sample of 9 objects is representative
of local ULIRGs that harbour warm, AGN-driven outflows. Details of the individual objects in the sample and their spectra are presented in 
the Appendix B.

\subsection{Observations}\label{sect:obs}

\subsubsection{VLT/Xshooter}

\begin{center}
\begin{table*}
\centering
\caption{VLT/Xshooter observation details for the ULIRG sample discussed in this paper. `Object': object designation in the IRAS Faint Source Catalogue Database. z$_{NED}$: redshift of the object as presented in the NASA/IPAC Extragalactic Database (NED). `Night': date the object was observed. `EXP': exposure time for the observations (s). `Air mass': full air mass range of the observations. `Slit PA': slit position angle of the observations. `Aperture': the size of the extraction aperture for the spectrum in arcseconds. `Scale': the pixel scale for our adopted cosmology (kpc/''). `GA$_V$': galactic extinction in A$_V$ using the re-calibrated Galactic extinction maps \citep{sf11} of \citet{sfd98}.}
\begin{tabular}{lcccccccc}
\hline
Object	&	z$_{NED}$	&	Night	&	EXP	(s)	&	Air	mass	&	Slit PA ($\degree$)	&	Aperture ('')	&	Scale	(kpc/'')	&	GA$_V$ (Mag.)	\\	
\hline																							
F12072-0444	&	0.1284	&	12-05-2013	&	8$\times$650	&	1.066-1.109	&	40	&	0.696	&	2.222	&	0.120	\\
F13305-1739	&	0.1484	&	13-05-2013	&	8$\times$600	&	1.119-1.253	&	70	&	0.870	&	2.506	&	0.195	\\
F13443+0802SE	&	0.1353	&	12-05-2013	&	8$\times$600	&	1.183-1.214	&	170	&	1.044	&	2.320	&	0.066	\\
F13451+1232W	&	0.1217	&	13-05-2013	&	8$\times$600	&	1.252-1.425	&	20	&	1.044	&	2.120	&	0.092	\\
F14378-3651	&	0.0676	&	12-05-2013	&	8$\times$600	&	1.024-1.098	&	35	&	1.218	&	1.255	&	0.198	\\
F15130-1958	&	0.1087	&	19-05-2013	&	8$\times$600	&	1.013-1.160	&	113	&	1.044	&	1.917	&	0.385	\\
F15462-0450	&	0.0998	&	13-05-2013	&	4$\times$600	&	1.114-1.213	&	30	&	0.696	&	1.776	&	0.590	\\
F16156+0146NW	&	0.1320	&	13-05-2013	&	8$\times$600	&	1.116-1.198	&	30	&	0.870	&	2.262	&	0.235	\\
F19254-7245S	&	0.0617	&	19-05-2013	&	8$\times$600	&	1.494-1.532	&	10	&	0.696	&	1.141	&	0.235	\\

\hline																			
\end{tabular}
\label{tab:obs}
\end{table*}
\end{center}

The 9 ULIRGs in our VLT/Xshooter programme were observed in visitor mode during the nights of the 12 \& 13th of May 2013, and the observations were completed with further observations on the 19th of May 2013 in service mode. To avoid wavelength-dependent slit losses due to differential refraction by the Earth's atmosphere,  the targets were observed with their slit position angles at the parallactic angle of the centre of the observations. Telluric standard star observations were taken immediately after the science observations of each source at  air masses that matched the those of the centres of the observations of the ULIRGs. The ULIRGs were observed in nodding mode using the standard ABBA pattern, with a 30 arcsecond spatial offset to aid sky subtraction. The instrumental configuration was a 1.3$\times$11 arcsecond slit for the UVB arm, a 1.2$\times$11 arcsecond slit for the VIS arm, and a 1.2$\times$11 arcsecond slit for the NIR arm for the science targets. 


\subsubsection{HST/STIS}
  
In addition to the Xshooter spectroscopy presented in this paper, we also use data taken with the Space Telescope Imaging Spectrograph (STIS) installed on the HST to estimate the radii of the outflow regions (see $\S$\ref{sect:extent}). HST/STIS spectroscopic observations exist for two of the ULIRGs studied in the this paper -- F12072-0444 and F16156+0146NW. These objects were observed under the HST observing programs 8190 (PI:Farrah) and 12934 (PI:Tadhunter).  While F12072-0444 was observed using the  G430L and G750L gratings, with the 52X0.2'' slit aligned along position angle of 93.4$\degree$, F16156+0146NW was observed using the G750L grating, with the 52X0.1'' aligned along position angle 53.9$\degree$.

The main spectroscopic reduction steps were performed by the Space Telescope Science Institute (STScI) STIS pipeline {\it calstis} which bias-subtracted, flat-field corrected, cosmic ray rejected, wavelength calibrated and flux calibrated the spectra. We then used IRAF packages to improve the bad pixel and cosmic ray removal, as well as combine individual dithered spectra to improve the signal to noise. 

\subsection{Seeing Estimates}\label{sect:seeing}

\begin{center}
\begin{table*}
\centering
\caption{Seeing estimates for the Xshooter observations. `DIMM' is the mean DIMM seeing over the period of the observations as indicated in the headers of the observations and `1D' are the estimates from the extracted spatial profiles of the stars. For the `DIMM' measurements we present the values, followed by the standard errors on the mean (`$\pm$'). For the `1D' measurements we present the values (`Seeing'), followed by the standard errors (`$\pm$') and the difference between the average values for each acquisition image (`$\Delta$'). `N$_{stars}$' indicates the number of stars per acquisition image used in the estimates. `Final' presents the final FWHM$_{1D}$ seeing estimates and uncertainties which are applied throughout the remaining analysis. With the exception of `N$_{Stars}$', all entries are in units of arcseconds.}
\begin{tabular}{l|cc|ccc|cc}
\hline
Object	&	DIMM &		&	1D &     &       & N$_{Stars}$ & Final	\\
	&	Seeing	&	$\pm$	& Seeing & $\pm$ & $\Delta$ & &FWHM$_{1D}$		\\
\hline															
F12072-0444	&	0.82	&		0.02	&	0.74,0.71	&	0.02,0.02	&	0.03	&	3	&	0.73$\pm$0.03	\\
F13305-1739	&	0.93	&		0.04	&	1.21,1.11	&	0.04,0.03	&	0.10	&	2	&	1.16$\pm$0.10	\\
F13443+0802SE	&	0.67	&		0.02	&	0.71,0.76	&	0.01,0.01	&	0.05	&	4	&	0.74$\pm$0.05	\\
F13451+1232W$^a$	&	0.94	&		0.04	&	0.95,0.88	&	0.03,0.04	&	0.07	&	1	&	0.92$\pm$0.07	\\
F14378-3651	&	1.19	&		0.10	&	0.82,0.90	&	0.01,0.02	&	0.08	&	10	&	0.86$\pm$0.10	\\
F15130-1958	&	1.34	&		0.09	&	0.81,0.95	&	0.02,0.03	&	0.14	&	3	&	0.88$\pm$0.14	\\
F15462-0450$^b$	&	0.77	&		0.02	&	0.81	&	0.01	&	0.01	&	4	&	0.81$\pm$0.02	\\
F16156+0146NW	&	0.77	&		0.01	&	0.63,0.65	&	0.02,0.01	&	0.02	&	4	&	0.64$\pm$0.02	\\
F19254-7245S	&	1.03	&		0.02	&	1.02,0.99	&	0.05,0.06	&	0.03	&	10	&	1.01$\pm$0.03	\\
\hline
\end{tabular}
\begin{tablenotes}
\item[a] $^a$ There was only a single star on the acquisition images. Therefore there is no value for the $\sigma$ of the measurements.
\item[a] $^b$ F15462-0450 had only 1 set of observations and therefore only 1 acquisition image was available.
 \end{tablenotes}
\label{tab:seeing}
\end{table*}
\end{center}

To quantify the radial extents of the warm gas outflows, it is important to have accurate estimates of the seeing FWHM for our Xshooter observations. We have done this by making 1-dimensional Gaussian fits to spatial slices in the y-axis (slit) direction to the g' filter images of stars present in the acquisition images for the observations of the individual ULIRGs. The spatial slices were integrated over the same range of pixels in the x-direction as the slit width. This method naturally takes into account the integration of the 2-dimensional seeing disk in the direction perpendicular to the long axis of the slit. 

For comparison, we also present seeing estimates from the Differential Image Motion Monitor (DIMM) channel of the MASS-DIMM instrument\footnote{The DIMM at Paranal continually monitors the seeing using observations of stars close to the zenith with a filter centred at $\sim$5000\AA\ .}. However, note that, since the DIMM seeing estimates were not made in the same direction and at the same airmass as the target observations, the median DIMM seeing only gives a indication of the general seeing quality at the time of the observations. On the other hand, the DIMM measurements do provide a useful indication of the variation of the seeing over the observation period, which we quantify as the standard error in the mean DIMM seeing over the period.

The seeing estimates are compared in Table \ref{tab:seeing}. It is clear that the variations in the seeing conditions were significant during the observations for F14378-3651 and F15130-1958. In addition, when comparing the DIMM seeing FWHM estimates to the 1D estimates they often differ significantly: up to 0.53 arcsec in the most extreme case (F15130-1958). We will apply the 1D estimates of the seeing FWHM when we consider the spatial extents of the outflows in $\S$\ref{sect:extent}. \\

\subsection{Data Reduction}\label{sect:dr}

\begin{figure*}
\centering
\includegraphics[scale=0.8]{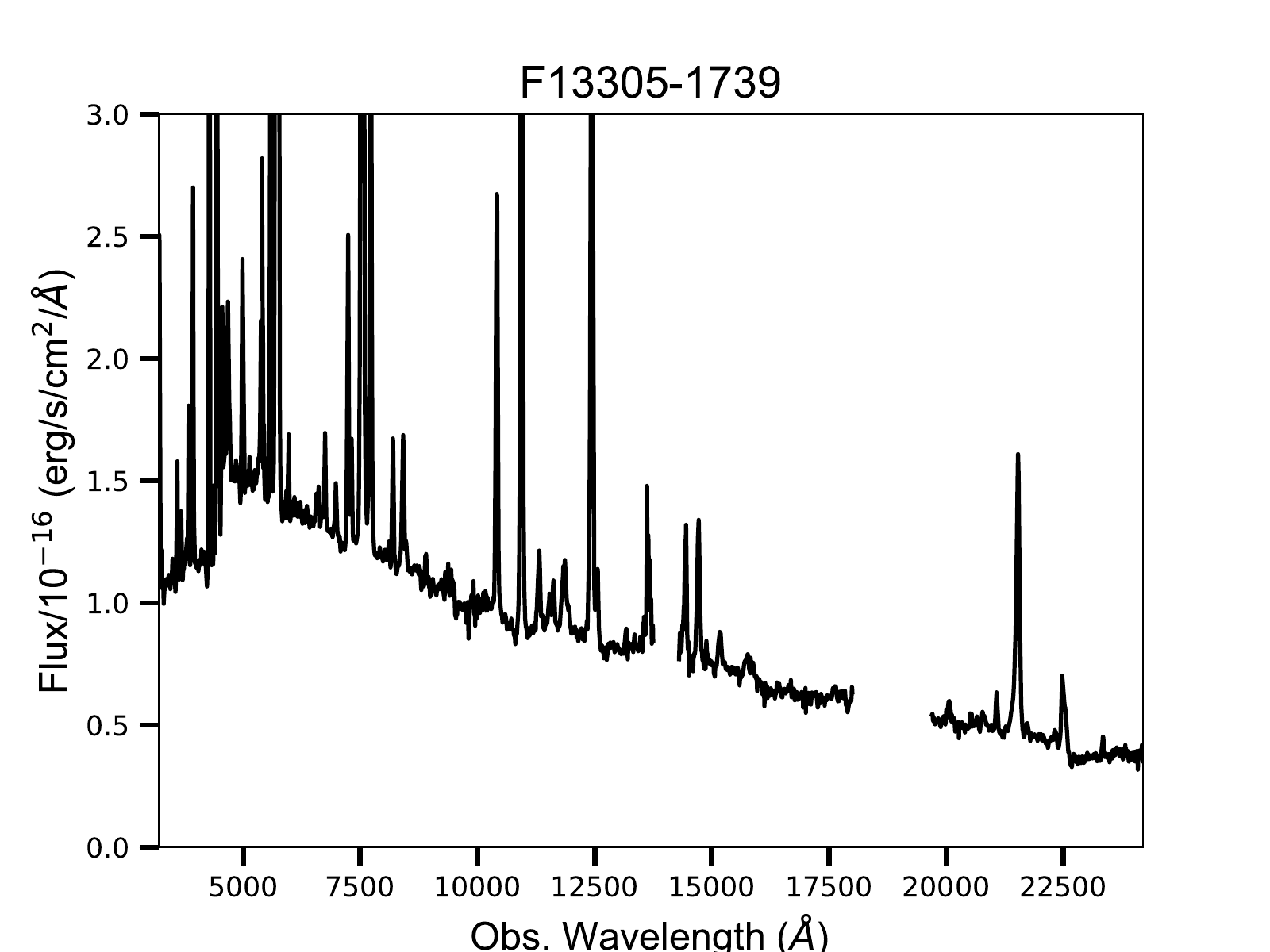}
\caption{The UVB-VIS-NIR spectrum for F13305-1739. The flux scale is measured in units of 10$^{-16}$ erg s$^{-1}$ cm$^{-2}$ \AA $^{-1}$, and the observed wavelength is measured in units of \AA .}
\label{fig:spec}
\end{figure*}

\begin{figure*}
\centering
\includegraphics[scale=0.48]{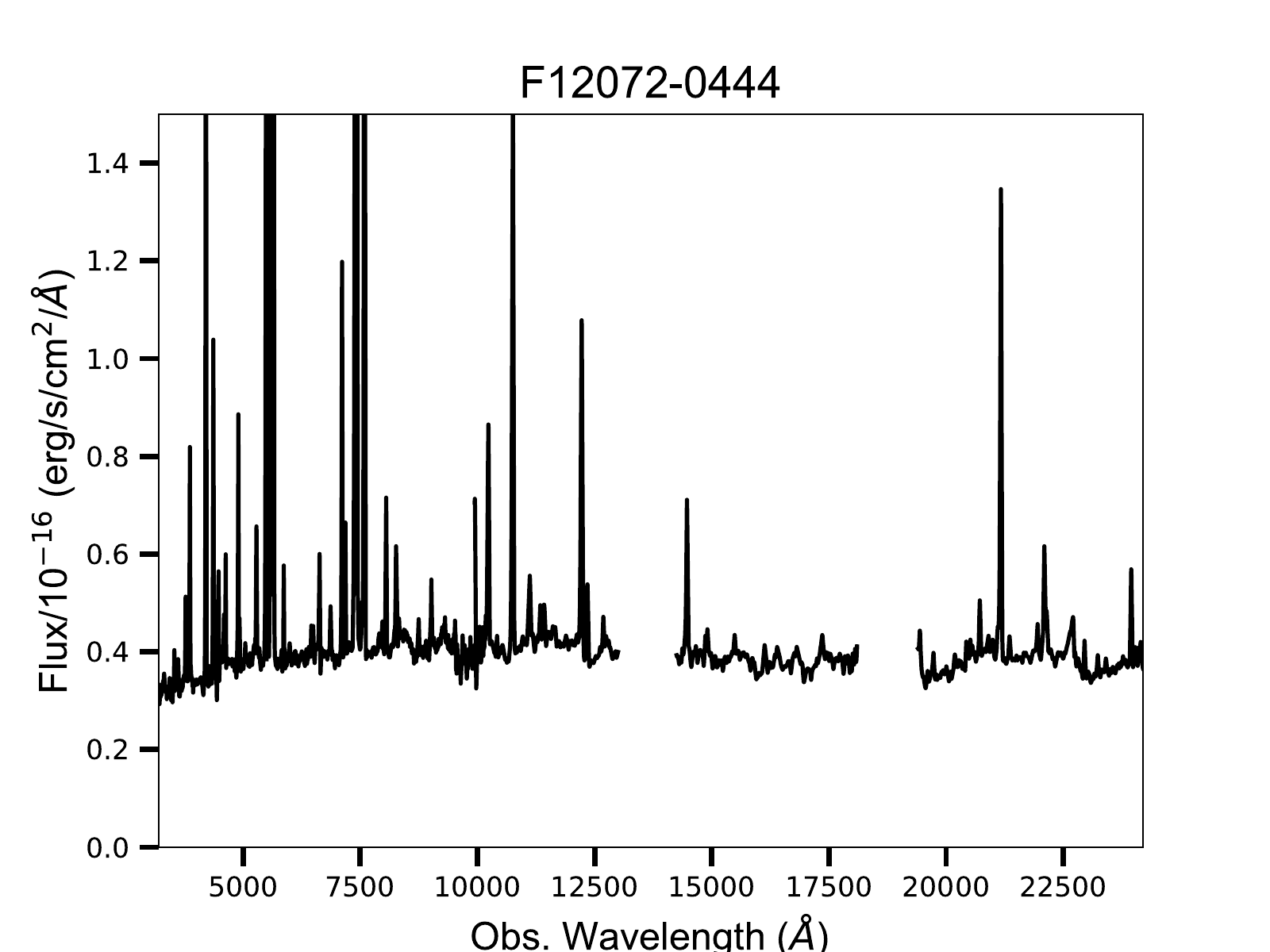}
\includegraphics[scale=0.48]{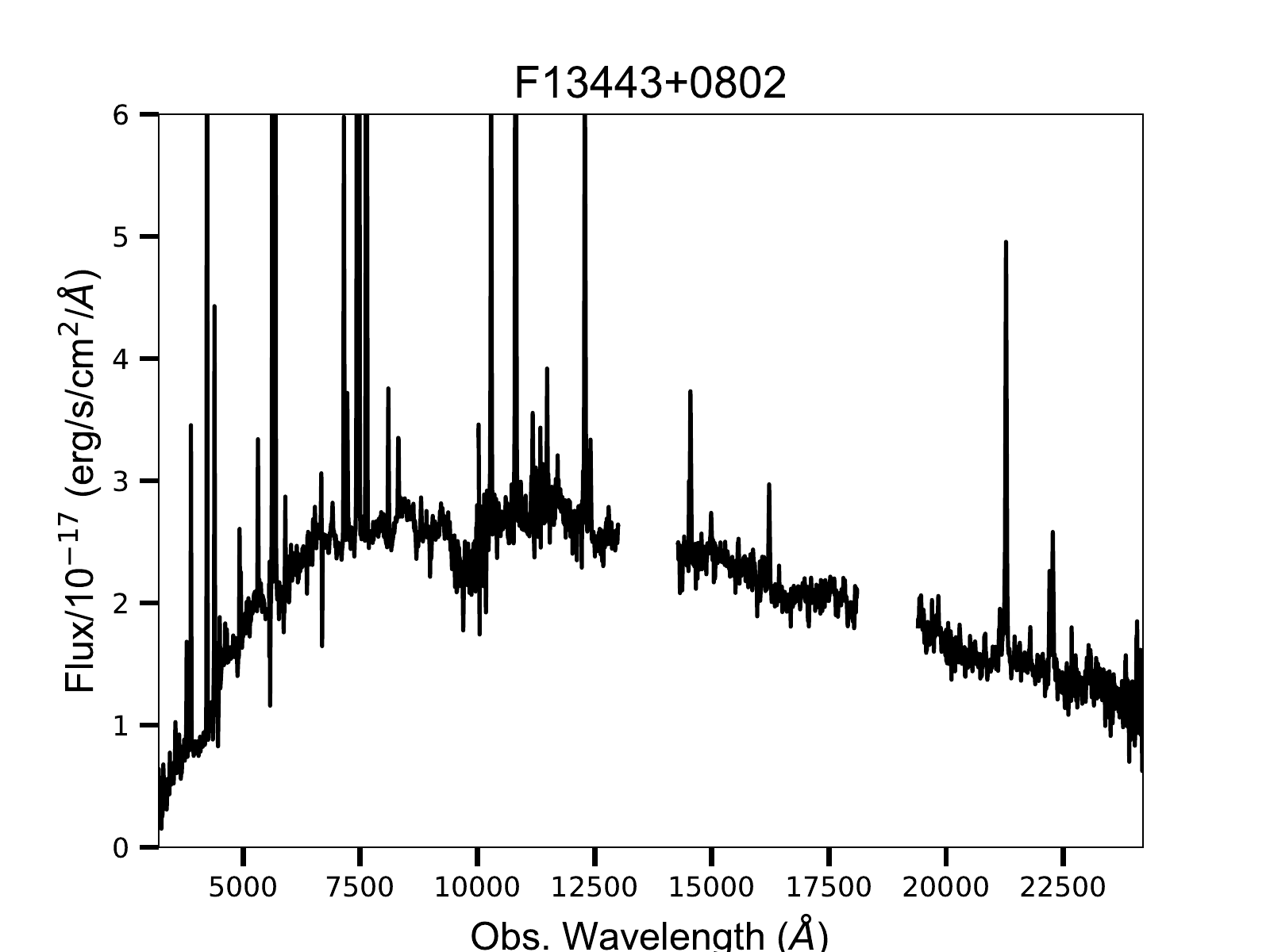}
\includegraphics[scale=0.48]{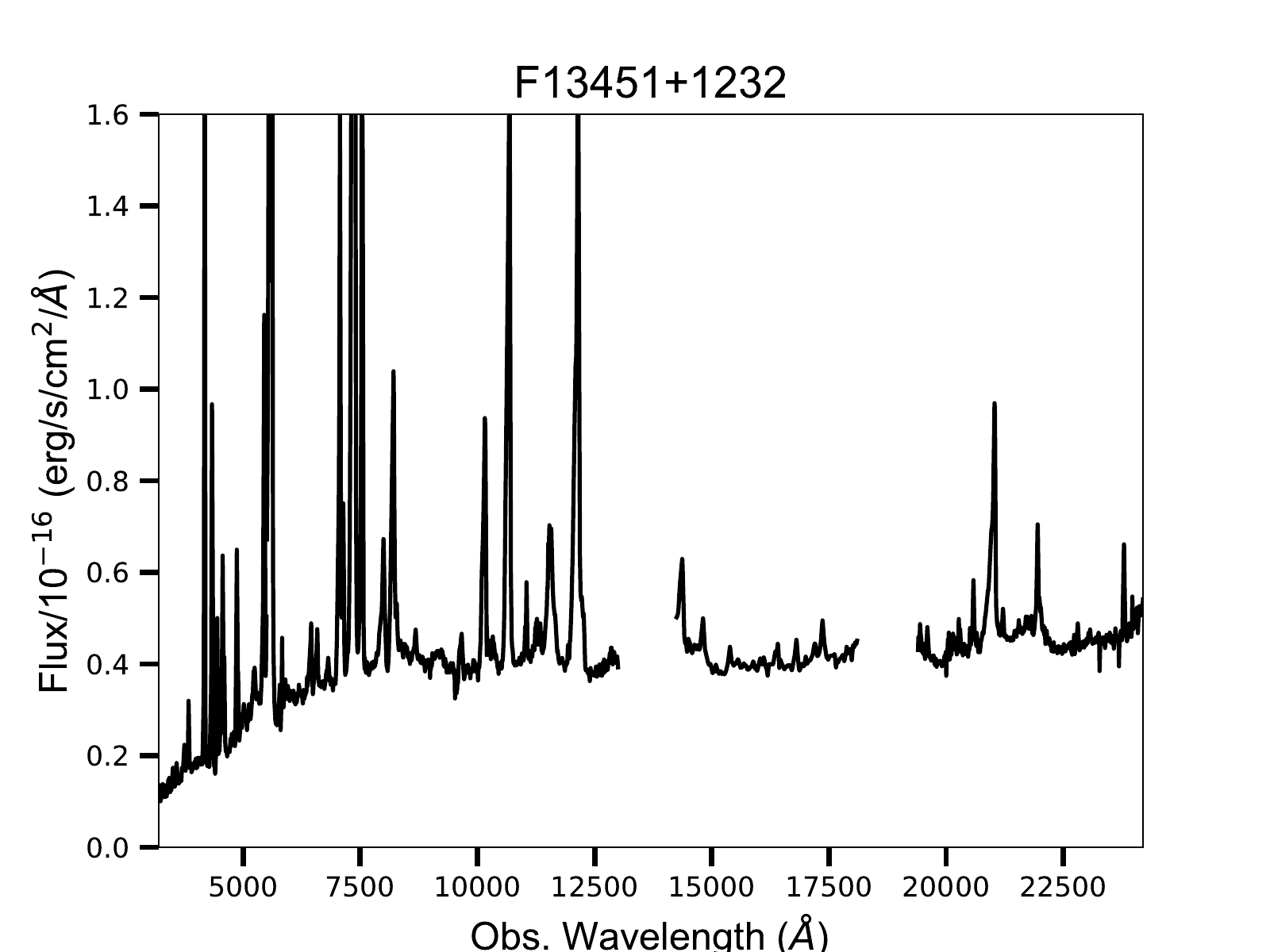}
\includegraphics[scale=0.48]{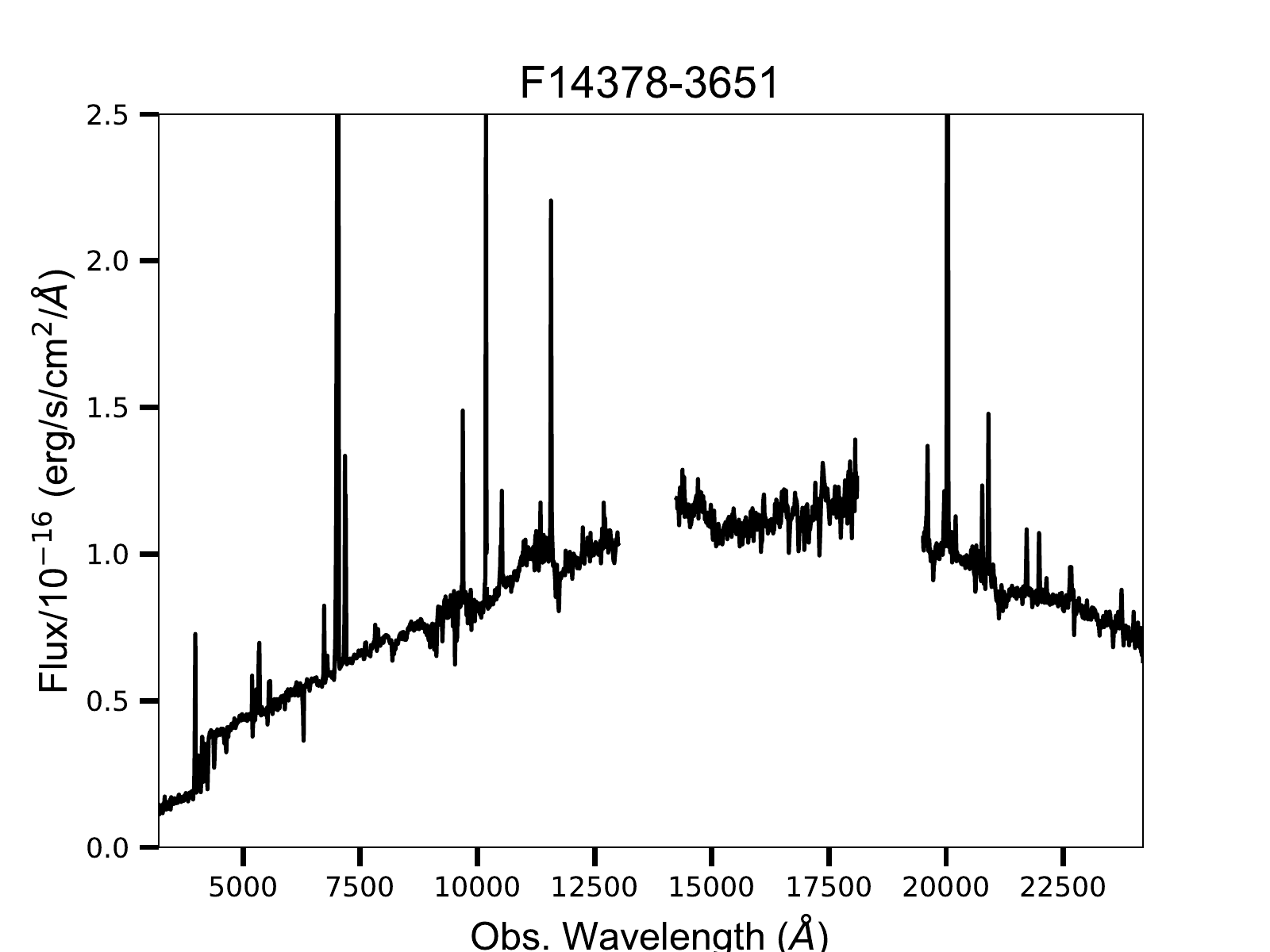}
\includegraphics[scale=0.48]{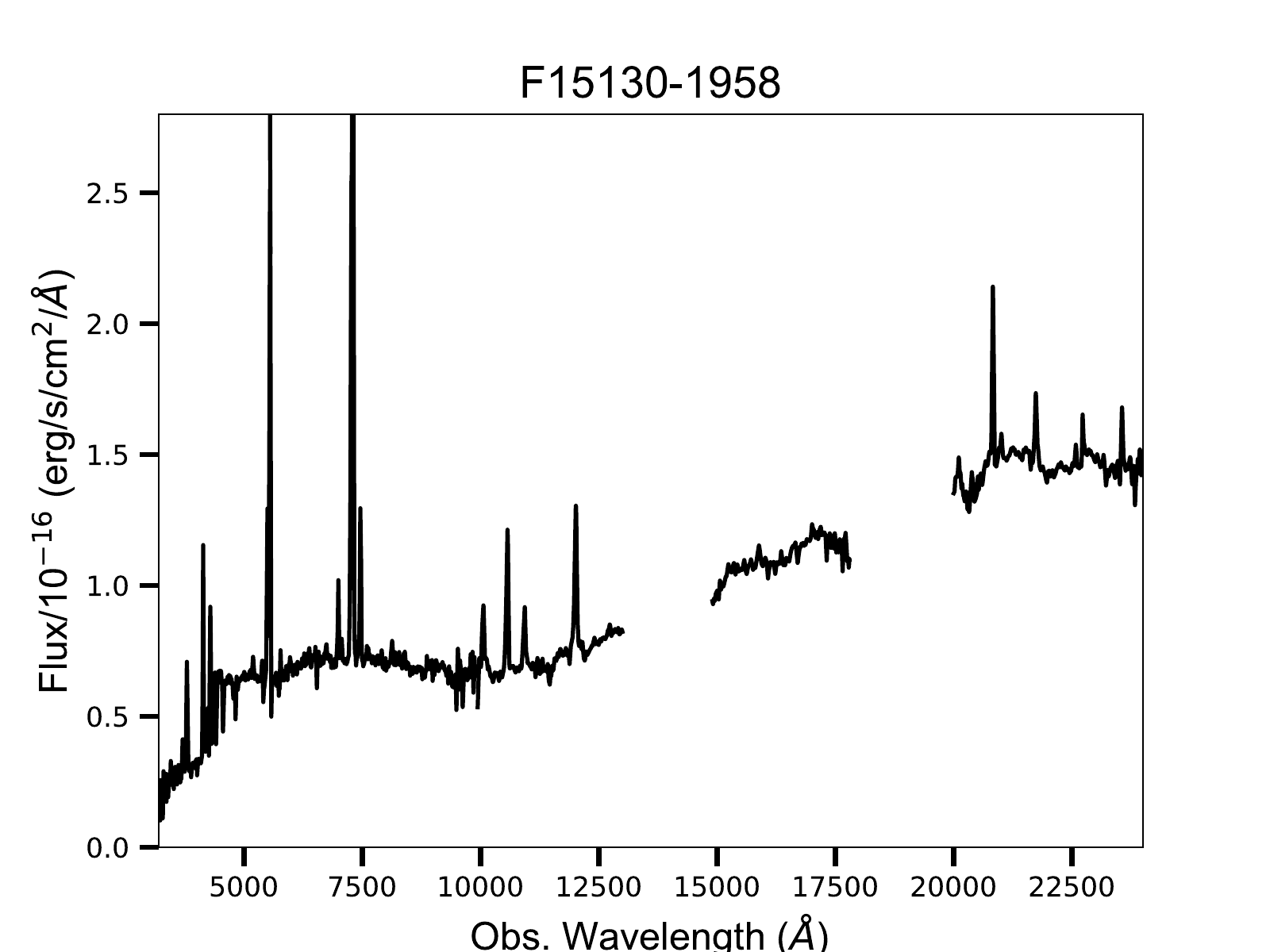}
\includegraphics[scale=0.48]{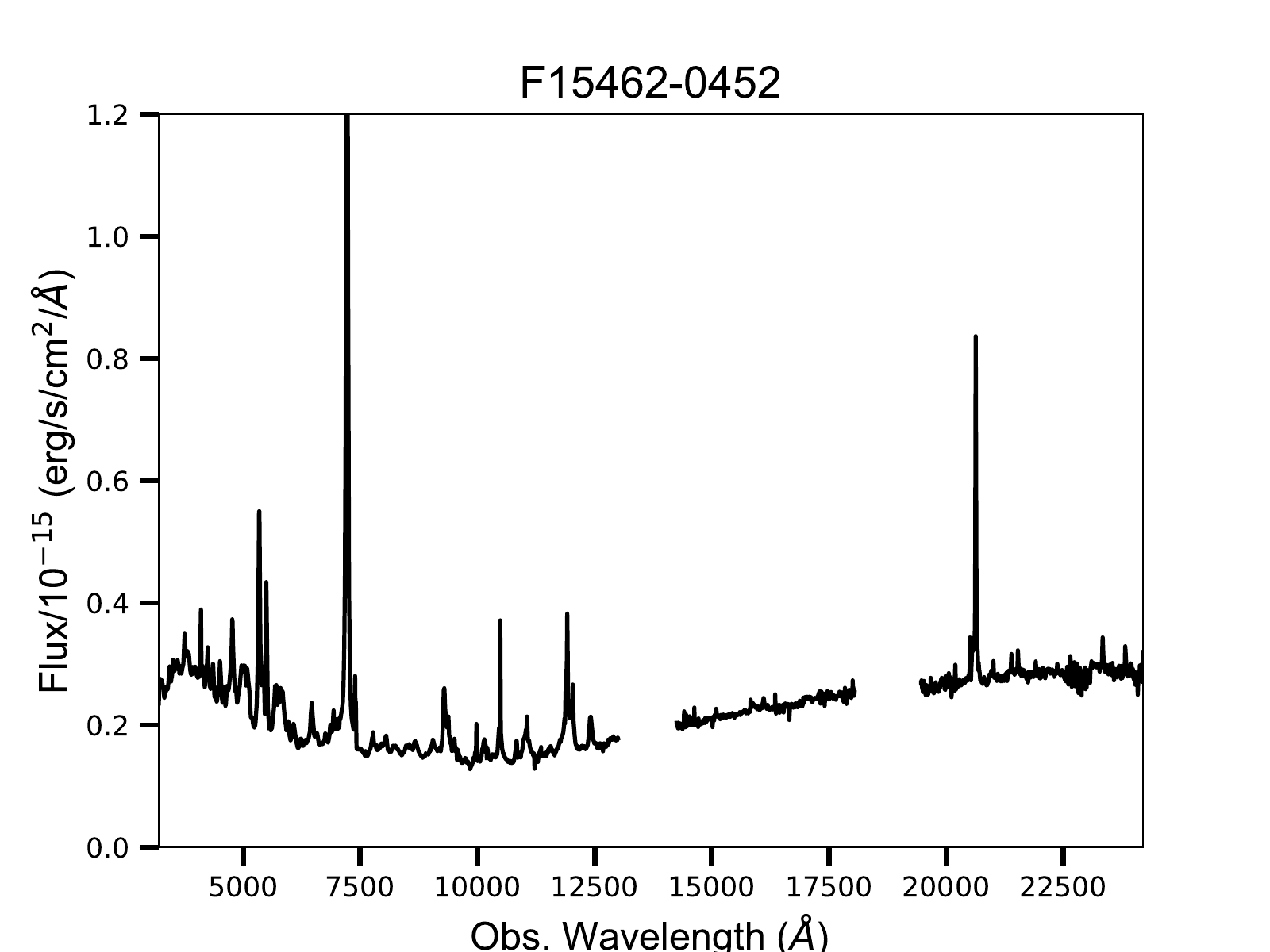}
\includegraphics[scale=0.48]{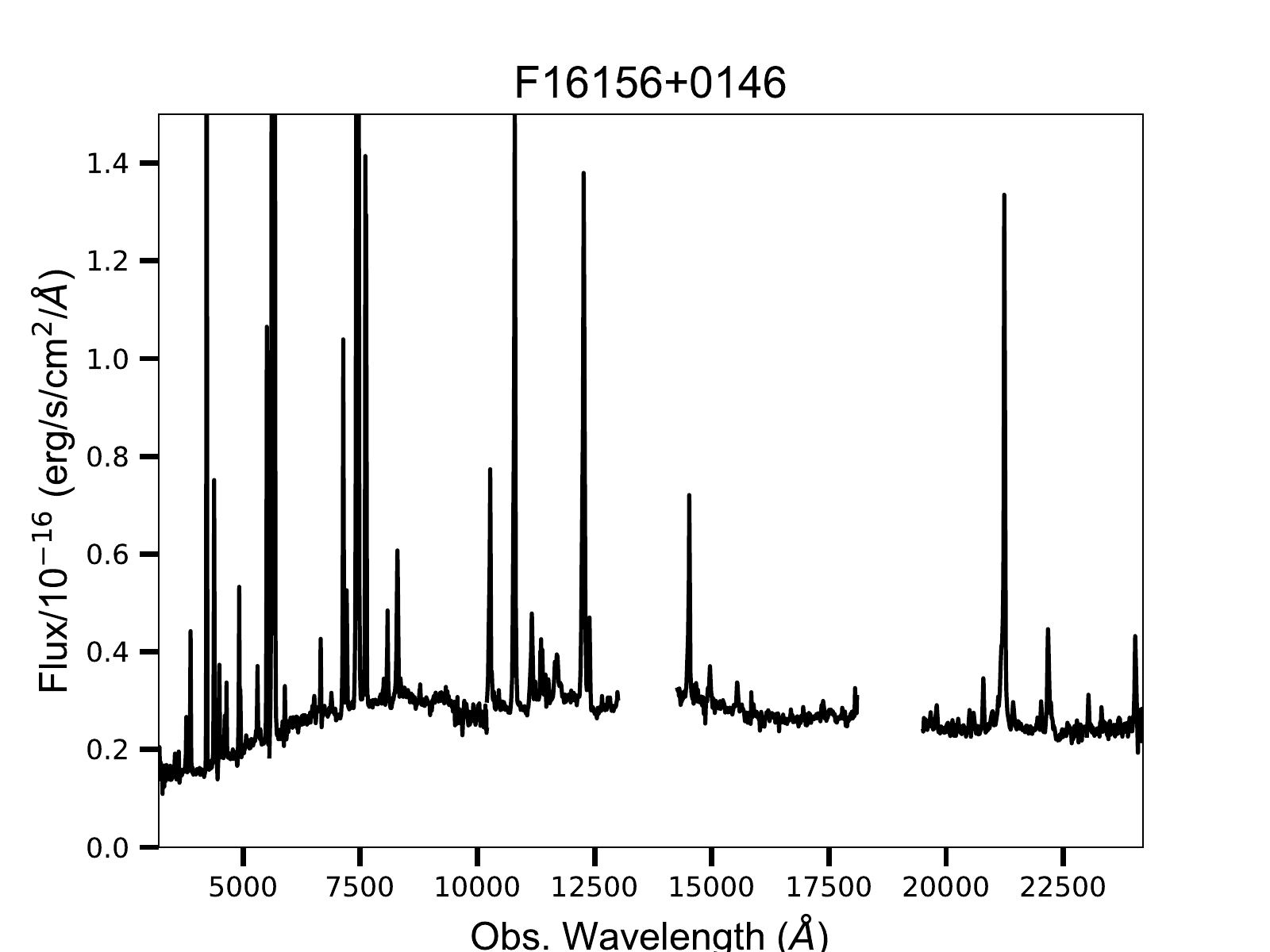}
\includegraphics[scale=0.48]{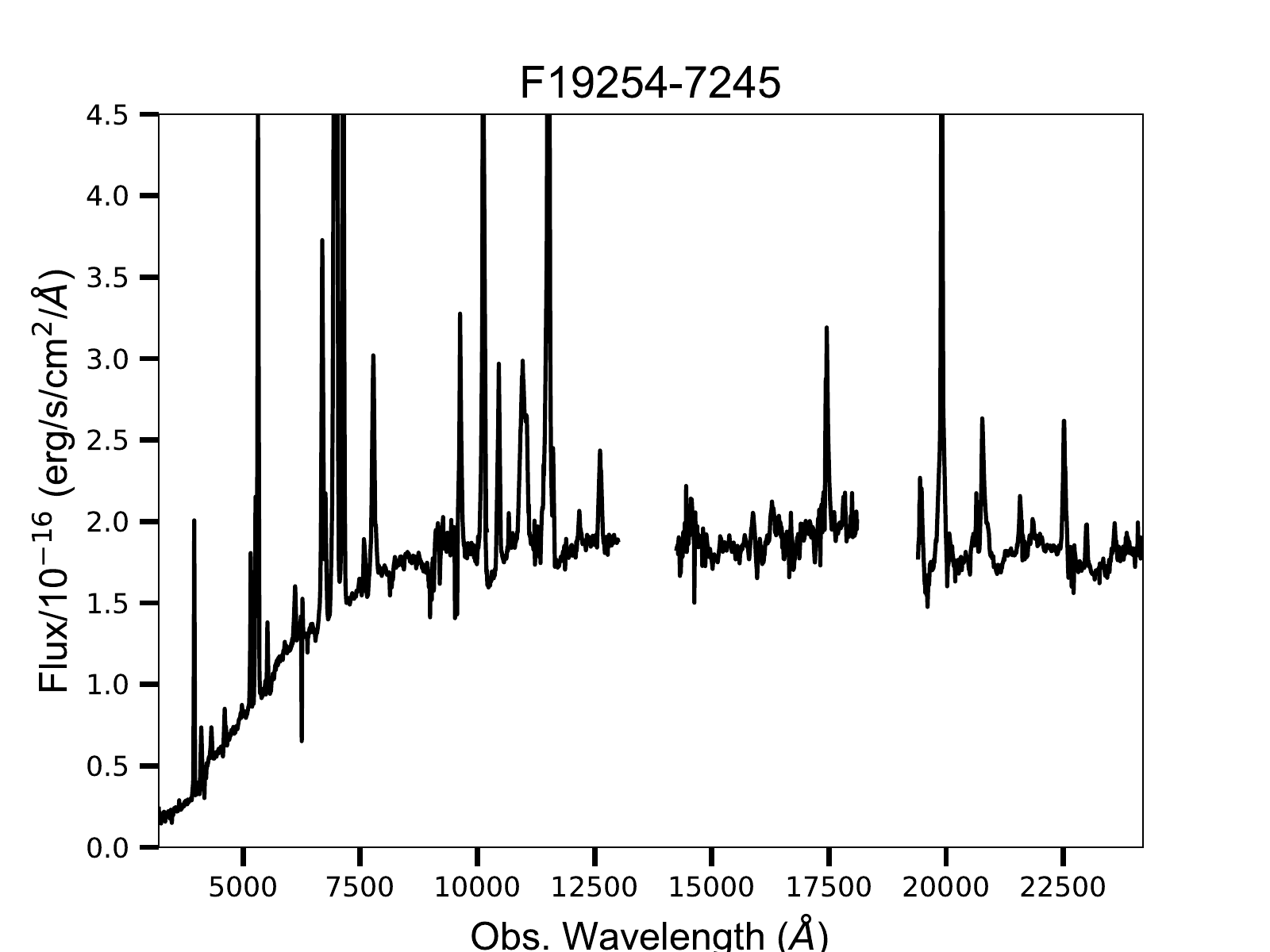}
\caption{The UVB-VIS-NIR spectrum for the remaining ULIRGs studied in this paper.}
\label{fig:specs}
\end{figure*}

For the initial stages of the data reduction process we used the Xshooter pipeline ESOREX (version 6.6.1) in physical mode \citep{freudling13}. This produced two-dimensional bias-subtracted, flat-field corrected, order merged and wavelength calibrated spectra for the ULIRG sample, as well as the flux and telluric standard stars. Note that for all objects, following the order merging in the K-band, there are two glitches in wavelength ranges 20915.4-21009.0\AA\  and 22720.8-22828.2\AA\  where the orders have not been adequately merged and straightened. These issues leads to some minor flux losses over these wavelength ranges at the extraction aperture stage.

We then used IRAF and the STARLINK FIGARO packages to perform second order corrections to improve the bad pixel, cosmic ray and night sky line interpolations/subtractions. The second order sky subtraction step was performed by using FIGARO to extract 10-pixel wide spectra above and below the target nucleus on the 2D image of the spectrum (avoiding any extended emission from the target), averaging and then subtracting them from the 2D image. While these second order sky subtractions were reasonably successful in the visual, J- and K-bands, often there were weak features left over from the subtraction in the H-band. 

The atmospheric absorption features were removed by dividing by the spectrum of a telluric standard star taken close in time and airmass to the observations of the ULIRGs. We also corrected for residual tilts in the two-dimensional spectra using the APALL routine in the NOAO package in IRAF. Before extraction, the spectra were corrected for Galactic extinction using the re-calibrated Galactic extinction maps \citep{sf11} of \citet{sfd98} with the \citet{cardelli89} extinction curve. Finally, 1D spectra were extracted.  The extraction aperture sizes were chosen to maximize the extracted nuclear emission, while minimizing the sky emission, thus achieving the best possible S/N. The sizes of the extracted apertures in the along-the-slit direction are given in Table 1.

We determined the absolute accuracy of the wavelength calibration and the instrumental width by measuring the line centres and FWHM of night sky emission sky lines from the data products before the flux calibration step. For the three nights of observations, we found that the wavelength calibration was accurate to better than 5 km s$^{-1}$ for the UVB and VIS arms, and to within better that 10 km s$^{-1}$ for the NIR arm.  Averaged across the three nights we found instrumental FWHM -- a measure of the typical spectra resolution --  of 72.7$\pm$1.5 km s$^{-1}$, 45.8$\pm$0.8 km s$^{-1}$, 69.2$\pm$1.3 km s$^{-1}$, 70.1$\pm$0.9 km s$^{-1}$ and 72.3$\pm$1.3 km s$^{-1}$ for the UVB, VIS, NIR-J, NIR-H and NIR-K arms respectively, as calculated at the centres of the 
different wavelength regions. 

To test the accuracy of the flux calibration we reduced the ULIRGs F12072-0444 and F13305-1739 using all the flux standard stars observed for their respective nights (see Appendix A). Over the full (UVB+VIS+NIR) wavelength range we find relative flux calibration accuracies of $\pm$8\% and $\pm$6\% for the nights of 12-05-2013 and 13-05-2013 respectively. For the night of 12-05-2013 we used the calibration curves from the standard stars EG274 and LT3218, whereas for the night of 13-05-2013 we used the calibration curves from the standard stars EG274, LT3218 and LTT987. For the objects observed on the night of 19-05-2013 (F15130-1958 and F19254-7245S) we used the master response curve provided by ESO because no standard stars were observed on that night during our run. The relative flux calibration accuracy of the master response curve is $\pm$10\%. In Figure \ref{fig:spec}  we present an example of a fully flux calibrated UV-VIS-NIR spectrum for one object -- F13305-1739 -- on an expanded scale, whereas calibrated spectra for the remaining 8 objects are presented in  Figure \ref{fig:specs}. Note that throughout this paper, when calculating the uncertainty on flux measurements from the spectra, we include both the uncertainty in the flux calibration and the uncertainties in the Gaussian fits to the emission-line profiles.

The full (UVB-VIS-NIR) spectra of the ULIRG sample objects (Figures 1 \& 2) reveal a remarkable variety in both continuum shapes and emission-line properties (from type 2 to type 1 AGN). The spectra range from objects in which the continua are dominated by relatively young, unreddened stellar populations (e.g. F13305-1739) that boost the flux at UV wavelengths, to those in which the continua appear highly reddened (e.g. F19254-7245S). 

\subsection{Host galaxy redshifts}\label{sect:z}

In order to measure the velocities of the warm outflows relative to the host galaxy rest frames, it is important to determine accurate redshifts for the host galaxies. Typically, the spectroscopic redshifts of AGN are determined using prominent emission-lines (e.g. H$\beta$, [OIII]$\lambda\lambda$4959,5007). However, the emission-lines profiles can be highly complex, leading to inaccurate redshift estimates. Indeed, the emission-lines may be dominated by highly blueshifted outflow components. For example, in PKS1549-79 the entire [OIII] emission-line profiles are blueshifted by $\sim$600 km s$^{-1}$ when compared to the [OII] emission-lines \citep{tadhunter01}. Therefore, basing the determination of the galaxy rest frames on the bright emission-lines could potentially lead to substantial underestimation
of the outflow velocities. To avoid these problems, our approach to determining the host galaxy redshifts is based on fitting single Gaussians profiles to the following stellar absorption features and averaging the results: 

\begin{itemize}

\item[-] higher order Balmer absorption lines, particularly 3771\AA\  and 3798\AA\  given that they avoid contamination from both emission-lines (e.g. [FeV], [FeVII] and [NeIII]) and ISM absorption lines (e.g. Ca H \& K);

\item[-] the MgIb$\lambda\lambda\lambda$5167,5173,5184 blend; and

\item[-] the CaII$\lambda\lambda\lambda$8498,8542,8662 triplet.

\end{itemize}

For F12072-0444 and F13451+1232W the equivalent widths of the stellar absorption features in the nuclear aperture are too low for them to be accurately measured. In these cases, we extracted spectra for extended apertures above and below the nucleus, measured the redshifts from the absorption features in these spectra, and then averaged the results from the two apertures. However, for F15462-0450 we did not detect any strong stellar features because its  type 1 AGN spectrum dominates the emission, even in the off-nuclear regions. Therefore we estimated the redshift for this object using the most redshifted narrow components of the emission-lines. This approach is justified because the redshifts of the reddest narrow components of the emission-lines for 6/8 of the 
remaining sample agree, within 3$\sigma$, with the stellar absorption line redshifts. The host galaxy redshifts for the rest of the sample are presented in Table \ref{tab:z}.

Interestingly, the most redshifted [OIII] kinematic components are significantly {\it blueshifted} relative to the stellar absorption lines in the cases of F12072-0444 and F15130-1958, by $-156\pm40$ km s$^{-1}$ and $-516\pm38$ km s$^{-1}$ respectively.  This is similar to the case of PKS1549-79 where the whole [OIII] profile is redshifted relative to the rest frame \citep{tadhunter01,holt06}, and suggests that the [OIII] profile is dominated by outflowing gas in these objects.

\begin{center}					
\begin{table}					
\centering					
\caption{Host galaxy redshifts redshifts for the ULIRGs. In all-but-one object, the redshifts were measured
using the stellar absorption lines. However, in the case of type 1 AGN F15462-0450 no stellar absorption features
were detected, and the narrow component of
the [OIII] emission-line profile was used to determine the host galaxy redshift. The `Line' column indicates which spectral features where used in the redshift determination. }
\begin{tabular}{lcl}					
\hline					
Object	&	z	&  Line		\\
\hline					
F12072-0444	&	0.12905$\pm$0.00013	& HI, CaII			\\
F13305-1739	&	0.14843$\pm$0.00011	& HI, CaII			\\
F13443+0802SE	&	0.13479$\pm$0.00015	& HI, CaII			\\
F13451+1232W	&	0.12142$\pm$0.00013	& MgIb			\\
F14378-3651	&	0.06809$\pm$0.00020	& HI		\\
F15130-1958	&	0.11081$\pm$0.00011	& HI, MgIb, CaII		\\
F15462-0450     &       0.099692$\pm$0.000028	& Narrow [OIII]     \\				
F16156+0146NW	&	0.13260$\pm$0.00026	& MgIb, CaII		\\
F19254-7245S	&	0.06165$\pm$0.00012	& HI, CaII			\\					
\hline					
\end{tabular}					
\label{tab:z}					
\end{table}					
\end{center} 					

\subsection{Stellar continuum subtraction}

For the purposes of measuring accurate emission-line fluxes, the underlying stellar continua were modelled and subtracted using the STARLIGHT spectral synthesis code \citep{cid05}\footnote{Note that we used version 04 of the STARLIGHT code \citep{mateus06} with the \citet{bruzual03} solar metallicity stellar templates provided as part of the STARLIGHT download.}. In order to produce the most accurate possible fits to the stellar continua, we masked out any spectral features that are not related to starlight (e.g. AGN emission-lines, telluric features and poorly subtracted night sky lines), while leaving key stellar absorption features such as the higher order Balmer series convergence, G-band, Mg Ib and the CaII triplet. The optical spectra were fitted over the wavelength  range 3200-9000\AA , where 9000\AA\  is the longest wavelength the stellar templates covered. In addition, a normalizing window was chosen which is free of strong emission/absorption features. This window had a rest wavelength range of 4740-4780\AA\  in all cases.

\begin{figure*}
\centering
\includegraphics[scale=0.62, trim=0cm 0cm 0cm 0cm]{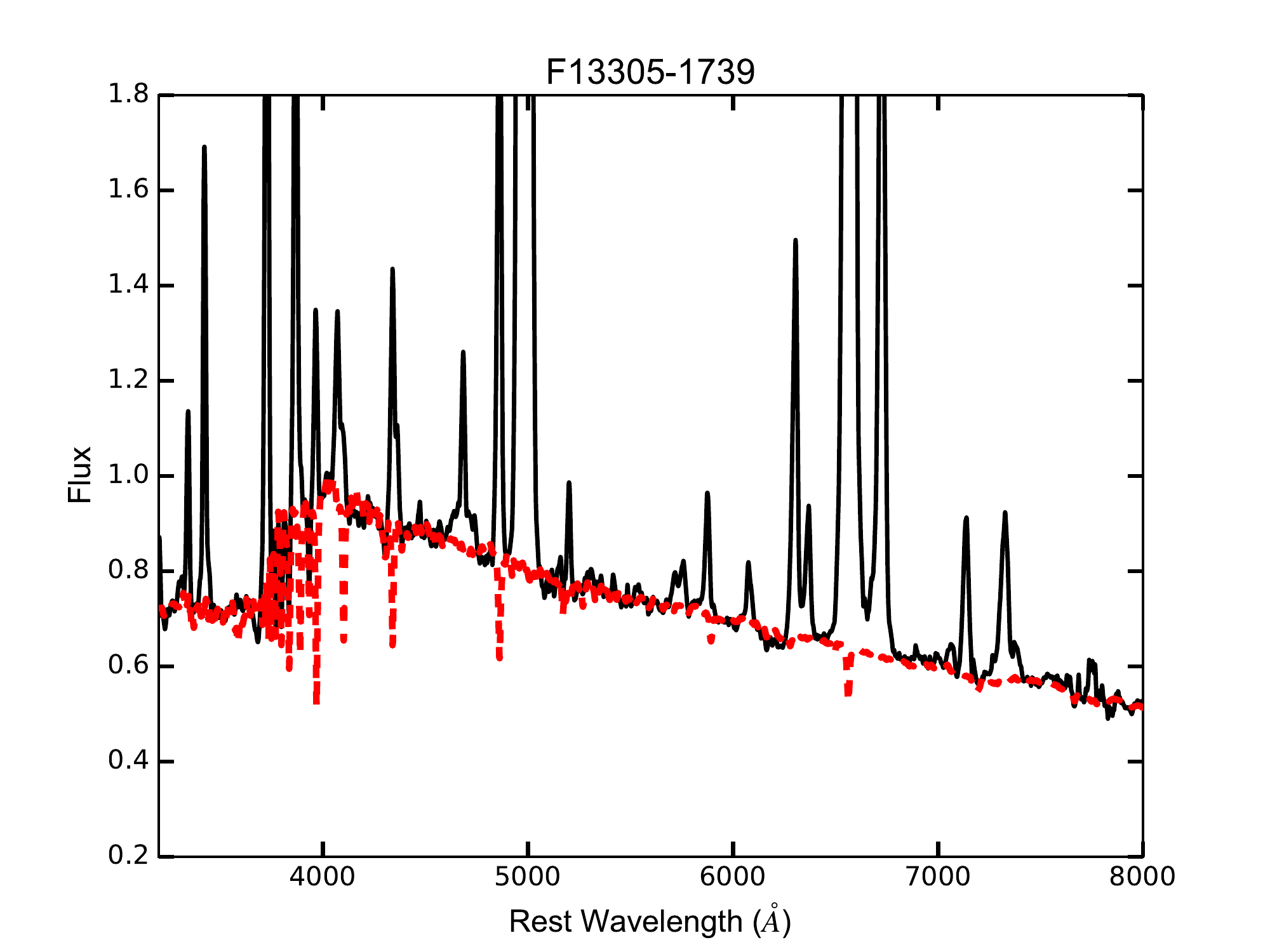}
\caption{A best fit STARLIGHT model for F13305-1739 ($\bar{\chi ^{2}}$=0.67). The dashed red line represents the stellar populations from the modelling, the black line represents the Xshooter spectrum. The flux scale is measured in units of 10$^{-16}$ erg s$^{-1}$ cm$^{-2}$ \AA $^{-1}$, and the observed wavelength is measured in units of \AA .}
\label{fig:sub}
\end{figure*}

An example of the continuum fitting for F13305-1739 is presented in Figure \ref{fig:sub}. To determine whether the continuum fits were suitable, we performed a visual inspection of fits to the Balmer series and Balmer break, Mg Ib and the CaII triplet stellar absorption features. Once we determined that the continuum fits were adequate, we subtracted the model results from the spectra. We could not successfully model the stellar population for F15462-0450. This is because its spectrum is dominated by the emission of its type 1 AGN rather than the stellar population of its host galaxy. Note that the purpose of the continuum fits was to subtract the stellar features and not to estimate the ages or metallicities of the stellar populations. This is because the contribution from scattered AGN light may give misleading ages for the young stellar populations. Nevertheless, in Table \ref{tab:spfits} we present the results from the STARLIGHT fits to the continua of the ULIRG sample. Note that we divide the stellar populations into three categories: a young stellar population (YSP) with ages $\tau_{age}$ $<$ 100 Myr, an intermediate stellar population (ISP) with ages 100 Myr $<$ $\tau_{age}$ $<$ 2 Gyr, and an old stellar population (OSP) with ages $\tau_{age}$ $>$ 2 Gyr. 

\begin{center}					
\begin{table}					
\centering					
\caption{Results from the STARLIGHT fits to the continua of the ULIRG sample as a light fraction over the rest wavelength interval 4740-4780$\AA$. The stellar populations are divided in to young stellar population (YSP) with ages $\tau_{age}$ $<$ 100 Myr, intermediate stellar population (ISP) with ages 100 Myr $<$ $\tau_{age}$ $<$ 2 Gyr, and old stellar population (OSP) with ages $\tau_{age}$ $>$ 2 Gyr.}
\begin{tabular}{lccc}					
\hline			
Name	&	YSP (\%)	&	ISP (\%)	&	OSP (\%)	\\
\hline							
F12072-0444	&	77.9	&	-	&	22.1	\\
F13305-1739	&	42.8	&	40.7	&	16.5	\\
F13443+0802SE	&	7.7	&	7.6	&	84.7	\\
F13451+1232W	&	36.8	&	-	&	63.2	\\
F14378-3651	&	19.9	&	73.6	&	6.5	\\
F15130-1958	&	15.2	&	66.1	&	18.7	\\
F16156+0146NW	&	46.8	&	-	&	53.2	\\
F19254-7245S	&	23.9	&	-	&	76.1	\\
\hline					
\end{tabular}					
\label{tab:spfits}					
\end{table}					
\end{center}

\subsection{Fitting the emission-lines}\label{sect:gfits}

\begin{figure*}
\centering
\includegraphics[scale=0.35]{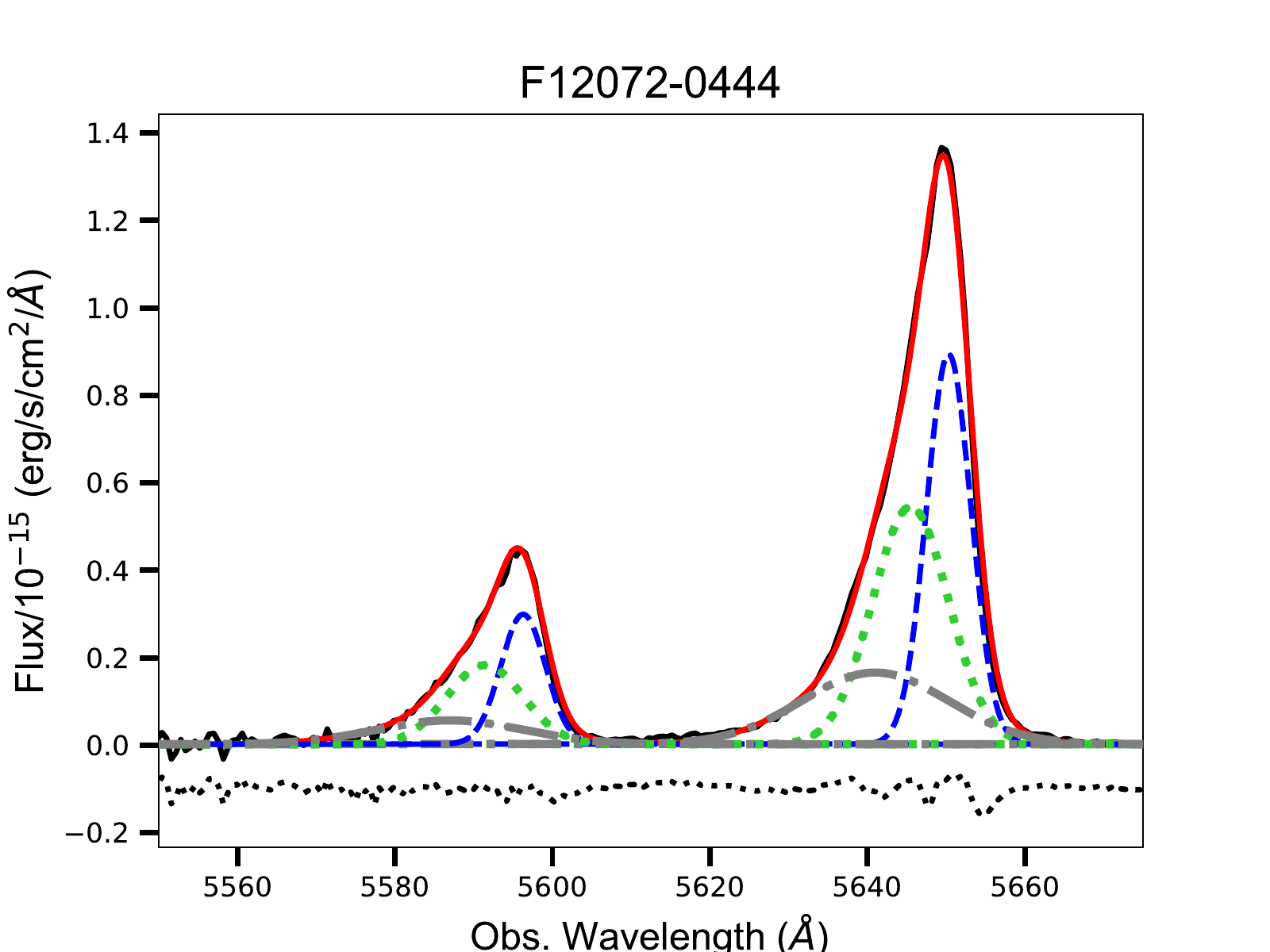}
\includegraphics[scale=0.35]{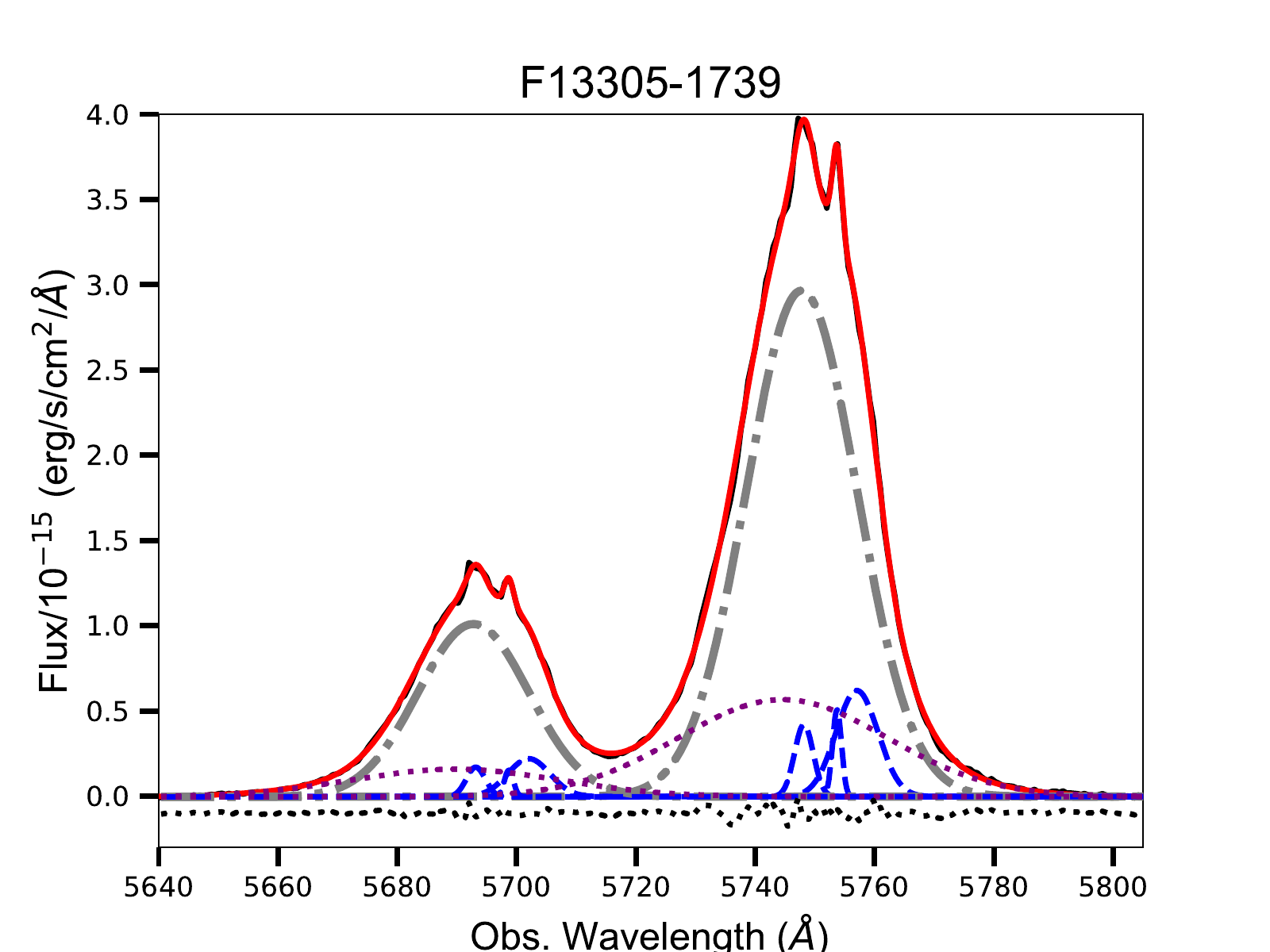}
\includegraphics[scale=0.35]{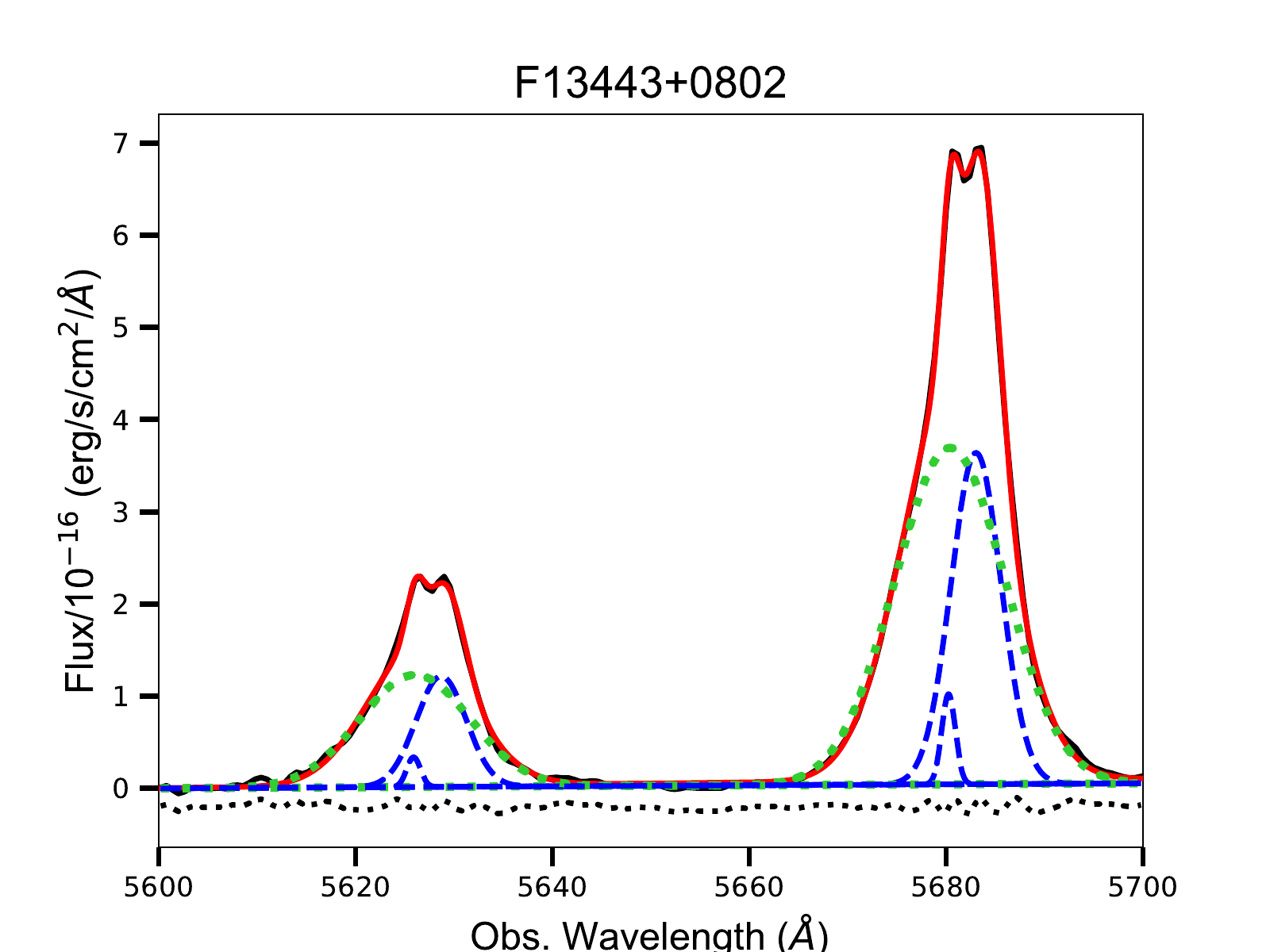}
\includegraphics[scale=0.35]{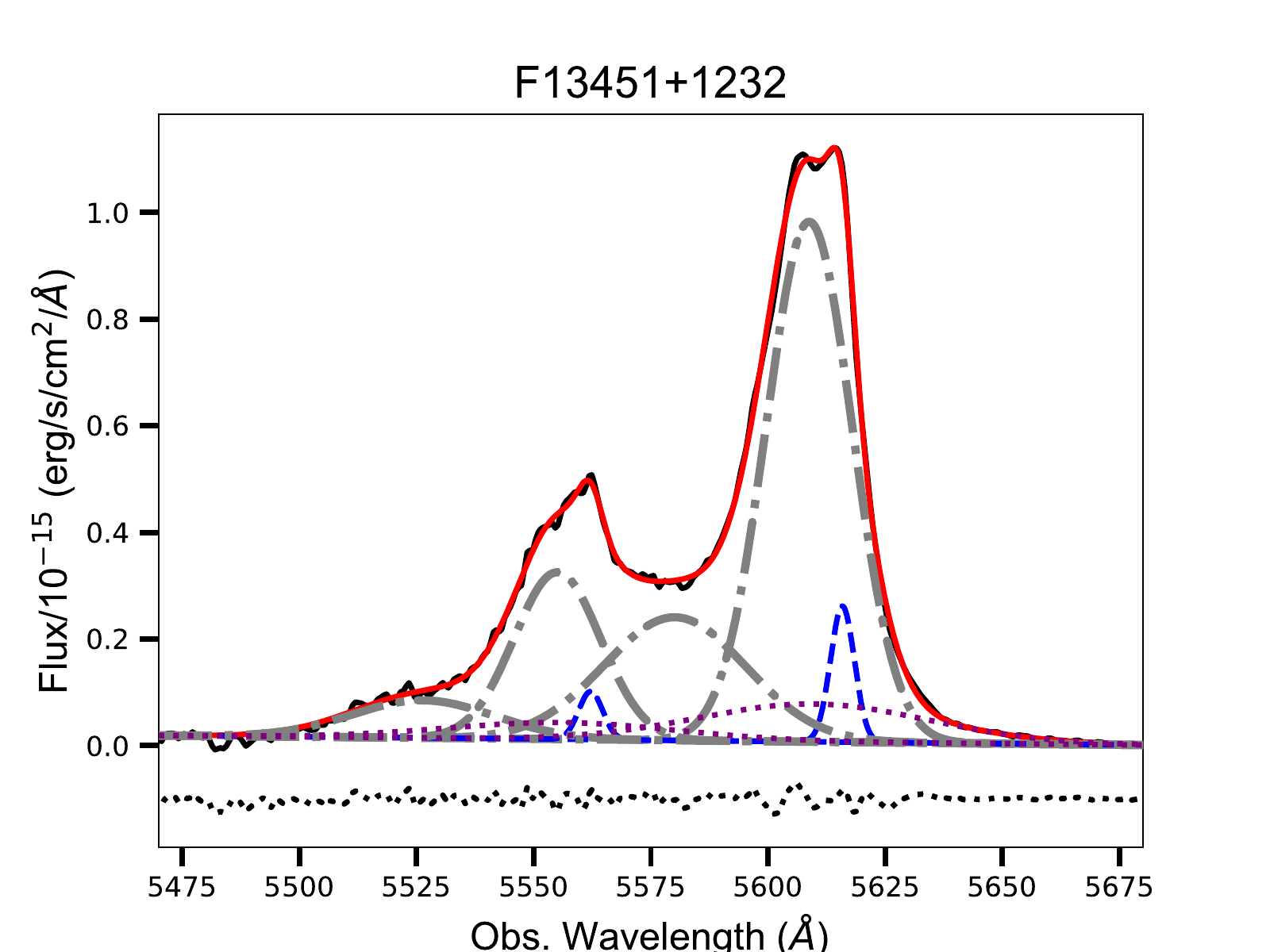}
\includegraphics[scale=0.35]{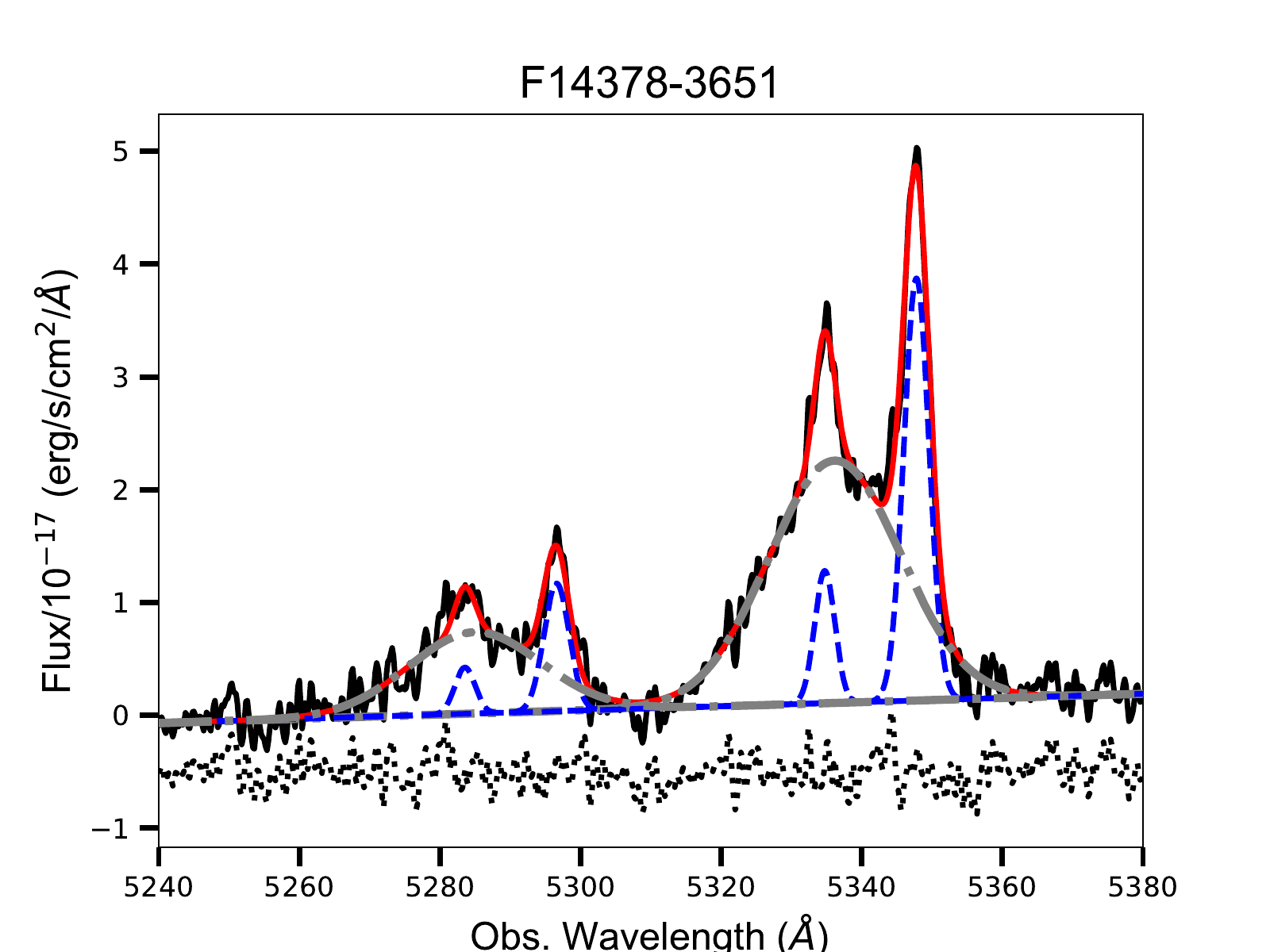}
\includegraphics[scale=0.35]{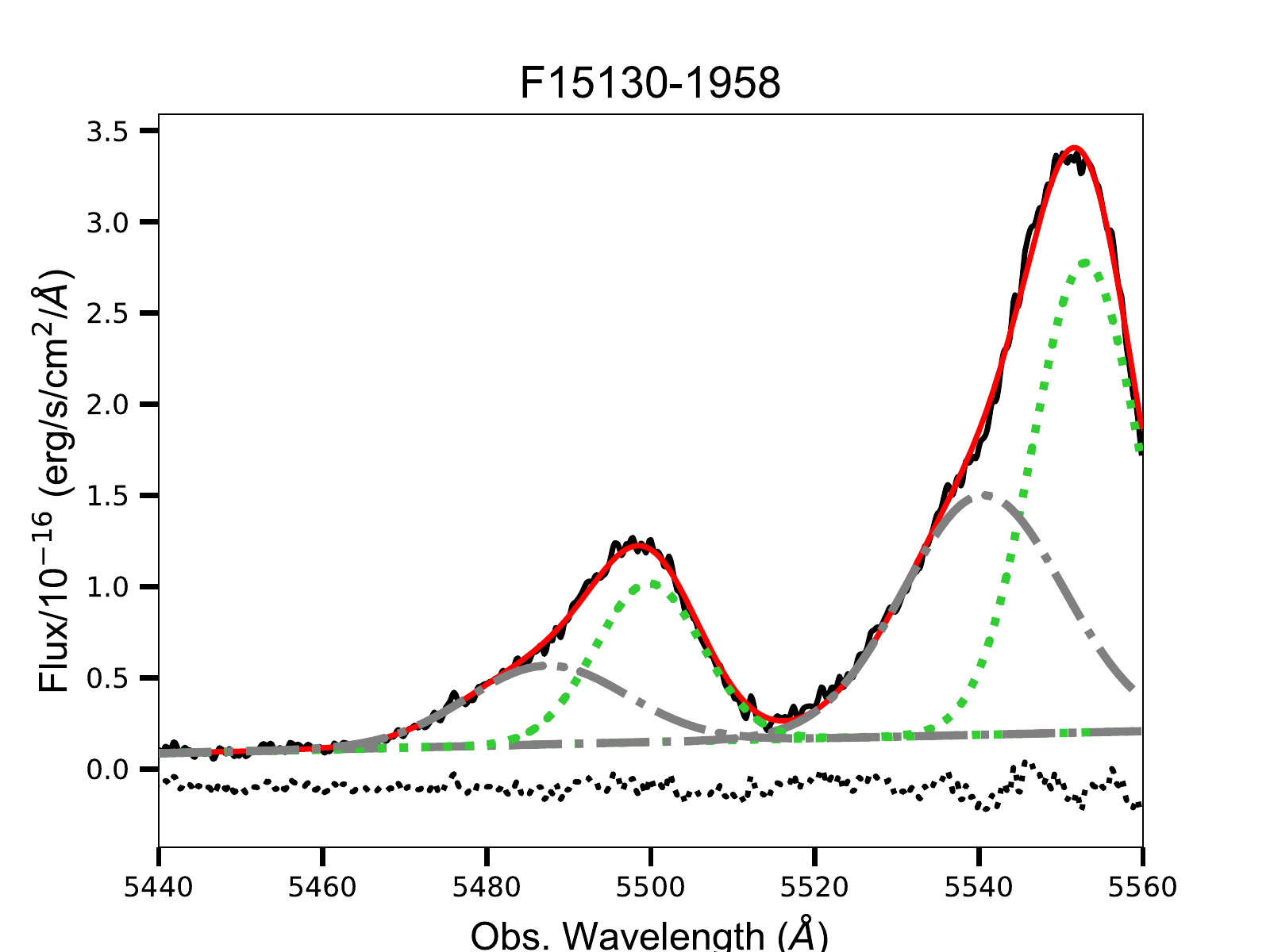}
\includegraphics[scale=0.35]{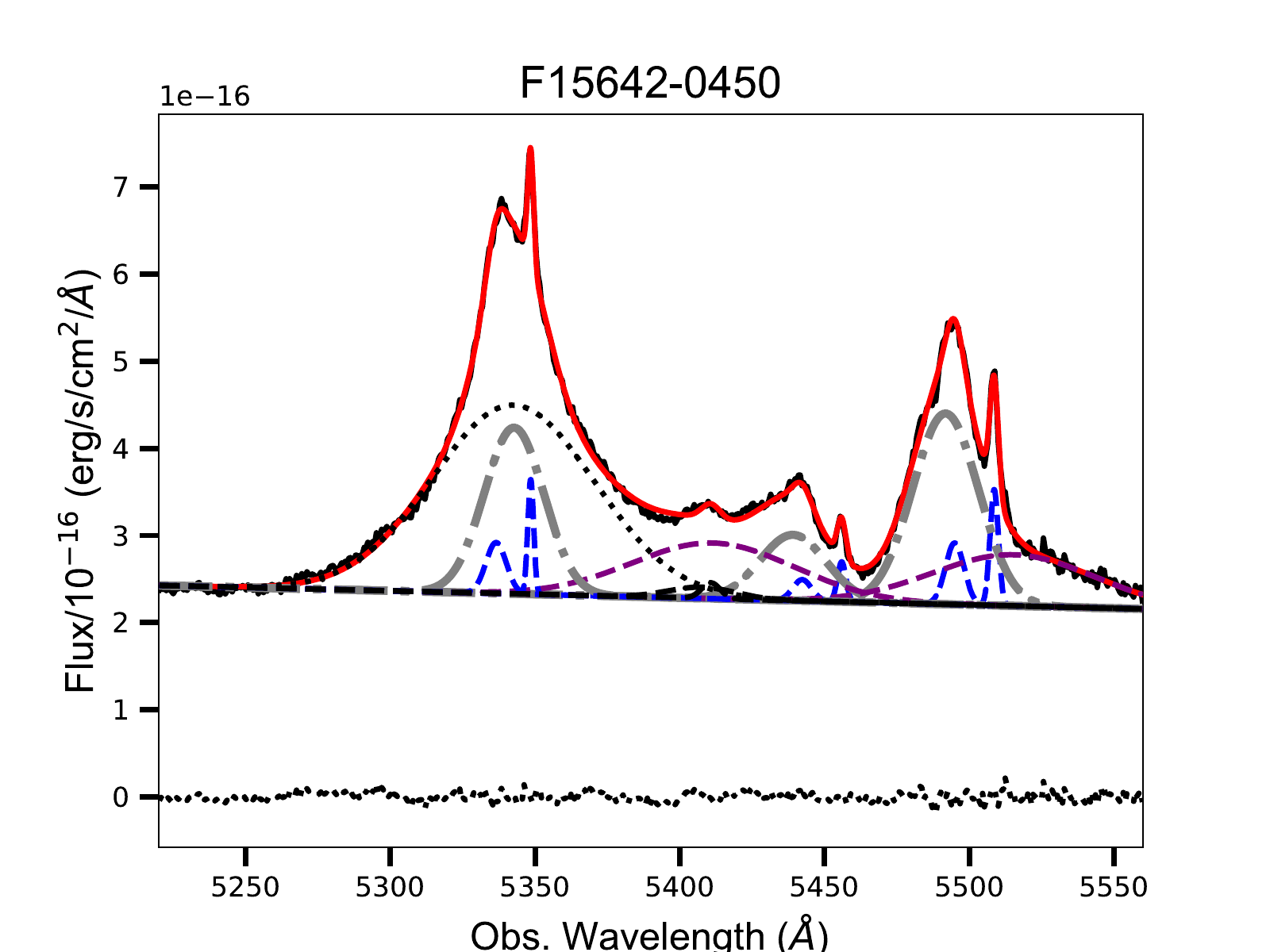}
\includegraphics[scale=0.35]{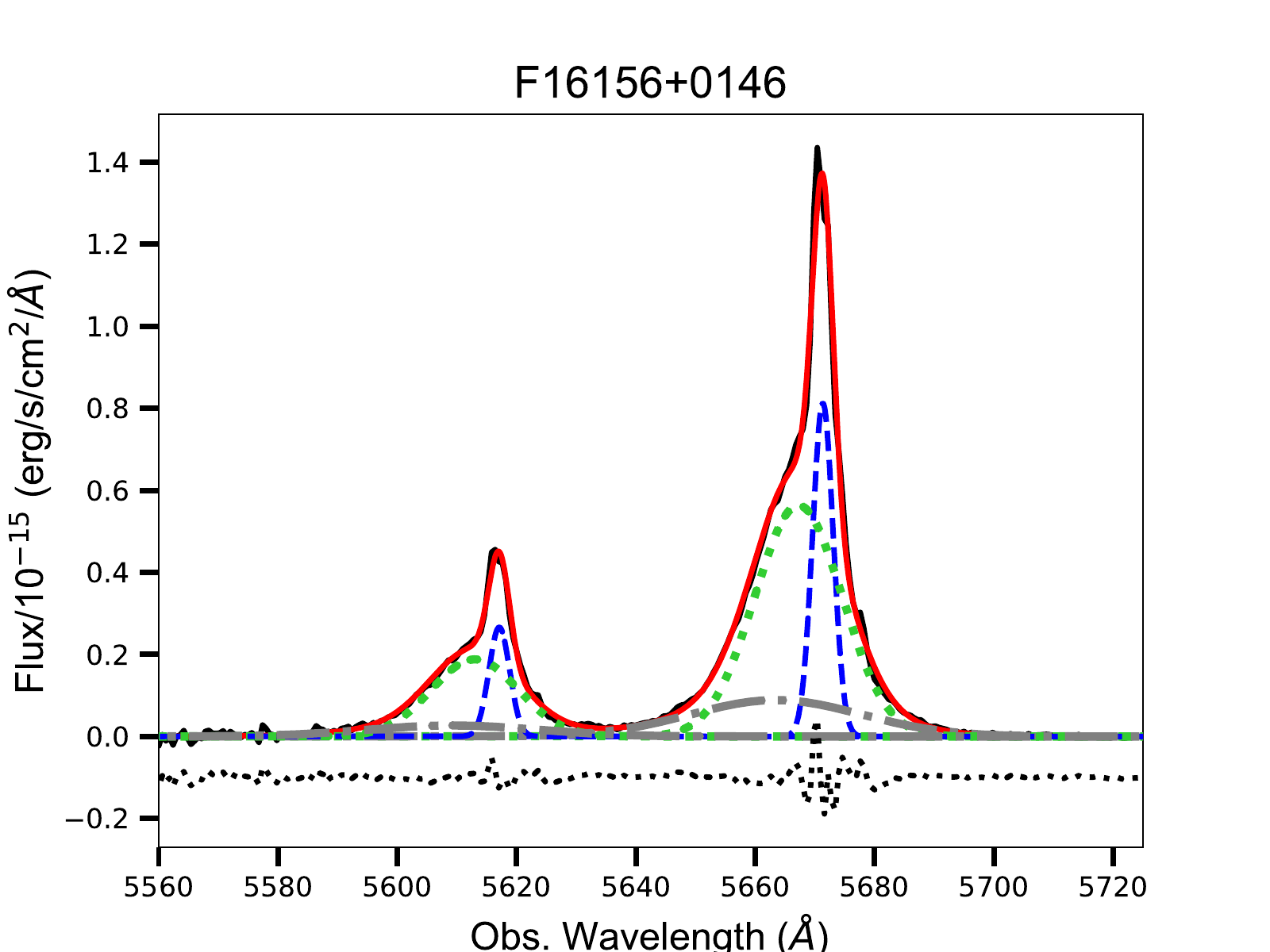}
\includegraphics[scale=0.35]{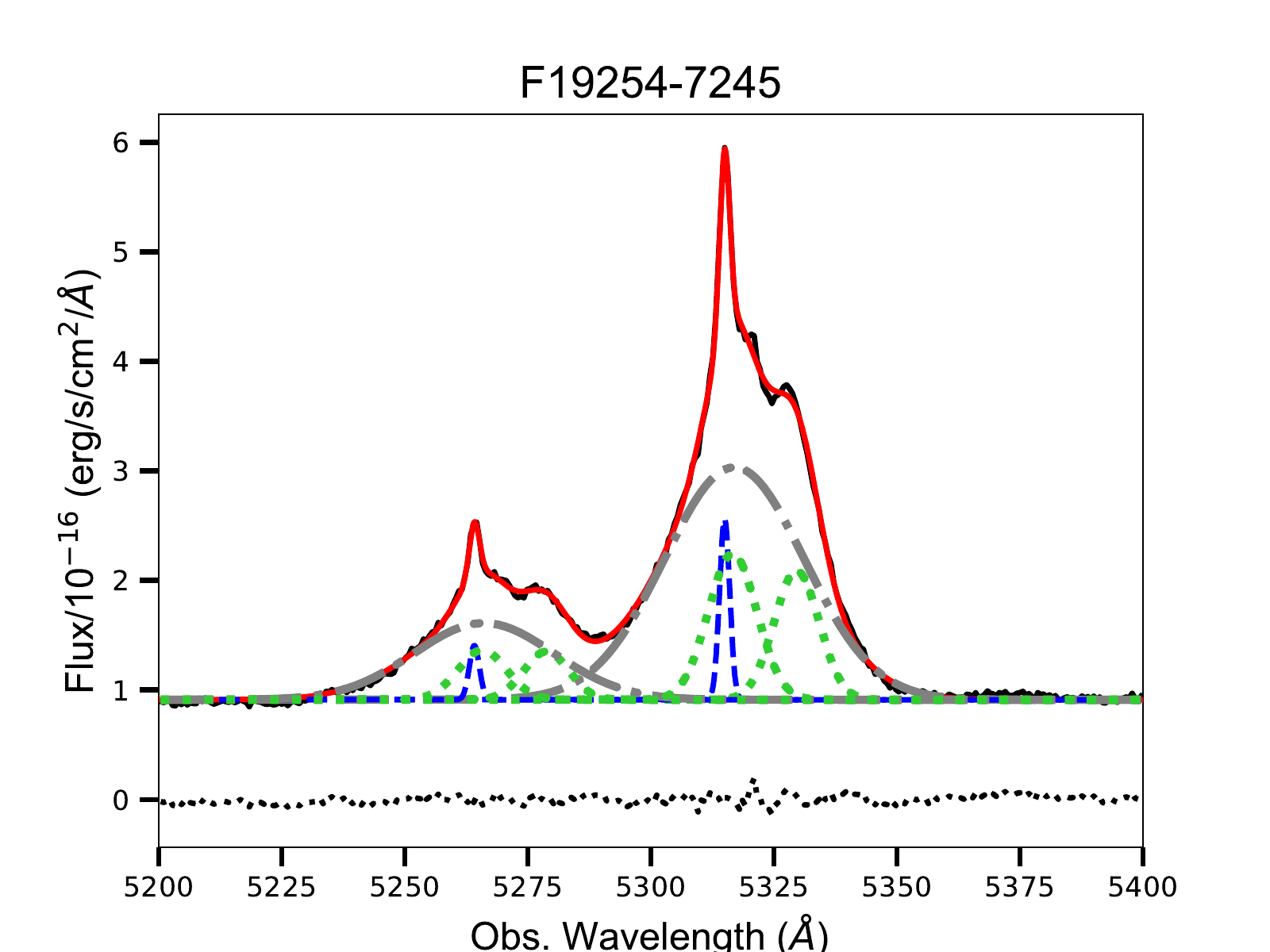}
\caption{[OIII]$\lambda\lambda$4959,5007 profiles of individual sources (solid black line). Overall Gaussian model fits are shown as solid red lines, narrow velocity components as dashed blue lines, intermediate velocity components as dotted green lines, broad velocity components as dashed and dotted grey lines, very broad velocity components as dotted purple lines, BLR as dashed black lines, FeII multiplets as dashed purple lines, ``narrow'' FeII$\lambda$4930 emission as black dashed lines and fitting residuals as dotted blue lines. Note that the entire H$\beta$, [OIII]$\lambda\lambda$4959,5007 and FeII range is shown for F15462-0450, and that the red wing of the [OIII]$\lambda$5007 profile for F15130-1958 is truncated in the plot because it falls on the join between the UVB and VIS spectra.}
\label{fig:allfits}
\end{figure*}

\begin{figure*}
\centering
\includegraphics[scale=0.45]{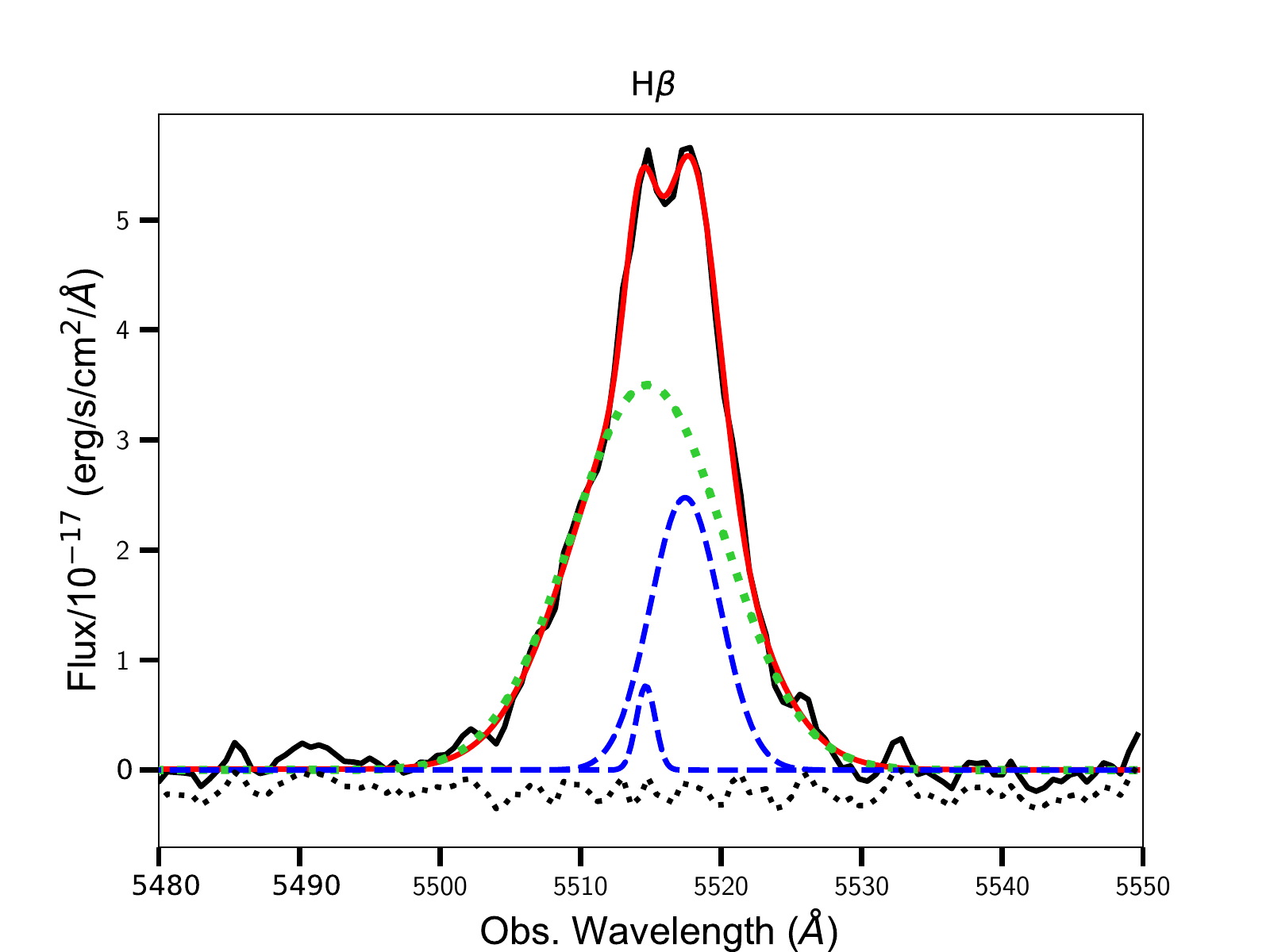}
\includegraphics[scale=0.45]{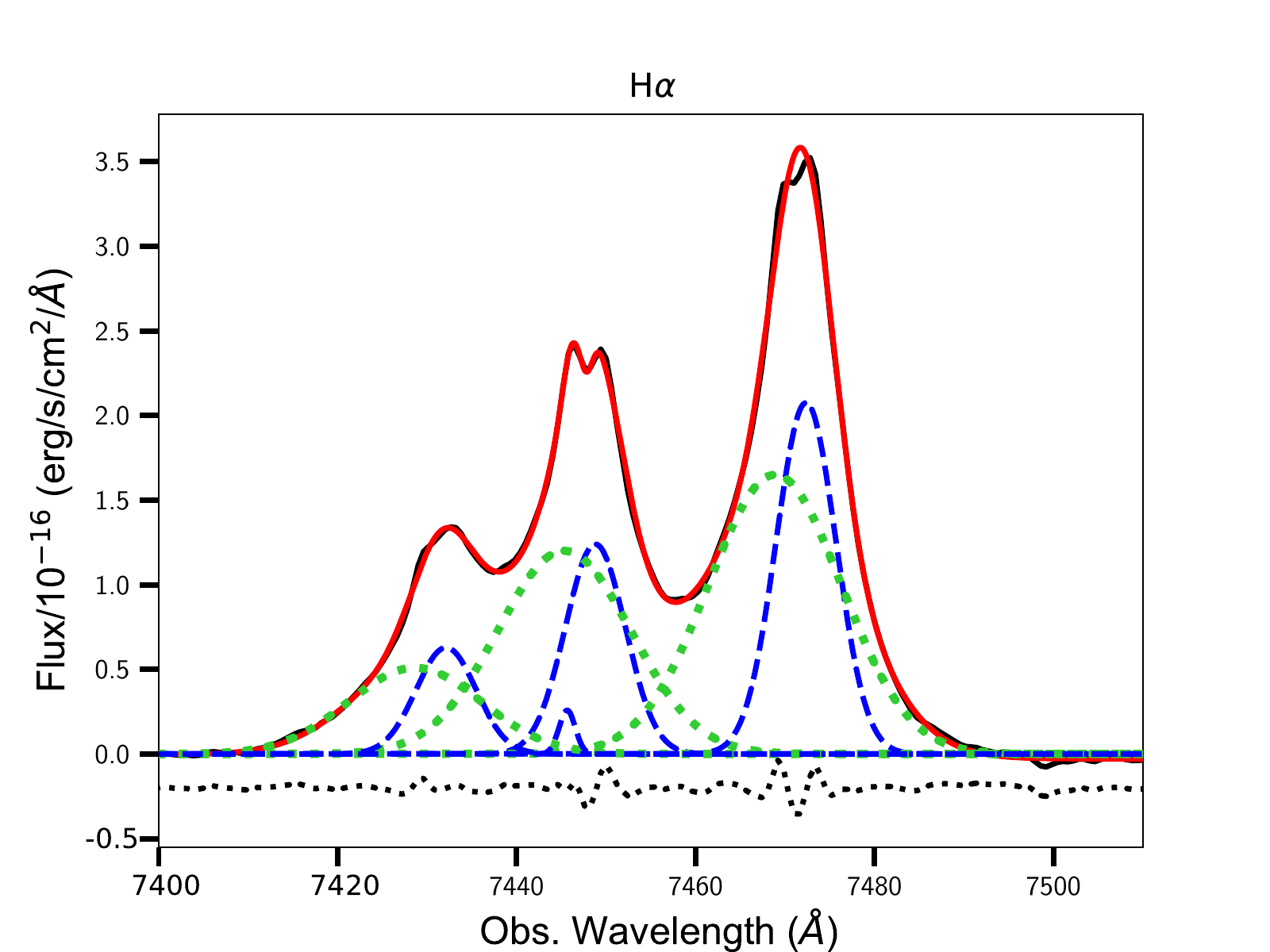}
\includegraphics[scale=0.45]{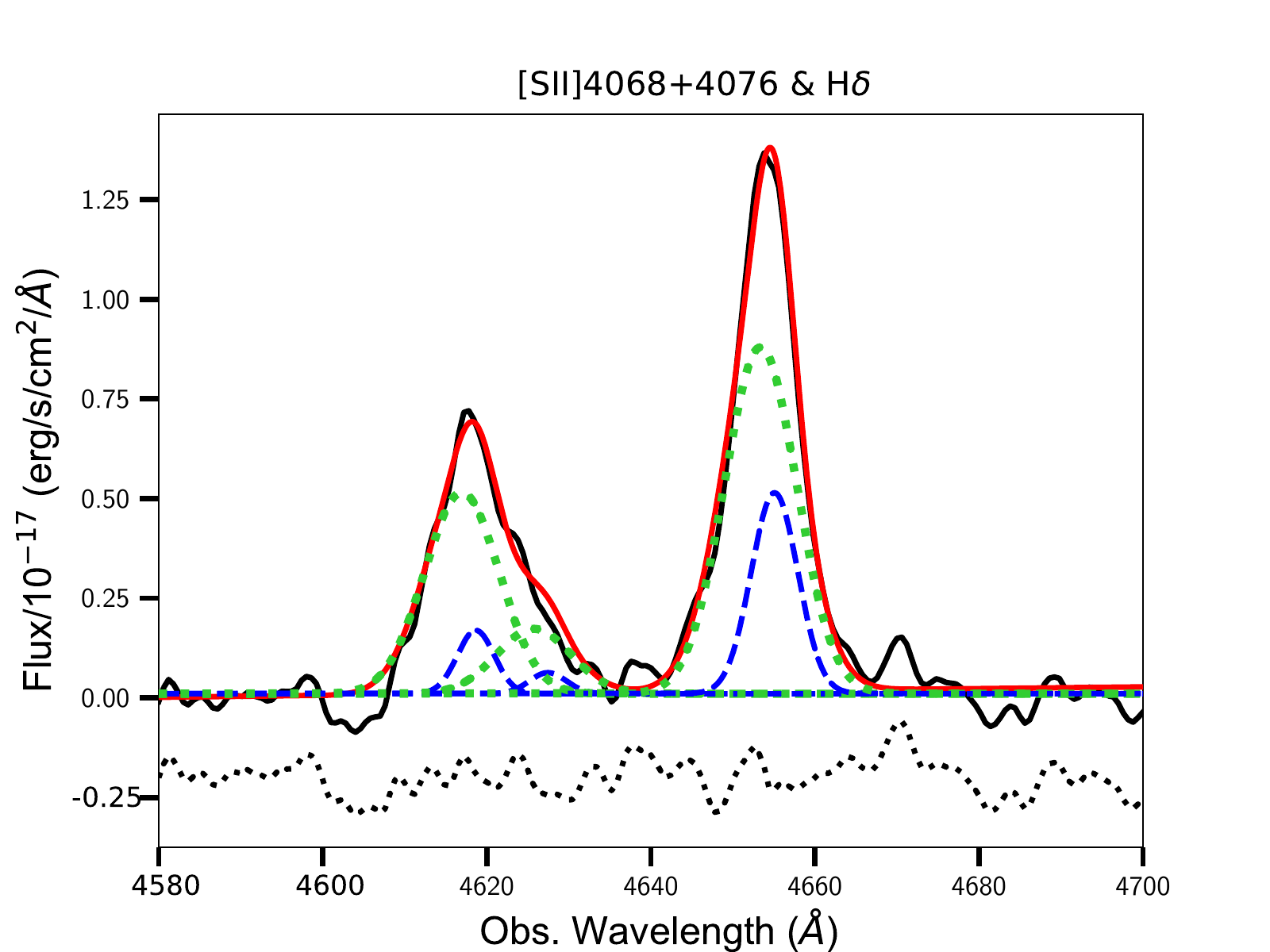}
\includegraphics[scale=0.45]{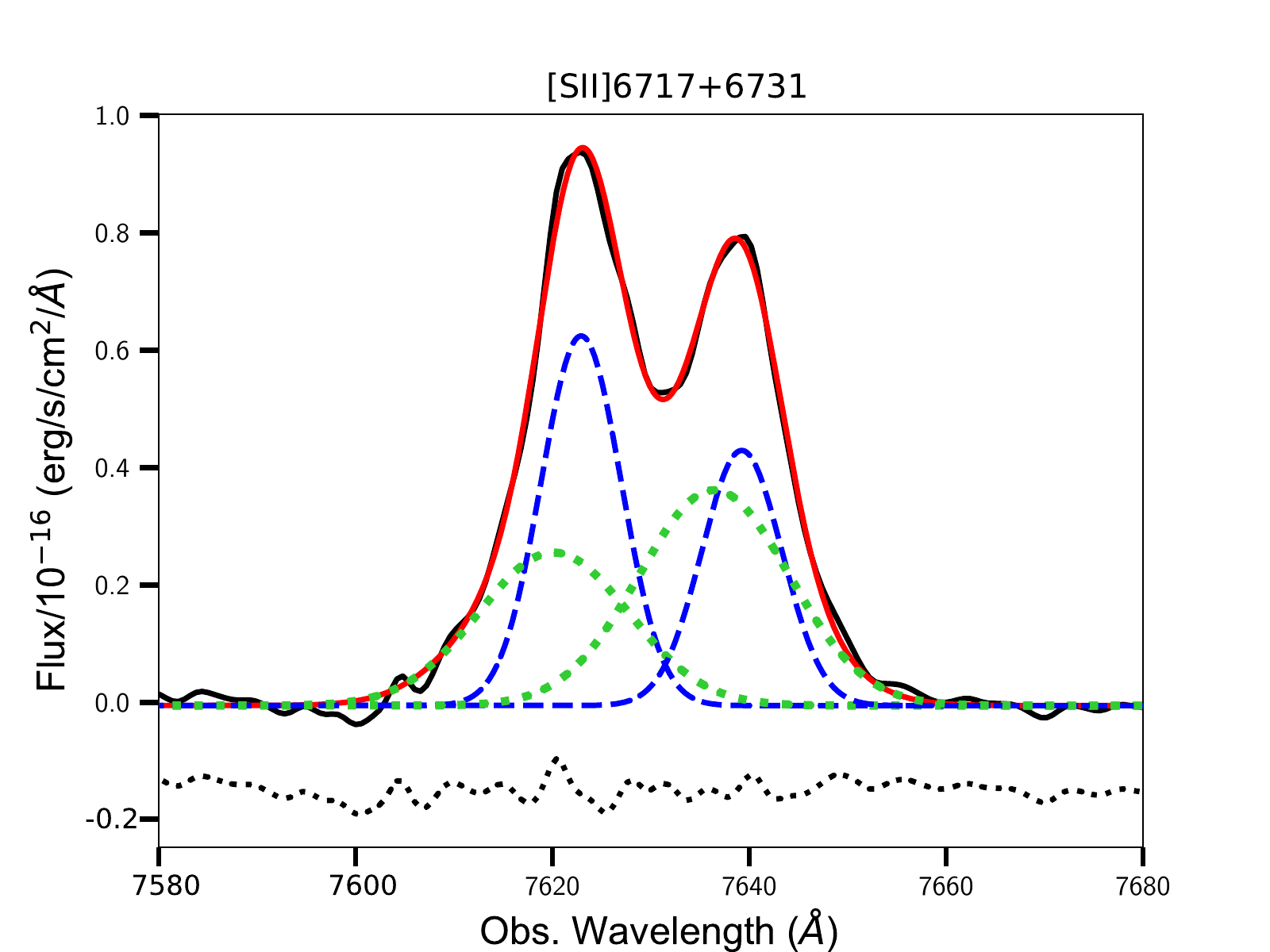}
\includegraphics[scale=0.45]{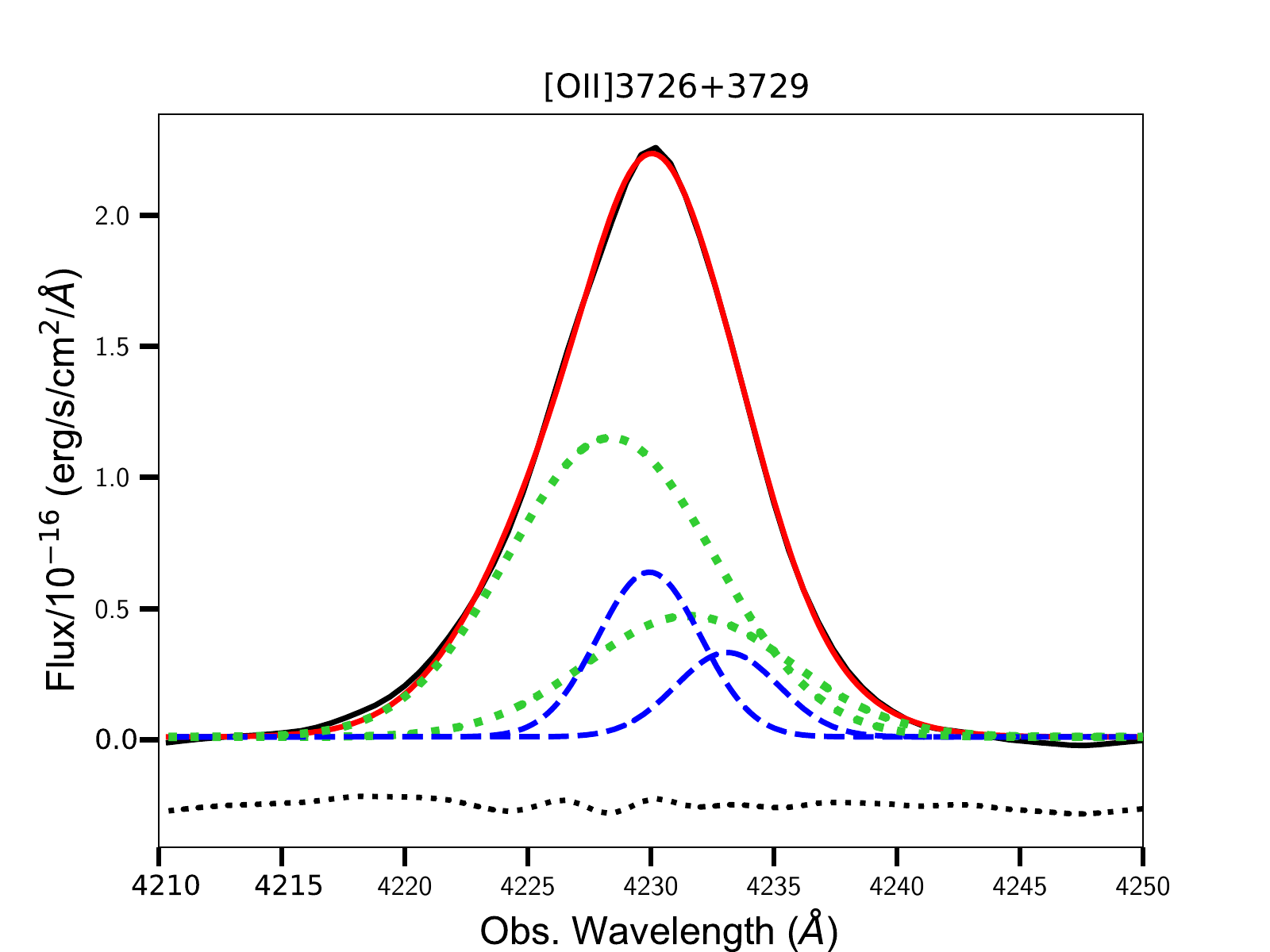}
\includegraphics[scale=0.45]{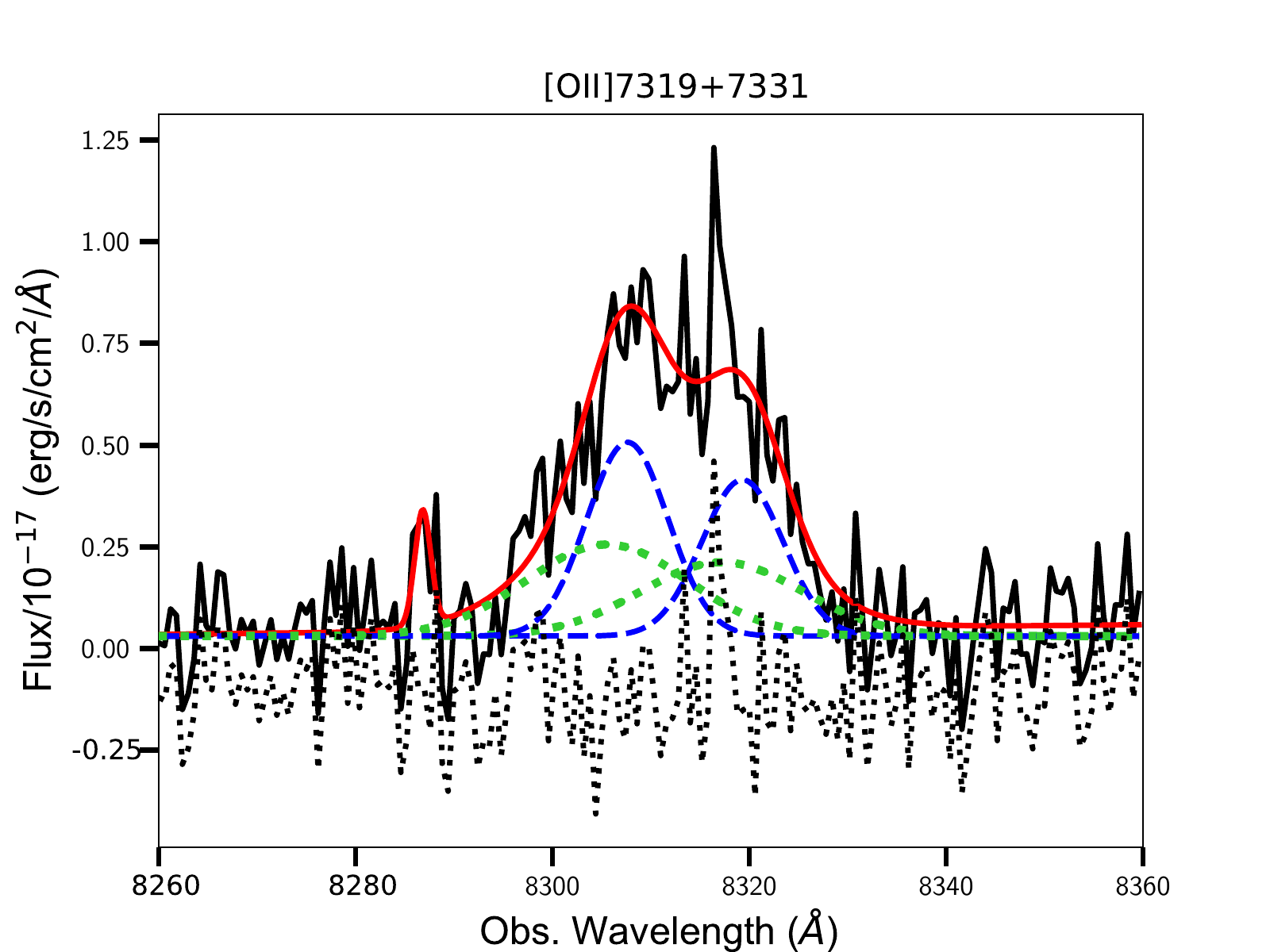}
\includegraphics[scale=0.45]{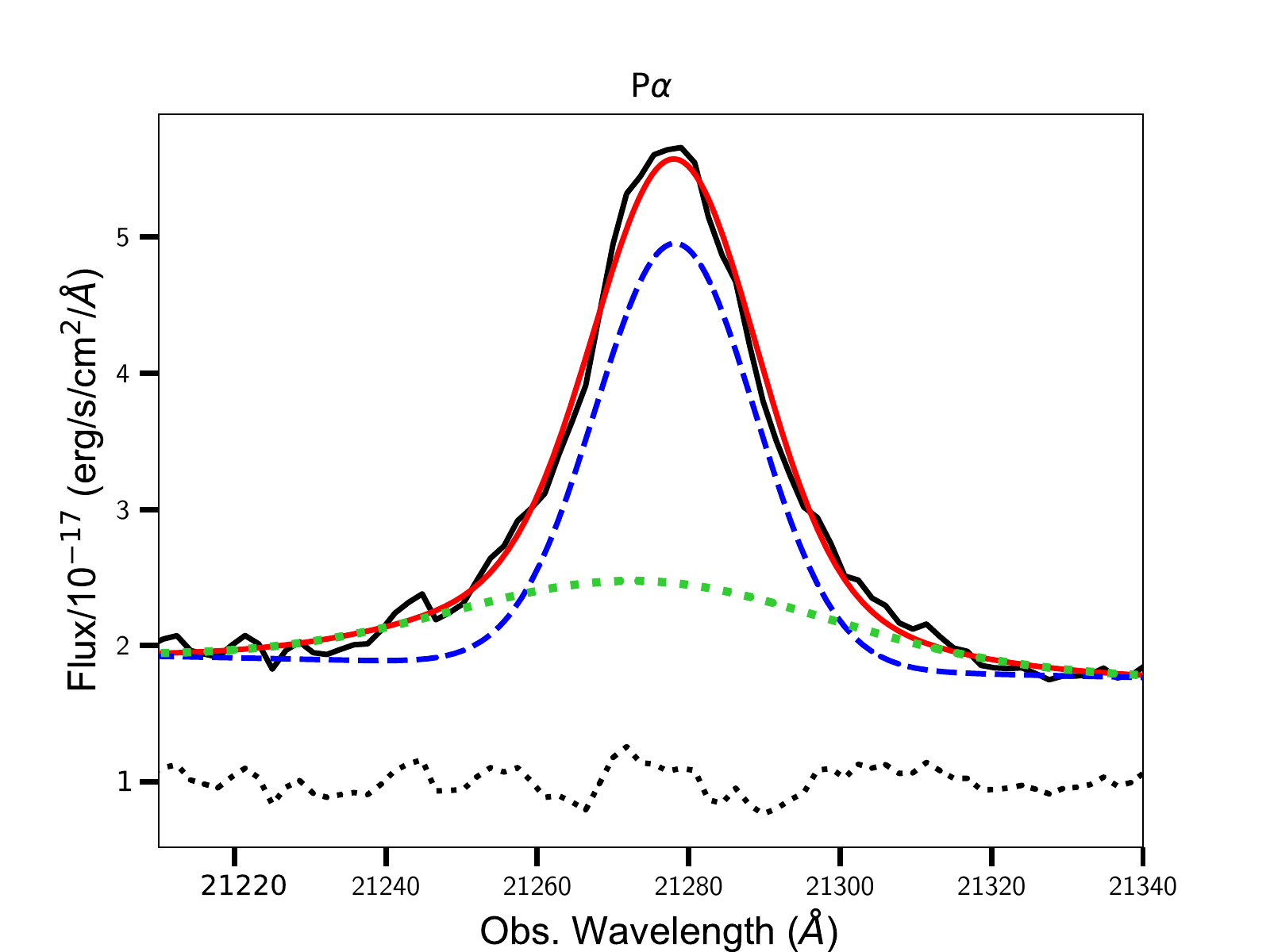}
\caption{The [OIII] model fits for F13443+0802SE applied to the H$\beta$, H$\alpha$, H$\delta$, [SII]$\lambda\lambda$4069,4076, [SII]$\lambda\lambda$6717,6731,  P$\alpha$, [OII]$\lambda\lambda$3726,3729 and [OII]$\lambda\lambda$7319,7330, emission-lines. The different line types correspond to those presented in Figure \ref{fig:allfits}. Note that for the [OII]$\lambda\lambda$7319,7330 an additional component has been included to fit a poorly subtracted cosmic-ray line.}
\label{fig:modfits}
\end{figure*}

The fluxes and kinematics of the emission-lines were measured by fitting multiple-Gaussian models using the Starlink package {\sc dipso}. Our approach was to use the minimum number of Gaussians necessary to provide an adequate fit the emission-line profiles without leaving significant features in the residuals.  We started by fitting the [OIII]$\lambda\lambda$5007,4959
doublet emission-lines,  which are generally among the brightest in the spectra, and whose  doublet separation is large enough to 
allow accurate measurement of all the kinematic components, including those that are highly blue- or redshifted. When fitting the [OIII]$\lambda\lambda$5007,4959 lines for a given kinematic component, we constrained the two lines to have the same FWHM, a 1:3.0 intensity ratio and a fixed separation set by atomic physics \citep{osterbrock06}. The fits to the [OIII]$\lambda\lambda$4959,5007 emission-lines of all the target objects are shown in Figure \ref{fig:allfits}, and the [OIII] Gaussian fitting parameters are given in Table \ref{tab:gfits}, where the velocity shifts ($\Delta$v) and line widths (FWHM) of the kinematic components are measured in the host galaxy rest frames (see $\S$\ref{sect:z}). In many cases multiple Gaussian components were required to fit the [OIII] line profiles. For consistency, we use the same scheme to label kinematic components as described in \citet{javier13}:

\begin{enumerate}
\item narrow (N): FWHM $<$ 500 km s$^{-1}$;
\item intermediate (I): 500 $<$ FWHM $<$ 1000 km s$^{-1}$;
\item broad (B): 1000 $<$ FWHM $<$ 2000 km s$^{-1}$;
\item very broad (VB): FWHM $>$ 2000 km s$^{-1}$.  
\end{enumerate} 

The kinematic parameters determined from the multiple-Gaussian [OIII] fits were then used to fit key emission-lines throughout the spectra in order to derive the line fluxes. Intrinsic velocity widths for the Gaussian components were obtained by subtracting the instrumental Gaussian width in quadrature from each of the components' FWHM. The instrumental widths for the observations were determined by averaging the FWHM of the sky-lines of the observations in each band (see $\S$\ref{sect:dr}). The kinematic parameters derived from the [OIII] fits were then used to attempt to fit all the other strong lines in the spectra.

We focussed on the [NII]$\lambda\lambda$6549,6583, [OII]$\lambda\lambda$3726,3729, [OII]$\lambda\lambda$7319,7330, [SII]$\lambda\lambda$4069,4076, [SII]$\lambda\lambda$6717,6731 emission-lines and, where observed/detected the hydrogen lines H$\alpha$, H$\beta$, P$\alpha$, P$\beta$ and Br$\gamma$. Note that the [NII]$\lambda$$\lambda$6548,6584 doublet, which is often blended with H$\alpha$, was held at a 1:3.0 ratio and doublet line separation from atomic physics \citep{osterbrock06}. 

When fitting all the recombination lines for the type 1 object F15462-045,  an additional broad component was fitted to account for the broad line region (BLR) emission. Based on the fit to the H$\beta$ line, the FWHM for this component is 4100$\pm$90 km s$^{-1}$. In addition, strong FeII emission is present in F15462-0450. This was fitted using the following constraints: the FeII emission-lines in the same multiplets (F, S or G) were  required to have  the  same  intrinsic  velocity  width  as  the  broad components of the H$\beta$ emission-line, and their predicted intensity ratios were constrained following the approach outlined in \citet{kova10}. In addition to the FeII multiplets, `N2' and `B' components for a widely reported \lq\lq narrow" component of FeII emission (FeII$\lambda$4930; \citealt{vandenberk01}) were required to produce an adequate fit (black dashed line). The overall fit to these emission features is shown in Figure \ref{fig:allfits}.

We give examples of the [OIII] model fits to other emission-lines for F13443+0802SE in Figure \ref{fig:modfits}. 
In general the fits were successful. However, there were instances where the [OIII] model did not adequately describe the profiles of the hydrogen recombination lines and/or the various [OII] and [SII] diagnostic blends (as indicated by `N' in the `H$\alpha$' and `[OII]/[SII]' columns in Table \ref{tab:gfits}). In the cases where the [OIII] model did not fit the [OII] and [SII] blends, we fixed the narrow components of the emission blends using the [OIII] model, and then constrained the broad components of the blends using fits to the broad component of the [OII]$\lambda\lambda$3726,3727 blend. Each emission-line in the trans-auroral blend required just one broad component with the exception of F13451+1232W which required two.  Note that, in the case of the hydrogen recombination lines not fitted by the [OIII] model, all the hydrogen lines were constrained to have the same kinematic components as H$\beta$. The kinematic parameters for these emission-lines are presented in Table \ref{tab:gfits}. 

In addition, when fitting the [OII] and [SII] blends, we constrained the intensity ratios of different components within each blend in the following ways.

\begin{enumerate}

\item For the [OII]$\lambda\lambda$3726,3729, [SII]$\lambda\lambda$4069,4076\footnote{We calculated the theoretical [SII](4069/4076)ratios using {\it CLOUDY} photoionization models (C13.04; \citealt{ferland13}). This ratio was found to lie in the range 3.01 $<$ [SII](4069/4076) $<$ 3.28 from the lowest to highest density limit.} and [SII]$\lambda\lambda$6717,6731 doublets we ensured that the intensity ratios fell within their theoretical values \citep{osterbrock06} and therefore we were not modelling unphysical values. Note that this was an unnecessary step for the [OII]$\lambda\lambda$3726,3729 doublet because all fits gave sensible values. However, in the cases of F12072-0444, F13443+0802SE, F13451+1232W and F16156+0146NW it was necessary to fix the narrow components to the lowest theoretical ratio for the [SII]$\lambda\lambda$4069,4076 doublet, therefore assuming the lowest density. 

\item For the [OII]$\lambda\lambda$7319,7330 doublet\footnote{Note that each component of the [OII]$\lambda\lambda$7319,7330 doublet is itself double: [OII]$\lambda$7319 comprises components at 7319 and 7320\AA, while  [OII]$\lambda$7330 comprises components at 7330 and 7331\AA. However, since the smaller doublet spacing is much smaller that the widths in \AA\,
of the individual kinematic components we are considering, we modelled [OII]$\lambda\lambda$7319,7320 and [OII]$\lambda\lambda$7330,7331 as single lines.} we fixed the 7319/7330 intensity ratio at 1.24. This ratio does not vary with density.

\end{enumerate}

\begin{figure}
\centering
\includegraphics[scale=0.38, trim=0.1cm 4.6cm 0.1cm 3.6cm]{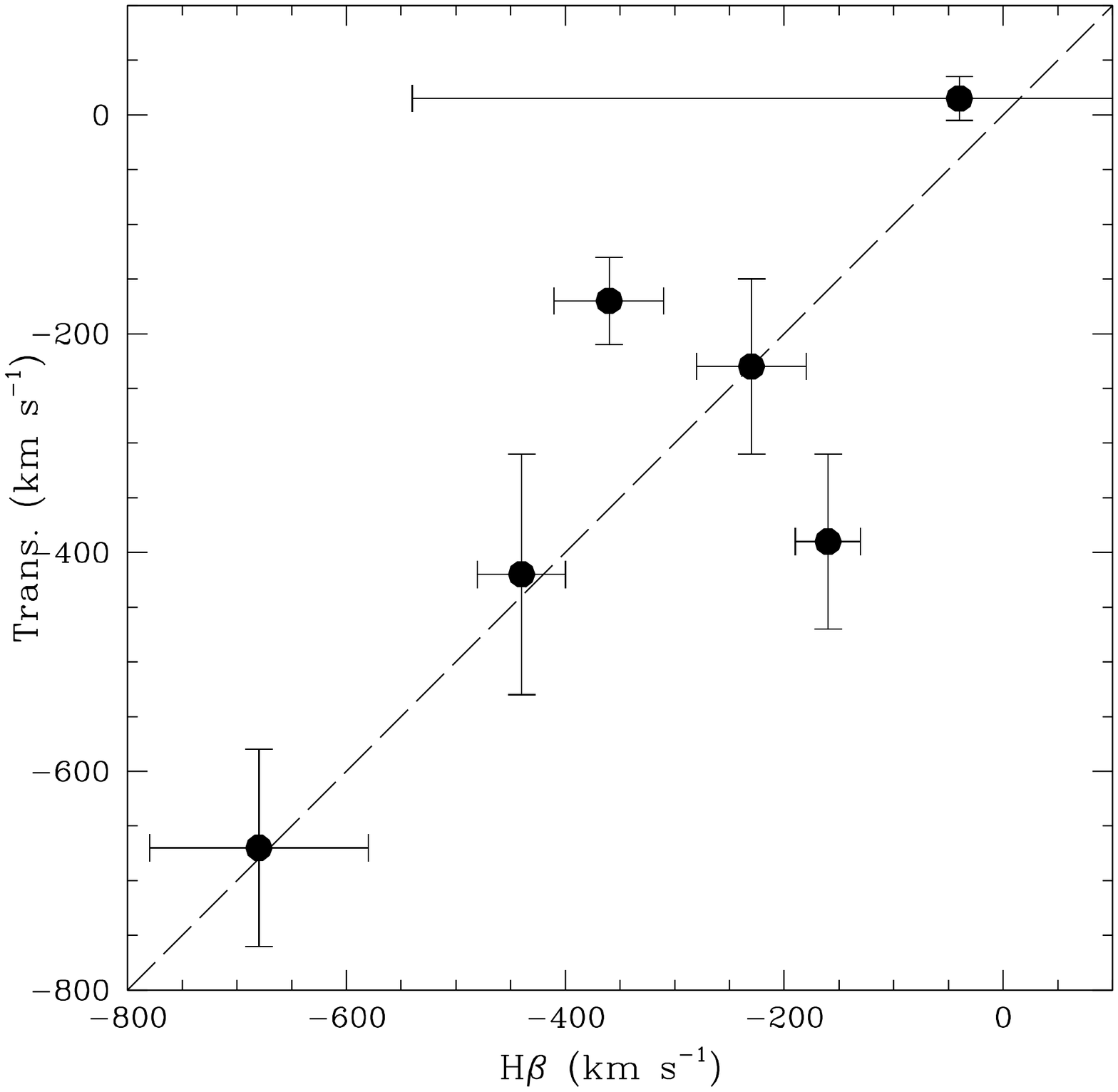}
\caption{A comparison of the velocity shifts of the broad components for the trans-auroral blends and H$\beta$ emission-lines. The dotted line indicates the one-to-one ratio.}
\label{fig:shiftcomp}
\end{figure}

\begin{figure}
\centering
\includegraphics[scale=0.38, trim=0.1cm 4.6cm 0.1cm 3.6cm]{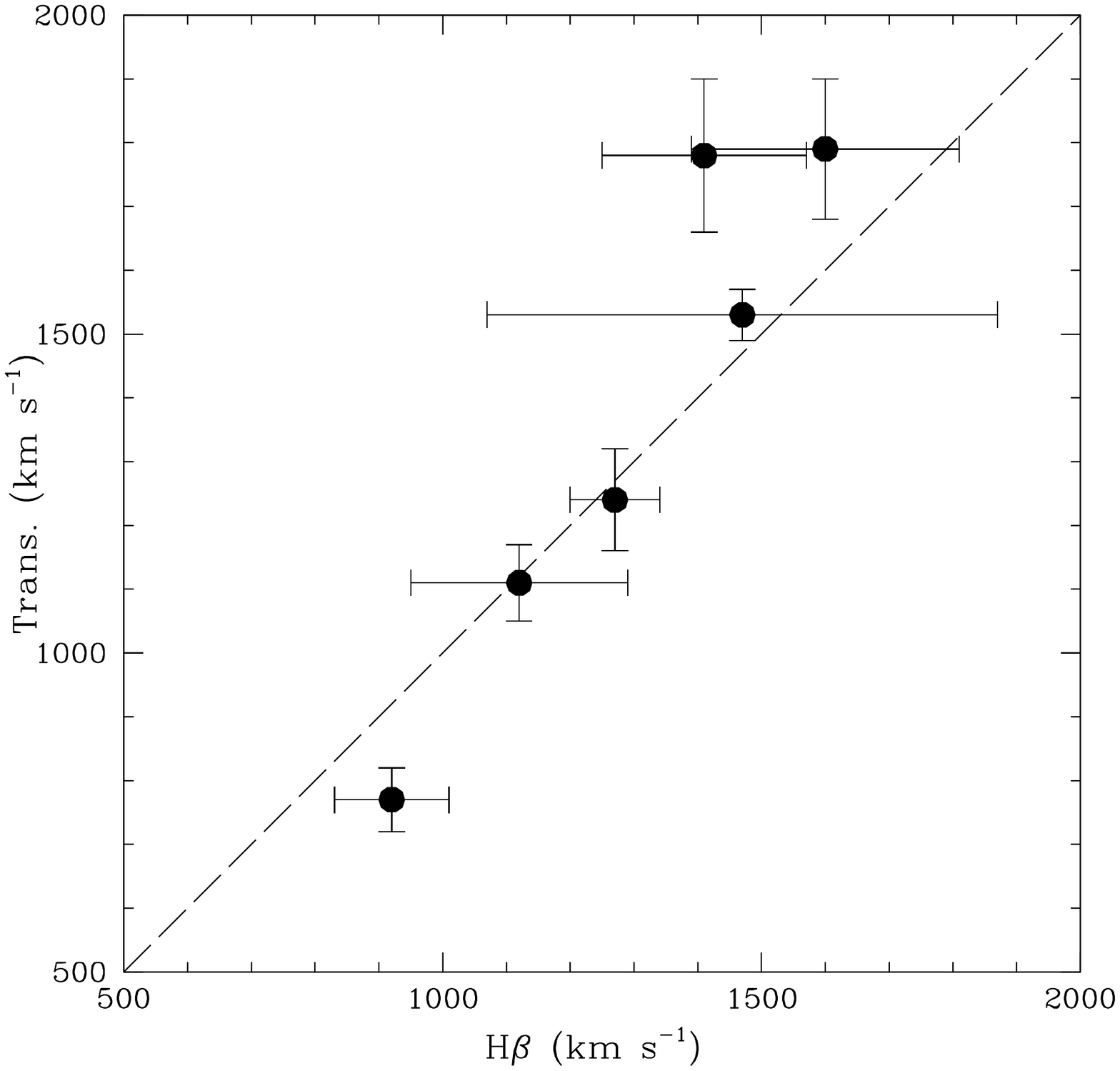}
\caption{A comparison of the velocity widths of the broad components for the trans-auroral blends and H$\beta$ emission-lines. The dotted line indicates the one-to-one ratio.}
\label{fig:fwhmcomp}
\end{figure}

It is important to confirm that the kinematics of the broad components of the trans-auroral lines follow those of the hydrogen lines, since both types of emission-lines are involved in
estimating the masses and kinetic powers of the outflows (see $\S$4).  In Table \ref{tab:kinecomp}, and Figures 
\ref{fig:shiftcomp} \& \ref{fig:fwhmcomp}, we compare the velocity shifts and FWHM for the broad components of the trans-auroral blends to those of the flux weighted H$\beta$ broad kinematic components for the objects where the [OIII]  model did not successfully fit the trans-auroral blends. While there are no significant differences in the FWHM of the emission-lines -- confirming the similarity of the kinematics -- we find that the velocity shift for F15462-0450 differs significantly ($>$3$\sigma$) when comparing the trans-auroral blends to the H$\beta$ emission. However, the determination of the precise kinematics of the H$\beta$ outflow component is complicated in this object by the presence of the BLR.      

In addition to these cases, some fits were not successful in the following cases.

\begin{itemize}

\item While two narrow components were required to fit the [OIII]$\lambda\lambda$4959,5007 emission-lines in F13443+0802SE, the blueshifted narrow component (`N2' in Table \ref{tab:gfits}) was only detected in the [OIII]$\lambda\lambda$4959,5007, H$\beta$ and H$\alpha$ emission lines. However, the remaining components of the Gaussian model (N1 and I) were used to successfully fit the rest of the optical emission-lines in this object. 

\item The broad components of the [OIII] model were not detected in the P$\alpha$ and Br$\gamma$ recombination lines of F14378-3651.

\item While two narrow components were required to fit the [OIII]$\lambda\lambda$4959,5007 emission-lines in F14378-3651, the blueshifted narrow component (`N2' in Table \ref{tab:gfits}) was only detected in these emission-lines. However, the remaining components of the Gaussian model (N1 and B) were used to successfully fit the rest of the optical emission-lines in this object.

\item The [NII] emission-lines in the H$\alpha$+[NII] blend for F16156+0146NW could not be fitted using the [OIII] model. The `I' and `B' components were replaced with a single broad component with $\Delta$v=-216$\pm$20 km s$^{-1}$ and FWHM=1160$\pm$40 km s$^{-1}$.

\item The P$\alpha$ and P$\beta$ recombination lines for F16156+0146NW show an additional blueshifted, narrow component ($\Delta$v=-118$\pm$15 km s$^{-1}$; FWHM=138$\pm$5 km s$^{-1}$) that is not detected in emission-lines at shorter wavelengths.

\end{itemize} 

In Appendix $\S$\ref{sect:specdes} we provide more details about the spectra of the individual sources.

\begin{center}
\begin{table*}
\centering
\caption{The [OIII] model parameters used when fitting the emission-lines of the spectra. `Comp.' refers to the velocity component defined in $\S$\ref{sect:gfits}. All $\Delta$v and FWHM velocity components are in units of km s$^{-1}$. `H$\alpha$' and `[OII]/[SII]' indicate whether the multi-component model successfully fitted the H$\alpha$+[NII] and trans-auroral blends. `[OIII] flux' and `H$\beta$ flux' give the [OIII]$\lambda$5007 and H$\beta$ fluxes respectively. Note that the line widths have been corrected for the instrumental width.}
\begin{tabular}{lccccccc}
\hline
Object	&	Comp.	&	$\Delta$v	&	FWHM	&	H$\alpha$	&	[OII]/[SII] & [OIII] flux & H$\beta$ flux	\\
	&		&	(km s$^{-1}$)	&	(km s$^{-1}$)	&	& & (erg s$^{-1}$ cm$^{-2}$) & (erg s$^{-1}$ cm$^{-2}$)	\\
\hline
F12072-0444	&	N	&	-156$\pm$40	&	346$\pm$25	&	Y$_{a}$	&	N	&	(6.32$\pm$0.32)E-15	&	-	\\
	&	I	&	-413$\pm$44	&	604$\pm$26	&		&		&	(6.58$\pm$0.34)E-15	&	-	\\
	&	B	&	-652$\pm$59	&	1160$\pm$100	&		&		&	(3.78$\pm$0.22)E-15	&	-	\\
	&	N$_{H\beta}$	&	-66$\pm$38	&	346	&		&		&	-	&	(1.17$\pm$0.06)E-15	\\
	&	I$_{H\beta}$	&	-323	&	604	&		&		&	-	&	(4.63$\pm$0.35)E-16	\\
	&	B$_{H\beta}$	&	-526	&	1160	&		&		&	-	&	(5.93$\pm$0.45)E-16	\\
F13305-1739	&	N1	&	+195$\pm$40	&	80$\pm$9	&	Y	&	N	&	(9.63$\pm$0.82)E-16	&	(1.02$\pm$0.21)E-16	\\
	&	N2	&	-94$\pm$34	&	188$\pm$14	&		&		&	(1.67$\pm$0.14)E-15	&	(3.93$\pm$0.47)E-16	\\
	&	N3	&	+370$\pm$35	&	432$\pm$23	&		&		&	(5.50$\pm$0.46)E-15	&	(5.10$\pm$0.72)E-16	\\
	&	B	&	-115$\pm$34	&	1140$\pm$90	&		&		&	(6.90$\pm$0.58)E-14	&	(5.00$\pm$0.49)E-15	\\
	&	VB	&	-279$\pm$40	&	2150$\pm$210	&		&		&	(2.50$\pm$0.25)E-14	&	(3.43$\pm$0.63)E-15	\\
 F13443+0802SE 	&	N1	&	+60$\pm$48	&	350$\pm$6	&	Y	&	Y	&	(2.42$\pm$0.17)E-15	&	(1.53$\pm$0.10)E-16	\\
	&	N2	&	-57$\pm$45	&	73$\pm$14	&		&		&	(1.81$\pm$0.13)E-16	&	(1.29$\pm$0.21)E-17	\\
	&	I	&	-50$\pm$46	&	698$\pm$42	&		&		&	(5.29$\pm$0.28)E-15	&	(4.99$\pm$0.25)E-16	\\
F13451+1232W	&	N	&	+69$\pm$46	&	319$\pm$6	&	Y	&	N	&	(1.64$\pm$0.13)E-15	&	(4.07$\pm$0.33)E-16	\\
	&	B1	&	-311$\pm$60	&	1160$\pm$150	&		&		&	(2.25$\pm$0.18)E-14	&	(1.97$\pm$0.16)E-15	\\
	&	B2	&	-1841$\pm$98	&	1900$\pm$150	&		&		&	(8.87$\pm$0.72)E-15	&	(4.65$\pm$0.41)E-16	\\
	&	VB	&	-262$\pm$40	&	3100$\pm$200	&		&		&	(4.48$\pm$0.40)E-15	&	(9.85$\pm$1.02)E-16	\\
F14378-3651	&	N1	&	+3$\pm$62	&	236$\pm$5	&	Y	&	Y	&	(1.67$\pm$0.09)E-16	&	(2.47$\pm$0.13)E-16	\\
	&	N2	&	-729$\pm$64	&	177$\pm$13	&		&		&	(4.19$\pm$0.35)E-17	&	-	\\
	&	B	&	-650$\pm$85	&	1240$\pm$190	&		&		&	(5.14$\pm$0.27)E-16	&	(1.67$\pm$0.12)E-16	\\
F15130-1958	&	I	&	-516$\pm$38	&	828$\pm$38	&	Y$_{a}$	&	Y$_{a}$	&	(4.20$\pm$0.42)E-15	&	-	\\
	&	B	&	-1186$\pm$77	&	1250$\pm$190	&		&		&	(3.22$\pm$0.32)E-15	&	-	\\
	&	I$_{H\beta}$	&	52$\pm$30	&	828	&		&		&	-	&	(2.57$\pm$0.59)E-16	\\
	&	B$_{H\beta}$	&	-722	&	1250	&		&		&	-	&	(5.58$\pm$0.91)E-16	\\
F15462-0450	&	N1	&	-	&	186$\pm$9	&	N	&	N	&	(5.58$\pm$0.46)E-16	&	(3.81$\pm$0.32)E-16	\\
	&	N2	&	-742$\pm$10	&	411$\pm$17	&		&		&	(5.63$\pm$0.50)E-16	&	-	\\
	&	B	&	-915$\pm$16	&	1460$\pm$20	&		&		&	(6.14$\pm$0.49)E-15	&	-	\\
	&	N2$_{H\beta}$	&	-670$\pm$22	&	461$\pm$17	&		&		&	-	&	(5.05$\pm$0.47)E-16	\\
	&	B$_{H\beta}$	&	-322$\pm$34	&	1350$\pm$30	&		&		&	-	&	(4.75$\pm$0.39)E-15	\\
	&	BLR	&	-336$\pm$11	&	4100$\pm$90	&		&		&	-	&	(1.45$\pm$0.12)E-14	\\
F16156+0146NW	&	N	&	+36$\pm$78	&	343$\pm$12	&	Y	&	N	&	(5.75$\pm$0.46)E-15	&	(8.43$\pm$0.68)E-16	\\
	&	I	&	-181$\pm$79	&	910$\pm$110	&		&		&	(1.47$\pm$0.12)E-14	&	(1.32$\pm$0.11)E-15	\\
	&	B	&	-381$\pm$82	&	1780$\pm$210	&		&		&	(5.99$\pm$0.53)E-15	&	(2.38$\pm$0.41)E-16	\\
F19254-7245S	&	N	&	+30$\pm$38	&	118$\pm$11	&	N	&	N	&	(4.23$\pm$0.45)E-16	&	(1.44$\pm$0.16)E-15	\\
	&	I1	&	+855$\pm$80	&	604$\pm$37	&		&		&	(1.35$\pm$0.16)E-15	&	-	\\
	&	I2	&	+120$\pm$36	&	656$\pm$72	&		&		&	(1.65$\pm$0.20)E-15	&	-	\\
	&	B	&	+139$\pm$38	&	1890$\pm$130	&		&		&	(7.55$\pm$0.77)E-15	&	-	\\
	&	B1$_{H\beta}$	&	+137$\pm$52	&	1380$\pm$70	&		&		&	-	&	(4.60$\pm$0.47)E-15	\\
	&	B2$_{H\beta}$	&	-645$\pm$510	&	1790$\pm$430	&		&		&	-	&	(1.31$\pm$0.37)E-15	\\
\hline
\end{tabular}
\begin{tablenotes}
\item[a]$^a$ While the overall $\Delta$v of the emission blends for the H$\beta$ and trans-auroral components are not in agreement with the [OIII] emission blend, the velocity shifts between the individual components (e.g. `N', `I' and `B') are comparable to those in the [OIII] model. 
\end{tablenotes}
\label{tab:gfits}
\end{table*}
\end{center}

\begin{center}
\begin{table*}
\centering
\caption{A comparison of the trans-auroral blend and H$\beta$ broad component fitting parameters for the ULIRGs where the [OIII] fitting model did not adequately fit the trans-auroral blends. `Trans. $\Delta$v' gives the trans-auroral blend velocity shift. `Trans. FWHM' gives the velocity width of the trans-auroral blends. `H$\beta$ $\Delta$v' gives the H$\beta$ emission-line velocity shift. `H$\beta$ FWHM' gives the velocity width of the H$\beta$ emission-line. All values are given in units of km s$^{-1}$.}
\begin{tabular}{lcccc}
\hline									
Object	&	Trans. $\Delta$v	&	Trans. FWHM	&  	H$\beta$ $\Delta$v	&  	H$\beta$ FWHM	\\
	&	(km s$^{-1}$)	&	(km s$^{-1}$)	&	(km s$^{-1}$)	&	(km s$^{-1}$)	\\
\hline									
F12072-0444	&	-420$\pm$110	&	770$\pm$50	&	-440$\pm$40	&	920$\pm$90	\\
F13305-1739	&	-390$\pm$80	&	1780$\pm$120	&	-160$\pm$30	&	1410$\pm$160	\\
F13451+1232W$^a$ 	&	-670$\pm$90	&	1790$\pm$110	&	-680$\pm$100	&	1600$\pm$210	\\
F15462-0450	&	-170$\pm$40	&	1240$\pm$80	&	-360$\pm$50	&	1270$\pm$70	\\
F16156+0146NW	&	-230$\pm$80	&	1110$\pm$60	&	-230$\pm$50	&	1120$\pm$170	\\
F19254-7245S	&	+15$\pm$20	&	1530$\pm$40	&	-40$\pm$500	&	1470$\pm$400	\\
\hline									
\end{tabular}
\begin{tablenotes}
\item[a]$^a$ The kinematics for the trans-auroral blends in F13451+1232W are the flux weighted average of the two broad components.  
\end{tablenotes}
\label{tab:kinecomp}
\end{table*}
\end{center}

\section{The basic properties of the outflowing gas}

In this paper we aim to quantify the key properties (mass outflow rate and kinetic powers) of the AGN-induced warm outflows in the target ULIRGs. To calculate these properties, we require accurate determinations of the electron densities, spatial extents and kinematics of the outflowing gas, as well as the intrinsic reddening and emission-line luminosities (see the Introduction). In this section, we present the results obtained from the emission-line fitting using the  [OIII] kinematic model described in $\S$\ref{sect:gfits}, as well as results on the radial extents of the outflows. 

In what follows we concentrate on the properties of the outflowing gas, which we assume to be represented by the intermediate and broad kinematic components presented in Table \ref{tab:gfits} that are often significantly shifted from the host galaxy rest-frame. In most cases the outflowing gas 
is observed as blueshifted kinematic components, but in the case of F19254-7245S the broad and intermediate emission-line components are
systematically redshifted. Note that, where more than one blue or redshifted broad/intermediate component is present, we do not study the sub-components separately because it is unclear whether they are truly distinct in a physical sense, or represent parts of a continuous outflow in which the kinematics and physical conditions vary smoothly with radius; the multiple Gaussians are merely a convenient way of fitting the emission-line profiles.  

\subsection{The radial extents of the warm outflows} \label{sect:extent}

\begin{center}
\begin{table*}
\centering
\caption{Radial extent of the broad, outflowing [OIII]$\lambda$5007 gas. `[OIII] range' indicates the velocity ranges  used to define the broad, outflowing [OIII]$\lambda$5007 gas. `FWHM$^{meas}_{[OIII]}$' is the radial extent of the broad component in arcseconds before being corrected for seeing. `FWHM$_{1D}$' is the 1D seeing estimate for the observations in arcseconds withe the uncertainty derived as described in the text. `Resolved' indicates whether the radial extent of the broad component is resolved. `D$_{[OIII]}$' is the diameter of the [OIII] outflow region, and R$_{[OIII]}$  the radial extent in kpc.}
\begin{tabular}{lcccccc}
\hline
Object	&	[OIII]	range	(km s$^{-1}$)	&	FWHM$^{meas}_{[OIII]}$	& FWHM$_{1D}$	&	Resolved	&	D$_{[OIII]}$	& R$_{[OIII]}$ \\
	&	&	(arcseconds)	&	(arcseconds)	&	(Y/N)	&	(arcseconds)	&	(kpc)	\\	
\hline																	
F12072-0444	&	 -1480 -- -400	&	 0.713$\pm$0.026	&	0.73$\pm$0.03	&	 N	&	 $<$0.37	&	 $<$0.41	\\
F13305-1739$^a$	&	-1460 -- -340	&	1.15$\pm$0.04	&	1.16$\pm$0.10	&	N	&	$<$0.89	&	$<$1.12	\\
F13443+0802SE	&	-1140 -- -350; +290 -- +1180	&	0.792$\pm$0.018	&	0.74$\pm$0.05	&	N	&	$<$0.49	&	$<$0.58	\\
F13451+1232W	&	-1590 -- -250	&	0.804$\pm$0.019	&	0.92$\pm$0.07	&	N	&	$<$0.66	&	$<$0.70	\\
F14378-3651	&	-2120 --  -880	&	0.861$\pm$0.029	&	0.86$\pm$0.10	&	N	&	$<$0.78	&	$<$0.49	\\
F15130-1958$^b$	&	-1550 -- 0	&	0.830$\pm$0.025	&	0.88$\pm$0.14	&	N	&	$<$0.96	&	$<$0.92	\\
F15462-0450	&	-1960 -- -160	&	0.719$\pm$0.022	&	0.81$\pm$0.02	&	N	&	$<$0.32	&	$<$0.28	\\
F16156+0146NW	&	-1620 -- -300	&	0.631$\pm$0.009	&	0.64$\pm$0.02	&	N	&	$<$0.28	&	$<$0.32	\\
F19254-7245S	&	+430 -- +1670	&	1.034$\pm$0.012	&	1.01$\pm$0.03	&	N	&	$<$0.44	&	$<$0.25	\\
\hline
\end{tabular}
\begin{tablenotes}
\item[a]$^a$ For this measurement we include the B and VB components of the Gaussian model (see $\S$\ref{sect:gfits}) and not the N1 component as it is likely related to the rotation disk of the host galaxy.
\item[b]$^b$ For F15130-1958 we use [OIII]$\lambda$4959 because [OIII]$\lambda$5007 is cut off at the end of the UVB arm.
\end{tablenotes}
\label{tab:spatial}
\end{table*}
\end{center}

\begin{center}
\begin{table*}
\centering
\caption{Radial extent of the broad, outflowing [OIII]$\lambda$5007 gas. `VLT (kpc)' is the radial extent  of the [OIII] outflow in kpc, as measured from the VLT/Xshooter spectrum. `HST/ACS (kpc)' indicates the the flux-weighted mean radial extent of the [OIII] gas from the centre of each galaxy in kpc, as measured using the HST imaging data presented in Tadhunter et al. (2017, in prep.). `HST/STIS (kpc)' is the radial extent of the [OIII], as estimated from  HST STIS data.}
\begin{tabular}{lccc}
\hline
Object	&	VLT (kpc)	&	HST/ACS (kpc)	&	HST/STIS (kpc)	\\
\hline
F12072-0444	&	$<$0.41	&	-	&	0.095$\pm$0.010	\\					
F13443+0802SE	&	$<$0.58	&	0.766$\pm$0.008	&	-	\\
F13451+1232W	&	$<$0.70	&	0.075$\pm$0.001	&	-	\\
F15130-1958	&	$<$0.92	&	0.064$\pm$0.008	&	-	\\
F16156+0146NW	&	$<$0.32	&	0.086$\pm$0.007	&	0.105$\pm$0.011	\\
\hline
\end{tabular}
\label{tab:spatialcomp}
\end{table*}
\end{center}

\noindent
To accurately quantify key properties of the outflows, such as their mass outflow rates and the kinetic powers, it is essential to determine their radial extents. Because we do not have high resolution HST images for all the ULIRGs, we used the 2D Xshooter spectra to estimate the outflow radii. Although visual inspection of our 2D spectra indicates  that the broadest and most kinematically disturbed emission-line components are strongly concentrated on the nuclei of the host galaxies, in some cases we detect narrower kinematic components ($FWHM < 500$~km s$^{-1}$) that are spatially extended  for stronger emission-lines such as [OII]$\lambda\lambda$3726,3729, H$\beta$, [OIII]$\lambda\lambda$5007,4959, [OI]$\lambda$6300, the H$\alpha$+[NII] blend, [SII]$\lambda\lambda$6717,6731, [SIII]$\lambda\lambda$9069,9531. This spatially extended gas is likely associated with the extended narrow line regions (NLR) of the galaxies, which are not necessarily outflowing. Therefore, in what follows we will concentrate on the spatial extents of the broad
and intermediate emission-line components. 

To determine the radial extent of the outflowing gas, we extracted spatial slices of the significantly blueshifted (or redshifted in the case of F19254-7245S) [OIII]$\lambda$5007 components from the 2D spectra over the velocity ranges given in Table \ref{tab:spatial}. These velocity ranges were chosen to avoid emission from the narrow [OIII]$\lambda$5007 components. We choose [OIII]$\lambda$5007 because it is generally the emission-line with the strongest outflow components, and does not suffer from blending
with other emission-lines as much as some other prominent features. Note that for F13443+0802SE, although no kinematic component was significantly shifted relative to the host galaxy rest-frame, we calculated the radial extent of the intermediate-velocity component presented in Table \ref{tab:gfits}. 

We also extracted spatial slices blueward and redward of [OIII]$\lambda$5007 that were centred on regions of emission-line-free continuum. These red and blue spatial continuum slices were 
averaged, scaled to have the same effective width in wavelength as the slices extracted for the [OIII] emission, then subtracted
from the [OIII] slices in order to derive the continuum-free spatial profiles of the outflowing gas. The resulting spatial profiles
were then fitted with single Gaussians to determine their FWHM.

The outflow regions were considered to be spatially resolved if the FWHM measured from the continuum-subtracted [OIII] spatial slices (FWHM$^{meas}_{[OIII]}$) were more than
3$\sigma$ larger than the 1D FWHM seeing estimates derived from the acquisition images (FWHM$_{1D}$: see Table \ref{tab:seeing}), where $\sigma$ represents the uncertainty in the seeing over the period of the observations of each object. The uncertainties returned by the 
Gaussian fits to the 1D spatial profiles derived from the stellar images in the acquisition images are generally much smaller than the variation in the seeing across the observation period for a given object. Therefore we took as our estimate of the uncertainty in the 1D seeing
FWHM (i.e. $\sigma$) the larger of the standard error in the mean DIMM seeing estimate and the difference between the seeing estimates 
derived from the each of the two acquisition images\footnote{This latter estimate was not available for F15462-0450 which had only one set of acquisition images.}. 

We found that the radial extents of the [OIII] outflows did not significantly exceeded the 1D seeing FWHM for any of the ULIRGs in our sample. We therefore determined conservative upper limits on the radial extents using:
\\

\noindent $FWHM_{[OIII]} < \sqrt{(FWHM_{1D} + 3\sigma)^2 - (FWHM_{1D})^2}$
\\ 

\noindent where $\sigma$ represents the uncertainty in the 1D seeing FWHM, and $FWHM_{[OIII]}$ represents the true spatial extent of the [OIII] outflow region; these upper limits on $FWHM_{[OIII]}$ were then converted to kpc, and divided by 2 to get upper limits on the radial extents of the outflows. Table \ref{tab:spatial} presents the measured R$_{[OIII]}$ (kpc) values. Taking into consideration that the $R_{[OIII]}$ measurements are upper limits, these estimates of the outflows are notably smaller than those reported in the literature for some studies of luminous AGN at both low and high redshifts (up to 15kpc; e.g. \citealt{alexander10}; \citealt{greene11}; \citealt{harrison12}; \citealt{harrison14}; \citealt{cresci15}; \citealt{mcelroy15}; \citealt{cristina17}), but are closer to those reported in \citet{montse16} and \citet{karouzos16} (a few hundred pc up to $\sim$2kpc).

In Table \ref{tab:spatialcomp} we compare the Xshooter radius estimates for  F13443+0802SE, F13451+1232W, F15130-1958 and F16156+0146NW with
radii derived by Tadhunter et al. (2017, in prep.) from [OIII] HST/ACS narrow-band images. In the cases of F13451+1232W, F15130-1958 and F16156+0146NW, the HST/ACS radii represent the instrumentally corrected HWHM measured for the compact cores. Whereas in the case of F13443+0802SE, the HST/ACS radius is the flux weighted mean radius. 

Reassuringly, the R$_{[OIII]}$ values as determined from the HST imaging for F13451+1232W, F15130-1958 and F16156+0146NW are smaller than the upper limits we calculate using the 2D spectra in this study. However, the measured radial extent for the [OIII] emission in F13443+0802SE as determined from the HST observations is significantly larger than the VLT/Xshooter upper limit. This is likely to be due to the fact that the HST/ACS narrow-band imaging includes emission from both the broad and narrow components. Therefore we recalculated the VLT/Xshooter R$_{[OIII]}$ value, this time including all the [OIII] emission (velocity range -1140 -- +1180 km s$^{-1}$) and find a seeing corrected radius of R$_{[OIII]}$ = 0.639$\pm$0.059 kpc, which agrees within 2$\sigma$ with the radius determined from the HST/ACS data.

In addition to the HST/ACS narrow-band imaging, HST/STIS spectroscopy observations (see $\S$2.2.2) are available for F12072-0444 and F16156+0146NW which have wavelength ranges that cover the [OIII]$\lambda\lambda$4959,5007 emission-lines. Interestingly, the [OIII]$\lambda\lambda$4959,5007 emission-line profiles for both objects in the STIS spectra are well fitted using only the `I' and `B' components from their respective models presented in Table \ref{tab:gfits}; there is no evidence for a significant contribution from narrow emission line components to the emission line profiles. This suggests that the narrow emission line components in these objects are emitted by spatially extended regions that are entirely outside the narrow HST/STIS slits used for the observations, or by regions that are so diffuse that little of their flux is intercepted by the slits.

In the case of the HST/STIS spectra, R$_{[OIII]}$ was determined using 1D spatial slices extracted from the 2D spectra in a similar manner to that described above for the Xshooter data. However, given that HST is a space-based telescope, the atmospheric seeing for the observations is not an issue. Rather, it was necessary to correct for the spatial line spread function of the observations. We estimated the spatial line spread function for the STIS data using observations of the standard star Feige 110\footnote{The proposal ID for this data is SM2/STIS 7100.} taken with an identical instrument setup by measuring the spatial FWHM 
of the standard star continuum emission, as estimated using a spatial slices through the standard star long-slit spectrum integrated over the same wavelength ranges as each of the the ULIRG [OIII] observations. We then compared the continuum-subtracted ULIRG [OIII] spatial FWHM to the standard-star continuum FWHM. In both cases, the [OIII] spatial FWHM exceeded the continuum FWHM (i.e. the line spread functions) by more than 3$\sigma$. Therefore we  subtracted the line spread functions in quadrature from the [OIII] spatial FWHM estimates, and then converted  to R$_{[OIII]}$ values as described above for the Xshooter observations. We note that the R$_{[OIII]}$ values estimated using HST/STIS {\it may}  under-estimate the true extents of the outflows, because the R$_{[OIII]}$ values are estimated from a single slit position: if the ionized emission is elongated in a particular direction, the slit may not be aligned along the full extent of the emission region.  

The R$_{[OIII]}$ values measured from the HST/STIS data are presented in Table \ref{tab:spatialcomp}. For F12072-0444 we find a value of R$_{[OIII]}$ = 0.095$\pm$0.010 kpc, which is smaller than the upper limit on R$_{[OIII]}$ measured for the broad and intermediate components ($<$0.41 kpc) using the VLT/Xshooter data.  Finally, for F16156+0146NW we find  R$_{[OIII]}$ = 0.105$\pm$0.011 kpc, which is smaller than the upper limit found using the Xshooter data ($<$0.32 kpc), but shows good agreement with the R$_{[OIII]}$ found using the HST/ACS data (0.086$\pm$-0.007 kpc). 

The comparisons of the R$_{[OIII]}$ upper limits, as determined using the Xshooter spectra, to those determined using the HST/ACS and HST/STIS data, suggests that our approach to determining R$_{[OIII]}$ from Xshooter spectra gives reasonable estimates of the radial extents of the outflow regions. 

\subsection{Reddening estimates} \label{sect:red}

\begin{center}													
\begin{table*}													
\centering													
\caption{The hydrogen emission-line ratios with respect to H$\beta$ and E(B-V) values for the total emission-line fluxes.  We use the \citet{calzetti00} reddening law to calculate E(B-V).}													
\begin{tabular}{lcccccccc}													
\hline																																
Object	&	H$\alpha$	&	E(B-V)	&	P$\beta$	&	E(B-V)	&	P$\alpha$	&	E(B-V)	&	Br$\gamma$	&	E(B-V)	\\
\hline																	
F12072-0444	&	6.4$\pm$1.1	&	0.68$_{-0.15}^{+0.13}$	&	1.0$\pm$0.2	&	0.59$_{-0.07}^{+0.06}$	&	3.2$\pm$0.5	&	0.62$_{-0.05}^{+0.04}$	&	-	&	-	\\
F13305-1739	&	-	&	-	&	0.50$\pm$0.11	&	0.84$_{-0.13}^{+0.10}$	&	1.4$\pm$0.3	&	0.90$_{-0.06}^{+0.05}$	&	-	&	-	\\
F13443+0802SE	&	4.4$\pm$0.7	&	0.35$_{-0.15}^{+0.13}$	&	0.61$\pm$0.13	&	0.44$_{-0.08}^{+0.06}$	&	1.7$\pm$0. 3	&	0.45$_{0.05}^{+0.04}$	&	-	&	-	\\
F13451+1232W	&	5.8$\pm$0.9	&	0.60$_{-0.14}^{+0.12}$	&	0.84$\pm$0.11	&	0.54$_{-0.05}^{+0.04}$	&	2.8$\pm$0.3	&	0.58$_{-0.03}^{+0.03}$	&	-	&	-	\\
F14378-3651$^{a,b}$	&	8.6$\pm$1.5	&	0.93$_{-0.16}^{+0.14}$	&	-	&	-	&	32.4$\pm$5.4	&	1.24$_{-0.05}^{+0.04}$	&	3.6$\pm$1.1	&	1.26$_{-0.10}^{+0.07}$	\\
F15130-1958	&	4.7$\pm$1.4	&	0.41$_{-0.29}^{+0.22}$	&	0.47$\pm$0.15	&	0.35$_{-0.12}^{+0.09}$	&	1.9$\pm$0.5	&	0.48$_{-0.07}^{+0.06}$	&	-	&	-	\\
F15462-0450$^c$	&	1.9$\pm$0.4	&	-0.37$_{-0.22}^{+0.17}$	&	0.27$\pm$0.04	&	0.17$_{-0.05}^{+0.04}$	&	0.30$\pm$0.04	&	-0.03$_{-0.04}^{+0.04}$	&	-	&	-	\\
F16156+0146NW$^d$	&	5.0$\pm$1.3	&	0.46$_{-0.26}^{+0.20}$	&	0.62$\pm$0.09	&	0.44$_{-0.05}^{+0.04}$	&	1.9$\pm$0.2	&	0.47$_{-0.03}^{+0.03}$	&	-	&	-	\\
F19254-7245S$^a$	&	8.3$\pm$1.1	&	0.90$_{-0.12}^{+0.10}$	&	-	&	-	&	8.6$\pm$2.7	&	0.81$_{-0.11}^{+0.08}$ & -	\\	-	&	-	\\
\hline
\end{tabular}
\begin{tablenotes}
\item $^a$ P$\beta$ was not detected in these objects. This is because at the redshifts of these objects, the wavelength of P$\beta$ coincided with the regions between the near-IR bands.
\item $^b$ The broad components for P$\alpha$ and Br$\gamma$ were not detected for F14378-3651.
\item $^c$ F15462-0450 is a type 1 AGN and therefore the contributions from the NLR and outflowing gas to the Hydrogen emission features are subject to degeneracies with the BLR contribution.
\item $^d$ The additional blueshifted narrow components for the P$\beta$ and P$\alpha$ recombination lines detected in F16156+0146NW (see $\S$\ref{sect:gfits}) have not been included in the calculations. 
\end{tablenotes}
\label{tab:extr}
\end{table*}
\end{center}

\begin{center}													
\begin{table*}													
\centering	
												
\caption{The hydrogen emission-line ratios with respect to H$\beta$ and E(B-V) values for the narrow emission-line fluxes. We use the \citet{calzetti00} reddening law to calculate E(B-V). Note that B and R refer to the blueshifted and redshifted narrow components respectively.}													
\begin{tabular}{lcccccccc}													
\hline																										
Object	&	H$\alpha$	&	E(B-V)	&	P$\beta$	&	E(B-V)	&	P$\alpha$	&	E(B-V)	&	Br$\gamma$	&	E(B-V)	\\
\hline																	
F12072-0444	&	7.1$\pm$1.3	&	0.76$_{-0.18}^{+0.15}$	&	1.1$\pm$0.2	&	0.63$_{-0.06}^{+0.05}$	&	4.1$\pm$0.7	&	0.68$_{-0.05}^{+0.04}$	&	-	&	-	\\
F13305-1739B$^e$	&	-	&	-	&	0.33$\pm$0.10	&	0.50$_{-0.12}^{+0.09}$	&	-	&	-	&	-	&	-	\\
F13305-1739R	&	-	&	-	&	0.41$\pm$0.14	&	0.69$_{-0.13}^{+0.10}$	&	1.2$\pm$0.2	&	0.78$_{-0.06}^{+0.05}$	&	-	&	-	\\
F13443+0802SEB$^e$	&	6.0$\pm$1.3	&	0.62$_{-0.21}^{+0.17}$	&	-	&	-	&	-	&	-	&	-	&	-	\\
F13443+0802SER	&	7.4$\pm$1.2	&	0.80$_{-0.15}^{+0.13}$	&	1.0$\pm$0.2	&	0.61$_{-0.08}^{+0.06}$	&	5.1$\pm$0.9	&	0.74$_{-0.05}^{+0.04}$	&	-	&	-	\\
F13451+1232W	&	5.7$\pm$0.6	&	0.57$_{-0.09}^{+0.08}$	&	0.37$\pm$0.08	&	0.27$_{-0.08}^{+0.06}$	&	0.81$\pm$0.09	&	0.24$_{-0.03}^{+0.03}$	&	-	&	-	\\
F14378-3651$^{a,b}$	&	12.3$\pm$2.0	&	1.23$_{-0.15}^{+0.13}$	&	-	&	-	&	54.3$\pm$8.7	&	1.38$_{-0.05}^{+0.04}$	&	6.0$\pm$1.0	&	1.34$_{-0.05}^{+0.04}$	\\
F15462-0450$^c$	&	5.5$\pm$0.6	&	0.55$_{-0.10}^{+0.09}$	&	1.1$\pm$0.1	&	0.62$_{-0.04}^{+0.04}$	&	4.4$\pm$0.5	&	0.70$_{-0.03}^{+0.03}$	&	-	&	-	\\
F16156+0146NW$^d$	&	5.9$\pm$0.6	&	0.61$_{-0.09}^{+0.08}$	&	0.35$\pm$0.07	&	0.25$_{-0.07}^{+0.06}$	&	2.0$\pm$0.2	&	0.48$_{-0.03}^{+0.03}$	&	-	&	-	\\
F19254-7245S$^a$	&	12.2$\pm$3.2	&	1.23$_{-0.20}^{+0.26}$	&	-	&	-	&	17.4$\pm$3.1	&	1.07$_{-0.05}^{+0.05}$	&	-	&	-	\\
\hline
\end{tabular}
\begin{tablenotes}
\item See notes in Table \ref{tab:extr}.
\item $^e$ The blueshifted narrow component was not detected in P$\alpha$ or  P$\beta$ 
\end{tablenotes}
\label{tab:extrn}
\end{table*}
\end{center}

\begin{center}													
\begin{table*}													
\centering													
\caption{The hydrogen emission-line ratios with respect to H$\beta$ and E(B-V) values for the combined broad/intermediate emission-line fluxes.  We use the \citet{calzetti00} reddening law to calculate E(B-V).}													
\begin{tabular}{lcccccccc}													
\hline													
Object	&	H$\alpha$	&	E(B-V)	&	P$\beta$	&	E(B-V)	&	P$\alpha$	&	E(B-V)	&	Br$\gamma$	&	E(B-V)	\\
\hline																	
F12072-0444	&	5.1$\pm$1.0	&	0.48$_{-0.18}^{+0.15}$	&	0.73$\pm$0.14	&	0.49$_{-0.07}^{+0.06}$	&	2.0$\pm$0.4	&	0.48$_{-0.05}^{+0.04}$	&	-	&	-	\\
F13305-1739	&	-	&	-	&	0.52$\pm$0.03	&	0.86$_{-0.06}^{+0.05}$	&	1.5$\pm$0.3	&	0.94$_{-0.06}^{+0.05}$	&	-	&	-	\\
F13443+0802SE	&	3.4$\pm$0.6	&	0.13$_{-0.15}^{+0.13}$	&	0.48$\pm$0.11	&	0.36$_{-0.08}^{+0.07}$	&	0.63$\pm$0.12	&	0.17$_{-0.06}^{+0.05}$	&	-	&	-	\\
F13451+1232W	&	5.9$\pm$0.9	&	0.61$_{-0.14}^{+0.12}$	&	0.89$\pm$0.12	&	0.56$_{-0.05}^{+0.04}$	&	3.0$\pm$0.3	&	0.60$_{-0.03}^{+0.03}$	&	-	&	-	\\
F14378-3651$^{a,b}$	&	5.6$\pm$1.0	&	0.57$_{-0.17}^{+0.14}$	&	-	&	-	&	-	&	-	&	-	&	-	\\
F15130-1958	&	4.7$\pm$1.4	&	0.41$_{-0.29}^{+0.22}$	&	0.47$\pm$0.15	&	0.35$_{-0.12}^{+0.09}$	&	1.9$\pm$0.5	&	0.48$_{-0.07}^{+0.06}$	&	-	&	-	\\
F15462-0450$^d$	&	3.2$\pm$0.4	&	0.08$_{-0.12}^{+0.10}$	&	0.13$\pm$0.02	&	-0.06$_{-0.06}^{+0.05}$	&	0.19$\pm$0.03	&	-0.16$_{-0.05}^{+0.04}$	&	-	&	-	\\
F16156+0146NW$^d$	&	4.3$\pm$1.1	&	0.34$_{-0.25}^{+0.19}$	&	0.77$\pm$0.11	&	0.51$_{-0.05}^{+0.03}$	&	1.9$\pm$0.2	&	0.47$_{-0.08}^{+0.03}$	&	-	&	-	\\
F19254-7245S$^a$	&	15.7$\pm$4.4	&	1.44$_{-0.28}^{+0.21}$	&	-	&	-	&	6.4$\pm$2.5	&	0.80$_{-0.13}^{+0.09}$	&	-	&	-	\\
\hline													
\end{tabular}													
\begin{tablenotes}
\item See notes in Table \ref{tab:extr}. 
 \end{tablenotes}
\label{tab:extbr}
\end{table*}
\end{center}	

\begin{center}
\begin{table*}
\centering
\caption{Comparison of the median E(B-V) values using the total emission-line fluxes (`Full E(B-V)'), narrow components only (`Narrow E(B-V)') and the broad components only (`Broad E(B-V)'). The uncertainties are the standard error on the median.}
\begin{tabular}{lccc}
\hline
Object	&	Total	E(B-V)	&	Narrow E(B-V)	&	Broad	E(B-V)	\\
\hline									
F12072-0444	&	0.618$\pm$0.028	&	0.681$\pm$0.039	&	0.481$\pm$0.004	\\		
F13305-1739$^f$	&	0.870$\pm$0.030	&	0.50$_{-0.12}^{+0.09}$;0.735$\pm$0.049	&	0.899$\pm$0.037	\\		
F13443+0802SE$^f$	&	0.435$\pm$0.032	&	0.62$_{-0.21}^{+0.17}$;0.742$\pm$0.058	&	0.172$\pm$0.045	\\		
F13451+1232W	&	0.577$\pm$0.018	&	0.270$\pm$0.106	&	0.599$\pm$0.015	\\		
F14378-3651$^{a,b}$	&	1.241$\pm$0.106	&	1.343$\pm$0.044	&	0.57$_{-0.17}^{+0.14}$	\\		
F15130-1958$^g$	&	0.413$\pm$0.036	&	-	&	0.413$\pm$0.036	\\		
F15462-0450	&	-0.027$\pm$0.158	&	0.621$\pm$0.044	&	-0.062$\pm$0.068	\\		
F16156+0146NW	&	0.457$\pm$0.009	&	0.483$\pm$0.106	&	0.466$\pm$0.052	\\		
F19254-7245S	&	0.854$\pm$0.043	&	1.152$\pm$0.079	&	1.122$\pm$0.320	\\		
\hline
\end{tabular}
\begin{tablenotes}
\item See notes in Table \ref{tab:extr}. 
\item $^f$ Narrow E(B-V) values are presented for blueshifted and redshifted cases respectively.
\item $^g$ No narrow component was detected in the recombination lines for F15130-1958.
 \end{tablenotes}
\label{tab:extcomp}
\end{table*}
\end{center}

\noindent
In order to determine accurate emission-line luminosities and hence gas masses for the dusty near-nuclear regions  of ULIRGs, it is important to estimate, then correct for, the dust extinction.
The superior wavelength coverage of Xshooter allows the detection of hydrogen emission-lines from the Balmer, Paschen and, depending on the redshift (z$<$0.108), the Brackett  series. This provides several hydrogen line ratios and, crucially, a long wavelength baseline over which to measure the  reddening of the emission regions in the ULIRG sample.

To calculate the reddening, we use the reddening law of \citet{calzetti00}, which is suitable for starburst galaxies such as ULIRGs, and assume intrinsic hydrogen line ratios derived from Case B recombination theory \citep{osterbrock06}. Ideally, the hydrogen ratios would be determined for each component of the multi-component Gaussian fits to the hydrogen emission-lines for each object. However, degeneracies in the fits sometimes make it difficult to separate the fluxes for the broad/shifted components. With this in mind, we have taken two approaches: first, using the integrated fluxes for the full hydrogen emission-line profiles;  and second, separating the the fluxes of the broad/intermediate and the narrow components, where these
components are defined in $\S$$\ref{sect:gfits}$. 

We do not include the H$\gamma$/H$\beta$ and H$\delta$/H$\beta$ ratios in our reddening estimates, because the H$\gamma$ and H$\delta$ lines are more likely 
to be affected by small errors in the subtraction of the underlying stellar absorption features, and the reddening estimates derived from then are generally more sensitive to small uncertainties in the ratios than the other hydrogen line ratios. This is best illustrated using an example: a measured H$\gamma$/H$\beta$=0.40 would lead to E(B-V)=0.744, but a 10\% decrease in this flux ratio, which is conceivable given the uncertainty on the flux calibration (see $\S$\ref{sect:dr}), would give H$\gamma$/H$\beta$=0.36 and  lead to E(B-V)=1.234 -- a change in reddening of $\Delta$E(B-V)=0.490. However, if we consider P$\beta$/H$\beta$=0.441, which leads to E(B-V)=0.744, a increase this ratio by 10\% to P$\beta$/H$\beta$=0.485 gives E(B-V)=0.815, a change of only $\Delta$E(B-V)=0.071. 

Tables \ref{tab:extr}-\ref{tab:extbr} present both the measured hydrogen line ratios with respect to H$\beta$ for the total (broad+intermediate+narrow), narrow, and combined broad/intermediate components respectively, as well as the the E(B-V) values derived from these ratios. Overall, considering the uncertainties, the reddening estimates derived from the different hydrogen lines
ratios show excellent agreement for particular kinematic components detected in individual objects. Table \ref{tab:extcomp} presents the average E(B-V) values for the total, narrow and broad components, obtained by taking the median of the values derived from  the different hydrogen line ratios for each object. The latter will be used to correct for the effects of dust reddening throughout the rest of this paper. Note that for F13305-1739 there are no reddening estimates using the H$\alpha$ recombination line. This is because the H$\alpha$+[NII] blend could not be fitted with the same kinematic model as the H$\beta$ emission-line. Nevertheless both P$\beta$ and P$\alpha$ were successfully fitted with the H$\beta$ model and therefore could be used to estimate the level of reddening. In addition, for the narrow components of F13305-1739 we present separate reddening estimates for both the blueshifted (component N2) and redshifted (components N1 and N3) emission components. Finally, for F15130-1958 no narrow (FWHM $<$ 500~km s$^{-1}$) emission component was detected and therefore there is no narrow component reddening value for this object.

\vglue 0.3cm\noindent
\subsubsection{Total flux}

Considering the estimates based on the total emission-line fluxes, as obtained by integrating over the full emission-line profiles, the ULIRG sample shows a wide range of reddening values. To determine whether the ULIRGs are more or less reddened when compared to other types of AGN, we compare the reddening values to those of the PG quasars \citep{boroson92}, a sample of \lq typical\rq\  blue-selected quasars, and the 2MASS-selected AGN of \citet{rose13}, a representative sample of AGN with red near-infrared colours (J-K$_S$ $>$ 2.0). See \citet{rose13} for a full discussion of both of these samples.  

The median reddening value for the ULIRG sample is E(B-V) = 0.46$\pm$0.07, which is comparable to the 2MASS-selected AGN, E(B-V)=0.53$\pm$0.11, but much higher than that of the PG quasars, E(B-V)=0.10$\pm$0.10.

\subsubsection{Narrow components}   

We also estimate the reddening for the narrow components alone. In Table \ref{tab:extrn} we present the narrow component hydrogen-line ratios with respect to H$\beta$, as well as the reddening values for the narrow components. Note that F13305-1739B\&F13443+0802SEB and F13305-1739R\&F13443+0802SER refer to the blueshifted and redshifted narrow components respectively. The range in reddening values is comparable to those obtained using total and broad hydrogen-line fluxes: 0.2 $<$ E(B-V) $<$ 1.4. In addition, the median value is E(B-V) = 0.63$\pm$0.06, which is similar to the median values derived from the total and broad profile fluxes, and also similar to the median obtained for 2MASS-selected AGN \citep{rose13}. 

In $\S$\ref{sect:gfits} we noted that F16156+0146NW showed an additional narrow, blueshifted component in the P$\alpha$ and P$\beta$ recombination lines. Given that this component was not detected in the shorter wavelength hydrogen lines, it is possible that this is a highly reddened component. We find P$\alpha$/P$\beta$=2.8$\pm$0.4 for this component, which suggests E(B-V)=1.1$_{-0.5}^{+0.4}$ using the reddening law of \citet{calzetti00}. This reddening value is higher than the reddening estimates obtained for the
total, narrow and broad emission components in this object (see Table \ref{tab:extcomp}), thus supporting the idea that this component is highly reddened, which is why it is not detected in the optical emission-lines covered by the UVB and VIS spectra.

\subsubsection{Broad/intermediate components}   

In addition to results from the total and narrow fluxes, it is interesting to estimate the reddening for the outflowing gas alone. In Table \ref{tab:extbr} we present both the hydrogen-line ratios with respect to H$\beta$, and the reddening values for the combined broad/intermediate component fluxes. The range of reddening values is similar to the total and narrow results: 0 $<$ E(B-V) $<$ 1.5. The median value is E(B-V) = 0.49$\pm$0.07, which is comparable to the median value for the total profiles, as well as the 2MASS-selected AGN \citep{rose13}. 

\subsubsection{Emission-line reddening summary}

Looking at the overall results summarised in Table \ref{tab:extcomp}, we find that there is evidence for significant reddening in both the broad and narrow emission-line components in most ULIRGs. The levels of dust extinction implied by these reddening estimates will have a major effect on the emission-line luminosities and mass outflow rates. For example, assuming a typical emission-line reddening of $E(B-V)=0.5$ and the Calzetti et al. reddening law, the extinction correction at the wavelength of the key H$\beta$ line
is more than a factor of $\times$8.

One surprising feature of our results is that we do not find a clear correlation between reddening and kinematics, in the sense that the
broad/intermediate components are more highly reddened than the narrow component. If the outflow regions were highly stratified, with negative velocity gradient, such that the broad/intermediate components were emitted closer to the AGN nuclei and therefore suffered a higher degree of reddening, one might expect that the broad/intermediate components would show a higher degree of reddening than the narrow components. Such a trend has been noted in the case of PKS1345+12 by  \citet{holt03}. However, we find no such trend in across our ULIRG sample: on average, the narrow components show a similar degree of reddening to the broader components, and in three objects --- F13443+0802SE, F14378-3651, and F15462-0450 --- the reddening measured for the narrow components is significantly higher than that measured for the broad components. 

One possible explanation for the lack of a correlation between disturbed emission line kinematics and reddening is that the narrow lines are emitted by extended disks of material that have quiescent kinematics, but nonetheless suffer significant extinction due to dust internal to the disks, whereas the broader components are emitted by  winds ejected out of the planes of the disks whose observed levels of extinction and reddening depend on orientation relative to the line of sight, as well as the degree to which the dust in the winds has been destroyed by the processes (e.g. shocks) that accelerated the gas. 

When comparing our results for F13451+1232W with those derived by \citet{holt03}, we find that our values, with the exception of the narrow component, are intermediate between those of the `intermediate' and `broad' velocity components of the H$\alpha$/H$\beta$ ratios (0.42 and 1.44 respectively in Holt et al. 2003).

\subsection{Bolometric luminosities} \label{sect:lbol}

\begin{center}
\begin{table*}
\centering
\caption{Estimates of the AGN luminosities of the ULIRG sample. `L$_{[OIII]}$-UNC' gives the logarithm of the uncorrected [OIII]$\lambda$5007 luminosity in units of erg s$^{-1}$. `L$_{[OIII]}$' gives the logarithm reddening corrected [OIII] luminosity in units of erg s$^{-1}$. `Xshooter/ISIS' gives the absolute ratio of the [OIII] luminosity using the Xshooter data of this work and the WHT/ISIS data of \citet{javier13}. `L$_{BOL-H}$' gives the logarithm of the bolometric luminosity as determined using the bolometric correction of \citet{heckman04} in units of erg s$^{-1}$. `L$_{BOL-L}$' gives the logarithm of the bolometric luminosity as determined using the bolometric correction of\citet{lamastra09} in units of erg s$^{-1}$. `H/L' gives the absolute ratio of `L$_{BOL-H}$' and `L$_{BOL-L}$'.  `H/ISIS' gives the absolute ratio of L$_{BOL}$ using the L$_{BOL-H}$ and the average L$_{BOL}$ using from the IRS spectroscopy presented in \citet{veilleux09}. `L/ISIS' gives the absolute ratio of L$_{BOL}$ using the L$_{BOL-L}$ and the average L$_{BOL}$ using from the IRS spectroscopy presented in \citet{veilleux09}.}
\begin{tabular}{lcccccccc}
\hline
Object	&	L$_{[OIII]}$-UNC	&	L$_{[OIII]}$	&	Xshooter/ISIS	&	L$_{BOL-H}$	&	L$_{BOL-L}$	&	H/L	&	H/IRS	&	L/IRS	\\
\hline																	
F12072-0444	&	41.83$\pm$0.04	&	42.94$\pm$0.05	&	1.27	&	45.38$\pm$0.04	&	45.60$\pm$0.05	&	0.60	&	0.34	&	0.56	\\
F13305-1739	&	42.75$\pm$0.03	&	43.64$\pm$0.03	&	1.29	&	46.30$\pm$0.03	&	46.29$\pm$0.03	&	1.02	&	-	&	-	\\
F13443+0802SE	&	41.56$\pm$0.07	&	42.33$\pm$0.07	&	-	&	45.10$\pm$0.07	&	44.99$\pm$0.07	&	1.29	&	-	&	-	\\
F13451+1232W	&	42.13$\pm$0.04	&	43.16$\pm$0.04	&	1.11	&	45.68$\pm$0.04	&	45.81$\pm$0.04	&	0.74	&	0.95	&	1.28	\\
F14378-3651	&	39.88$\pm$0.02	&	42.15$\pm$0.03	&	-	&	43.42$\pm$0.02	&	44.81$\pm$0.03	&	0.04	&	-	&	-	\\
F15130-1958	&	41.32$\pm$0.09	&	42.48$\pm$0.09	&	1.52	&	44.87$\pm$0.09	&	45.14$\pm$0.09	&	0.54	&	0.19	&	0.36	\\
F15462-0450	&	41.63$\pm$0.05	&	41.66$\pm$0.05	&	0.56	&	45.17$\pm$0.05	&	43.81$\pm$0.05	&	22.9	&	0.22	&	0.01	\\
F16156+0146NW	&	41.86$\pm$0.05	&	42.68$\pm$0.05	&	1.21	&	45.41$\pm$0.05	&	45.34$\pm$0.05	&	1.17	&	0.57	&	0.49	\\
F19254-7245S	&	40.96$\pm$0.02	&	42.54$\pm$0.02	&	-	&	44.51$\pm$0.02	&	45.20$\pm$0.02	&	0.20	&	-	&	-	\\
\hline
\end{tabular}
\label{tab:bol}
\end{table*}
\end{center}

\begin{center}
\begin{table*}
\centering
\caption{SMBH properties for 3 of the ULIRGs in this paper. `Log$_{10}$ M$_{SMBH}$' gives the black hole mass in M$_{\sun}$. `Log$_{10}$ L$_{EDD}$' gives the Eddington luminosity in erg s$^{-1}$. `$\lambda$$_{EDD}$-H' and `$\lambda$$_{EDD}$-L' give the Eddington ratio using L$_{BOL}$ as determined using the bolometric corrections of \citet{heckman04} and \citet{lamastra09} respectively. Note that the SMBH properties for F15462-0450$_{\sigma}$ use the \citet{dasyra06} results, and the properties for F15462-0450$_{H\beta}$ use the BLR H$\beta$ emission-line from the Xshooter spectroscopy.}
\begin{tabular}{lcccc}
\hline
Name	&	Log$_{10}$ M$_{SMBH}$ (M$_{\sun}$)	&	Log$_{10}$ L$_{EDD}$ (erg s$^{-1}$)	&	$\lambda$$_{EDD}$-H	&	$\lambda$$_{EDD}$-L	\\
\hline									
F14378-3651	&	7.66	&	45.81	&	$4.07\times10^{-03}$	&	$1.00\times10^{-01}$	\\
F15130-1958	&	7.92	&	46.06	&	$6.45\times10^{-02}$	&	$1.20\times10^{-01}$	\\
F15462-0450$_{\sigma}$	&	7.84	&	45.98	&	$1.55\times10^{-01}$	&	$6.76\times10^{-03}$	\\
F15462-0450$_{H\beta}$	&	7.83$\pm$0.03	&	45.93$\pm$0.03	&	0.17$\pm$0.05	&	0.0076$\pm$0.0003	\\
\hline
\end{tabular}
\label{tab:edd}
\end{table*}
\end{center}

Ultimately we want to place the kinetic powers ($\dot{E}$) derived for the warm outflow in ULIRGs (see $\S$4) in the context of the predictions of galaxy evolution models. In order to do this, it is important to compare them with the bolometric (radiative) luminosities  of the AGN ($L_{BOL}$). However, $L_{BOL}$ is difficult to determine for most of the ULIRGs in our sample which have highly extinguished type 2 AGN (the exception is F15462-0450).

\citet{javier13} reported that the majority of the ULIRGs in the present study have Seyfert-like luminosities based on their [OIII]$\lambda$5007 emission-line luminosities (L$_{[OIII]}$ $<$ 10$^{35}$ W); however, they did not correct these luminosities for the intrinsic dust extinction. Given that we have calculated E(B-V) for the ULIRGs in $\S$\ref{sect:red}, here we calculate [OIII] luminosities (L$_{[OIII]}$) for the ULIRGs that have been corrected for
extinction using our Xshooter results. 

To determine L$_{[OIII]}$ we use the integrated (broad, intermediate and narrow components) [OIII]$\lambda$5007 fluxes. Note, we assume here that there is a negligible contribution to [OIII]$\lambda$5007 from stellar-photoionized regions. The luminosities have been determined using the observed frame [OIII]$\lambda$5007 fluxes and the luminosity distances appropriate for the assumed cosmology. In addition, we correct the luminosities for the intrinsic reddening determined in $\S$\ref{sect:red} using the reddening law of \citet{calzetti00}. The uncorrected L$_{[OIII]}$ (L$_{[OIII]}$-UNC) and L$_{[OIII]}$ corrected for dust reddening for the ULIRG sample are presented in Table \ref{tab:bol}. From this it is already clear that many of the sources in our sample harbour luminous, quasar-like AGN: 8/9 of our sources would be classified as type 2 quasars based on their de-reddened [OIII] emission-line luminosities according to the criterion
of Zakamska et al. (2003: $L_{[OIII]} > 1.2\times10^{42}$ erg s$^{-1}$).

The [OIII] luminosities of many of the ULIRGs studied in this paper have been estimated by \citet{javier13}, who used long-slit WHT/ISIS data to investigate AGN-driven outflows in local ULIRGs. 6/9 of the ULIRGs in this paper overlap with the \citet{javier13} sample. When comparing the uncorrected [OIII]$\lambda$5007 emission-line luminosities, we find that 4/6 show an Xshooter/ISIS flux ratios of $\sim$0.9-1.3 (see Figure \ref{tab:bol}). The ratios for the remaining objects show significant differences -- F15130-1958 has a ratio of 1.56 and F15462-0450 has a ratio of 0.56 -- perhaps reflecting differences in the seeing between the two sets of observations. 

To determine the AGN bolometric luminosities of the ULIRGs, we use the [OIII]$\lambda$5007 emission-line luminosity as a proxy for total  AGN radiative power (e.g. see \citealt{heckman04}; \citealt{bian06}; \citealt{dicken14}). We adopt two approaches to determining L$_{BOL}$: (1) the bolometric correction of \citet{heckman04}, L$_{BOL}$=3500L$_{[OIII]}$, where L$_{[OIII]}$  has not been corrected for dust reddening, and (2) the bolometric correction factors of \citet{lamastra09} that use reddening corrected L$_{[OIII]}$, with bolometric correction factors of 87, 142 and 454 for reddening corrected luminosities of log$_{10}$ L$_{[OIII]}$ (erg s$^{-1}$) 38-40, 40-42 and 42-44 respectively. Other spectroscopic indicators, such as the L$_{5100}$ optical continuum luminosity are not suitable for most of our ULIRG sample because of a combination of reddening of the AGN continuum and host galaxy contamination. 

In Table \ref{tab:bol} we show the L$_{BOL}$ values calculated  using both the \citet{heckman04} method (L$_{BOL-H}$) and the \citet{lamastra09} method (L$_{BOL-L}$). The L$_{BOL-H}$ values are in the range 43.4 $<$ log$_{10}$ (L$_{BOL-H}$ erg s$^{-1}$) $<$ 46.3, whereas the L$_{BOL-L}$ are in the range 44.8 $<$ log$_{10}$ (L$_{BOL-L}$ erg s$^{-1}$) $<$ 46.3. Considering individual objects, in most cases the values agree within a factor of two; however, there are larger differences (factor $\sim$5 -- 25) in the cases of the two objects that show the highest levels of emission-line reddening (F14378-3651 and F19254-7245S) and the type 1 object F15462-0450. 

Given the high levels of extinction measured using the hydrogen line rations for F14378-3651 and F19254-7245S (see $\S$\ref{sect:red}), and also their continuum SEDs, it is likely that the \citet{lamastra09} correction is the most suitable method for these objects. 

In the case of F15462-0450, given that there is strong BLR emission present in its spectrum, and its continuum SED is typical of a type 1 AGN, the \citet{heckman04} correction is likely to be the most appropriate method. Indeed, as a test, we have also calculated L$_{BOL}$ for this object using the 5100 \AA\  monochromatic luminosity and the \citet{richards06} scaling correction (L$_{BOL}$=10.33$\lambda$L$_{\lambda}$). We find log$_{10}$ L$_{BOL}$ = 45.20$\pm$0.04, which is in excellent agreement with the luminosity calculated using the \citet{heckman04} method.

In addition to optical spectra, the AGN bolometric luminosities of 5/9 of our objects have been determined from Spitzer/IRS data by \citet{veilleux09} who used 6 methods to determine the AGN contribution to the total luminosities of the ULIRGs (see \citealt{veilleux09} for details). Here, we compare the average values derived from the latter methods. We find that the Xshooter\footnote{Here we combine the L$_{BOL}$ values using both the \citet{heckman04} and \citet{lamastra09} bolometric corrections.}/IRS luminosity ratios are in the range 0.01 $<$ Xshooter/IRS $<$ 1.3 (see Table \ref{tab:bol}), with the bolometric luminosities derived from the [OIII] luminosities in most cases lower than the Spitzer/IRS estimates.
Again, this highlights that the bolometric luminosities of the AGN in ULIRGs are highly uncertain.

Finally, in Table \ref{tab:edd} we compare the L$_{BOL}$ to the expected Eddington luminosities (L$_{EDD}$) for our sample in order to determine the Eddington ratios ($\lambda$$_{EDD}$). The SMBH masses (M$_{SMBH}$), and therefore L$_{EDD}$, were estimated by \citet{dasyra06} using the stellar velocity dispersions of the stars in the host galaxies. We find that 3 objects in our sample overlap with the sample studied in \citet{dasyra06}: F14378-3651, F15130-1958 and F15462-0450. Given that the BLR for F15462-0450 is clearly detected in its Xshooter spectrum, we also have a separate measure for M$_{SMBH}$. In this case, to determine M$_{SMBH}$ we used the properties of the BLR H$\beta$ emission-line, and adopt the M$_{SMBH}$-Balmer line relationship presented in \citep{greene05}, which is based on correlations between L$_{BOL}$, Balmer line luminosities and BLR line widths (see equation 7 in \citealt{greene05}). We find log$_{10}$ M$_{SMBH}$ (M$_{\sun}$) = 7.83$\pm$0.03, for F15462-0450, which is similar to that determined in \citet{dasyra06}. 

For objects with both M$_{SMBH}$ and L$_{BOL}$ estimates, we determined $\lambda$$_{EDD}$. With the exception of F14378-3651, the accretion rates derived in this way for the ULIRGs are comparable to those derived for PG quasars (0.23$\pm$0.14; \citealt{rose13}) and 2MASS-selected AGN (0.08$\pm$0.04; \citealt{rose13}). However, F14378-3651 appears to be accreting at a comparatively lower rate when using L$_{BOL-H}$ to describe the luminosity of the AGN. 

\subsection{Kinematics} \label{sect:kine}

From our [OIII] emission-line profile fits (see $\S$\ref{sect:gfits}), we find that the emission-line kinematics are complex, and for several of our sources we require more than one intermediate/broad Gaussian component to fit the wings to the line profiles However, given that we cannot always separate the various broad/intermediate components when determining the densities and reddenings of the outflow components, when estimating the mass outflow rates and kinematic powers of the outflows it is often convenient to have a single measure of the
velocity and FWHM of the outflowing gas. To this end, here we follow the approach of \citet{javier13} and calculate flux weighted mean velocity shifts and FWHM for the broad/intermediate Gaussian components of the [OIII]$\lambda$5007 profiles (see $\S$\ref{sect:gfits})\footnote{The flux weighted velocity shift is given by $v_{out}=\frac{\Sigma _{i}(F_{i}\times v_{out\mathit{i}})}{\Sigma F_{i}}$, where $v_{out\mathit{i}}$ are the individual velocity shifts of the broad and intermediate  components from the model fits and F$_{i}$ are the integrated fluxes for the components. The flux weighted velocity width is given by $FWHM=\frac{\Sigma _{i}(F_{i}\times FWHM_{i})}{\Sigma F_{i}}$, where  FWHM$_{i}$ are the individual velocity widths of the broad and intermediate components from the model fits.} . Note that the velocity shifts and FMHM are measured relative to the galaxy rest frames determined in $\S$\ref{sect:z}. Table \ref{tab:outc} presents weighted-v$_{out}$ and weighted-FWHM for the ULIRG sample.

A drawback of using the measured velocity shifts (flux weighted or otherwise) is that they do not account for projection effects.
Therefore they are likely to underestimate the true outflow velocities (see discussion in $\S$\ref{sect:outf} \& \ref{sect:outmax}). As argued in $\S$\ref{sect:outmax}, the velocity shift of the far wing of the [OIII]$\lambda$5007 emission-line relative to the host galaxy rest frame, as measured at 5\% of the total flux in the combined profile of the broad and intermediate Gaussian components from the model fits ($\Delta V_{5}$)\footnote{In the case  of F19254-7245S we use $\Delta V_{95}$ rather than $\Delta V_{5}$ because its [OIII] emission-line profile is dominated by
redshifted, rather than blueshifted components}, is likely to represent a more realistic estimate of the de-projected outflow velocities. Estimates of $\Delta V_{5}$
(or $\Delta V_{95}$ in the case of F19254-7245S) are presented in the final column of Table \ref{tab:outc}.

\begin{center}
\begin{table}
\centering
\caption{Flux weighted kinematics of the broad, outflowing [OIII]$\lambda$5007 gas of the ULIRGs. `FWHM' gives the flux weighted velocity width of the gas and `$\Delta$v' gives the flux weighted velocity shift of the gas relative to the rest-frame defined by the stellar absorption lines. Note that the kinematics are presented in units of km s$^{-1}$. `v$_{05}$'  gives the non-parametric velocity shifts at 5\% of the total flux (95\% in the case of F19254-7245S) for the outflowing components (km s$^{-1}$).}
\begin{tabular}{lccc}
\hline					
Object	&	FWHM	&	$\Delta$v	&   v$_{05}$   \\		
	&	(km	s$^{-1}$)	&	(km	s$^{-1}$) & (km	s$^{-1}$)	\\
\hline							
F12072-0444	&	800$\pm$120	&	-500$\pm$50	&	-1190$\pm$70	\\
F13305-1739	&	1410$\pm$160	&	-160$\pm$30	&	-1190$\pm$150	\\
F13443+0802SE	&	700$\pm$50	&	-50$\pm$13	&	-540$\pm$54	\\
F13451+1232W	&	1600$\pm$210	&	-680$\pm$100	&	-2570$\pm$120	\\
F14378-3651	&	1160$\pm$180	&	-720$\pm$70	&	-1490$\pm$150	\\
F15130-1958	&	1250$\pm$190	&	-1000$\pm$70	&	-1750$\pm$130	\\
F15462-0450	&	1460$\pm$20	&	-920$\pm$20	&	-1900$\pm$90	\\
F16156+0146NW	&	1120$\pm$170	&	-230$\pm$50	&	-1630$\pm$100	\\
F19254-7245S	&	1300$\pm$200	&	+180$\pm$70	&	-1300$\pm$90	\\	
\hline					
\end{tabular}
\label{tab:outc}
\end{table}
\end{center}

When comparing the weighted-v$_{out}$ and weighted-FWHM values with the values for the overlapping objects in the sample presented in \citet{javier13}\footnote{\citet{javier13} used the narrow component of their [OIII] fits to define the rest-frame of the system and therefore $\Delta$v.}, we find that all of the values agree within 3$\sigma$. 

For F15130-1958 we note that the whole [OIII] profile is shifted relative to the host galaxy rest-frame, which should lead to a significant difference in the calculated  weighted-v$_{out}$ of this work when compared to that in \citet{javier13}. However, v$_{out}$ in \citet{javier13} has a relatively large error bar (v$_{out}$=-606$\pm$186 km $^{-1}$) which leads to a 3$\sigma$ agreement with the weighted-v$_{out}$ calculated in this work. 

\subsection{Trans-auroral density and reddening estimates}

\subsubsection{Density and reddening determination}\label{sect:dens}

One of the main weaknesses of previous attempts to calculate the mass outflow rates and outflow kinematic powers in AGN has been the lack of accurate estimates of the electron density (n$_e$) of the outflowing gas. For AGN, the most readily available density-sensitive diagnostics at optical wavelengths are  the [SII]$\lambda\lambda$6716,6731 and [OII]$\lambda\lambda$3726,3729 doublet ratios \citep{osterbrock06}. However, these  are only sensitive to relatively low densities (10$^{2.0}$ $<$ n$_e$ $<$ 10$^{3.5}$ cm$^{-3}$) and therefore cannot be used for higher density clouds. In addition, the highly blueshifted or redshifted components often present in the forbidden line profiles of AGN lead to problems 
with blending, given that the doublet separations are relatively small ($\sim$15\AA\, and $\sim$3\AA\, in the rest frame for [SII] and [OII] respectively). The resulting degeneracies in the fits to the blends lead to substantial inaccuracies in the
measurement to the doublet ratios for individual kinematic components, with the problem particularly severe for components that are broad
and shifted. Indeed, the electron densities assumed in studies of warm outflows cover a wide range: $1 < n_{e} < 10^{5.0}$ cm$^{-3}$, a spread of 5 orders of magnitude (\citealt{javier13}; \citealt{liu13}; \citealt{carniani15}; \citealt{mcelroy15}; \citealt{montse16}).

The superior wavelength coverage and resolution of our Xshooter spectra allows us to use the technique based on the trans-auroral 
[OII]$\lambda\lambda$7319,7331 and [SII]$\lambda\lambda4$068,4076 lines to determine the densities and
reddening of the outflow regions simultaneously. This technique is based on the
following line ratios:
\\

\noindent $TR([OII]) = F(3726+3729)/F(7319+7331)$; 
\\

\noindent and
\\

\noindent $TR([SII]) = F(4068+4076)/F(6717+6731)$.
\\

On the one hand, the TR([OII]) and TR([SII]) ratios are sensitive to a wider range of densities 10$^{2.0}$ $<$ n$_e$ $<$ 10$^{6.5}$ cm$^{-3}$ \citep{appenzeller88} than the [SII](6731/6716) and [OII](3726/3729) ratios, while on the other the large wavelength  separation between the blends involved in the ratios allows us to simultaneously determine the reddening of the emission clouds. A further advantage is that this technique depends on the flux ratios {\it between} the emission-line blends rather than the ratios of individual lines  {\it within} the blends, and is therefore less affected by the small doublet separations. This technique was introduced by \citet{holt11} to study
the physical conditions in the warm outflow in F13451+1232W and updated in this work using {\it CLOUDY} (C13.04; \citealt{ferland13}) photoionisation code and the \citet{calzetti00} reddening law. 

The {\it CLOUDY} code was used to create plane-parallel, single-slab photoionization
models. We assumed that the photoionized gas is radiation bounded and has a solar composition, that the photoionizing continuum has a power-law shape ($\alpha$=-1.5; F$_{\nu}$ $\propto$ $\nu$$^{\alpha}$), and that the ionization parameter has a fixed value (U=0.005), although we note that the trans-auroral ratios are not particularly sensitive to these parameters \citep{holt11}. We varied the electron density in the range 2.0 $<$ log$_{10}$ n$_e$ $<$ 5.0, in intervals of $\Delta$log$_{10}$ n$_e$ = 0.1.

\begin{figure}
\centering
\includegraphics[scale=0.38, trim=0.1cm 4.6cm 0.1cm 3.6cm]{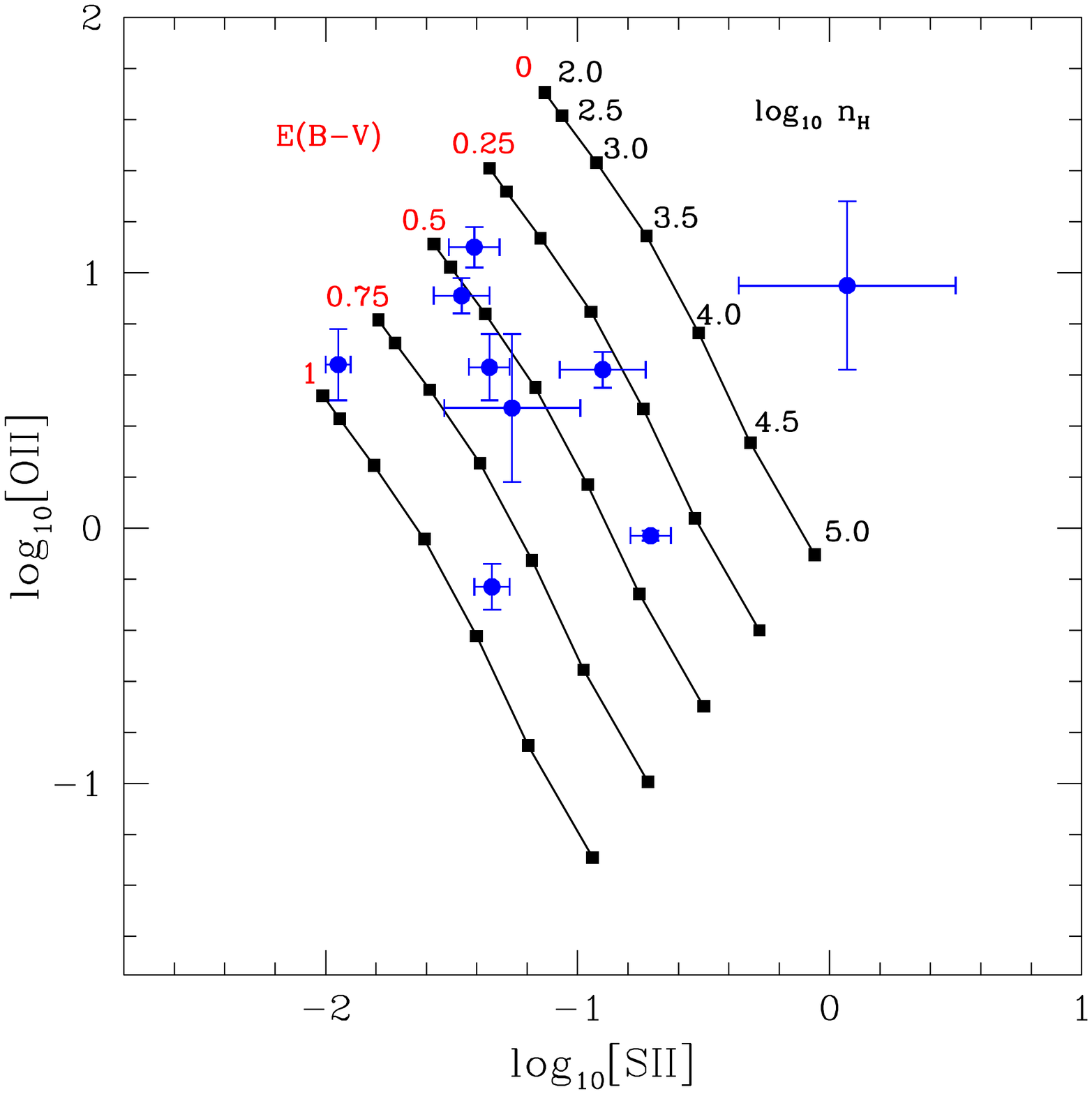}
\caption{Determining the densities and reddenings from the total fluxes of the trans-auroral [OII] and [SII] emission-lines. Log$_{10}$ (3726+3729)/(7319+7331) is plotted against log$_{10}$ [SII](4068+4076)/(6717+6731). The black squares joined by solid lines indicate the density sequence which spans the range 2.0 $<$ log$_{10}$ n$_e$ (cm$^{-3}$) $<$ 5.0. This indicated in black text. This sequence sweeps from the top-right of the plot to the bottom left in increasing E(B-V), which spans the range 0 $<$ E(B-V) $<$ 1.5. This is indicated in red text. The blue points indicate the trans-auroral ratios measured for the ULIRGs based on the total emission-line fluxes.}
\label{fig:trans}
\end{figure}

In Figure \ref{fig:trans} we plot log(TR([OII])) vs. log(TR([SII])) for the total (broad+intermediate+narrow) integrated fluxes measured for our ULIRG sample. We over-plot a grid of the ratios predicted by the photoionisation model, with this grid presented as sequences of varying electron density for a series of fixed E(B-V) values. We present the density and E(B-V) values obtained using the total flux trans-auroral ratios measured for the ULIRGs in Table \ref{tab:dens}. Taking into account the uncertainties in the flux ratios, we find n$_e$ in the range 10$^{2.0}$ $<$ n$_{e}$ $<$ 10$^{4.5}$ cm$^{-3}$, with an median density of log$_{10}$ n$_e$=3.45$\pm$0.26. This range spans 3 orders of magnitude, highlighting the importance of accurately estimating this parameter. For the reddening, we find a range of 0.18 $<$ E(B-V) $<$ 0.98, with an median value of E(B-V)=0.53$\pm$0.09. There is an outlier on Figure \ref{fig:trans}: the type 1 AGN F15462-0405. For this object the BLR H$\delta$ emission is blended with the [SII]$\lambda\lambda$4068,4076 doublet, and therefore it is difficult to separate the contributions from the [SII] doublet and H$\delta$, rendering the  TR([SII]) ratio measured for object less accurate. Nonetheless, taking into
account the uncertainties, the trans-auroral ratios measured for this object are consistent with with a low reddening for this object, in line  with the
results obtained using the ratios of the total hydrogen-line flux ratios (see $\S$\ref{sect:red}).

\begin{center}
\begin{table*}
\centering
\caption{The log$_{10}$ n$_e$ and E(B-V) values determined using the total emission-line fluxes for the ULIRG sample.}
\begin{tabular}{lcccc}
\hline							
Object	&	log$_{10}$[SII] & log$_{10}$[OII] &	Log$_{10}$ n$_e$ (cm$^{-3}$)	&	E(B-V)	\\
\hline							
F12072-0444	& -1.46$\pm$0.11	& 0.91$\pm$0.07	&	2.75$_{-0.25}^{+0.25}$	&	0.53$_{-0.13}^{+0.10}$	\\
F13305-1739	& -0.90$\pm$0.17	& 0.62$\pm$0.07	&	3.75$_{-0.30}^{+0.25}$	&	0.33$_{-0.15}^{+0.15}$	\\
F13443+0802SE	& -1.41$\pm$0.10	& 1.10$\pm$0.08	&	2.55$_{-0.50}^{+0.30}$	&	0.43$_{-0.05}^{+0.08}$	\\
F13451+1232W	& -0.71$\pm$0.08	& -0.03$\pm$0.02	&	4.45$_{-0.20}^{+0.20}$	&	0.38$_{-0.05}^{+0.08}$	\\
F14378-3651	& -1.95$\pm$0.05	& 0.64$\pm$0.14	&	$\sim$2.0-2.25	&	0.90$_{-0.08}^{+0.08}$	\\
F15130-1958	& -1.26$\pm$0.27	& 0.47$\pm$0.29	&	3.45$_{-0.60}^{+0.50}$	&	0.58$_{-0.25}^{+0.30}$	\\
F15462-0450$^{a}$ & 0.07$\pm$0.43	& 0.95$\pm$0.33	&	4.15$_{-0.50}^{+0.85}$	&	$<$0	\\
F16156+0146NW	& -1.35$\pm$0.08	& 0.63$\pm$0.13	&	3.15$_{-0.10}^{+0.20}$	&	0.58$_{-0.15}^{+0.10}$	\\
F19254-7245S	& -1.34$\pm$0.07	& -0.23$\pm$0.09	&	3.85$_{-0.10}^{+0.10}$	&	0.88$_{-0.05}^{+0.05}$	\\				
\hline			
\end{tabular}
\begin{tablenotes}
\item[a]$^a$ F15462-0405 is an outlier on Figure \ref{fig:trans}. The density value s consistent with the projection of the measured [OII] ratio with associated 1$\sigma$ errors onto the zero reddening curve.  
\end{tablenotes}
\label{tab:dens}
\end{table*}
\end{center}

\begin{figure}
\centering
\includegraphics[scale=0.38, trim=0.1cm 4.6cm 0.1cm 3.6cm]{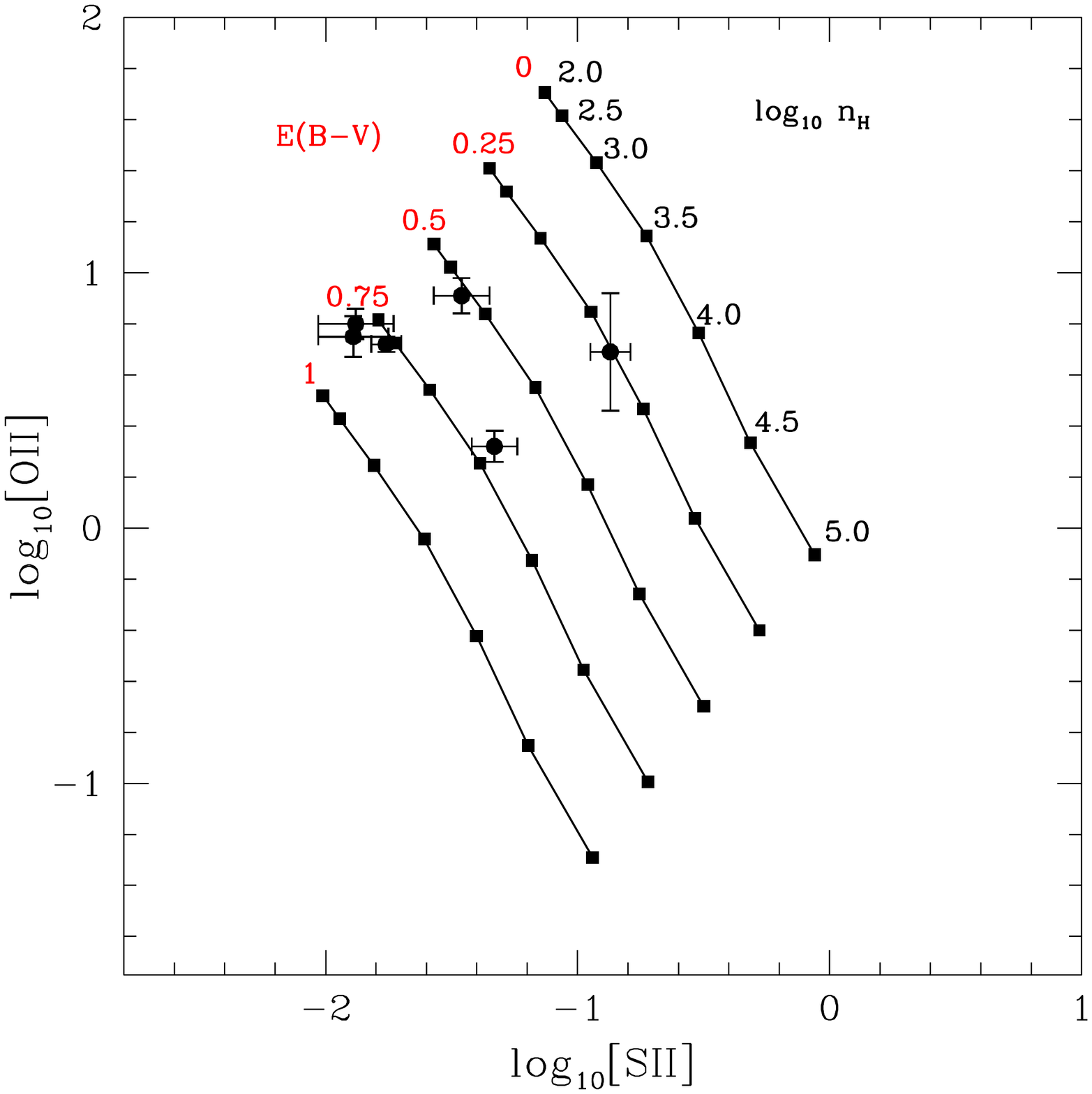}
\caption{Determining the densities and reddenings from the fluxes of the narrow components of the trans-auroral [OII] and [SII] emission-lines. Log$_{10}$ (3726+3729)/(7319+7331) is plotted against log$_{10}$ [SII](4068+4076)/(6717+6731). The overplotted gridlines are described in the caption of Figure \ref{fig:trans}. The black points indicate the trans-auroral ratios measured for the narrow components of the emission-line profiles of the ULIRGs.}
\label{fig:transn}
\end{figure}

In Figure \ref{fig:transn} we plot log(TR([OII])) vs. log(TR([SII])) for the narrow components only, with the resulting density and E(B-V) estimates presented in Table \ref{tab:densn}. Note that for F15130-1958, F15462-0450 and F19254-7245S the narrow components of the [SII]$\lambda\lambda$4068,4076 were not detected and therefore we do not include them in this analysis. Taking into account the error bars, we find electron densities in the range 10$^{2.00}$ $<$ n$_{e}$ $<$ 10$^{3.95}$ cm$^{-3}$, with an median density of log$_{10}$ n$_e$=2.45$\pm$0.31. 
In addition, we determine the n$_{e}$ for the narrow components using the unconstrained [SII] 6717/6731 intensity ratio alone (see Table \ref{tab:densn}; \citealt{osterbrock06}). For every object we find that the narrow components indicate a density range of 1.6 $<$ log$_{10}$ n$_e$ $<$ 2.8, with an median density of log$_{10}$ n$_e$=2.33$\pm$0.11. These densities are generally lower than those found using the trans-auroral density diagnostics, as expected given that the trans-auroral ratios are sensitive to higher density gas. 

\begin{center}
\begin{table*}
\centering
\caption{The log$_{10}$ n$_e$ values determined using the [SII] 6717/6731 intensity ratio and the trans-auroral diagnostics for the narrow components.}
\begin{tabular}{lcccc}
\hline		
Object	&	log$_{10}$[SII] & log$_{10}$[OII] & [SII]--n$_e$ (cm$^{-3}$)	&	Trans. n$_e$	(cm$^{-3}$)	\\
\hline									
F12072-0444	& -1.46$\pm$0.11	& 0.91$\pm$0.07	&		2.240$_{-0.033}^{+0.031}$	&	2.75$_{-0.25}^{+0.25}$		\\		
F13305-1739	& -0.87$\pm$0.08	& 0.69$\pm$0.23	&	2.477$_{-0.046}^{+0.044}$	&	3.75$_{-0.30}^{+0.20}$		\\		
F13443+0802SE	& -1.89$\pm$0.14	& 0.75$\pm$0.08	&	$<$1.669	&	$<$2.0	\\		
F13451+1232W	& -1.88$\pm$0.15	& 0.80$\pm$0.06	&	2.302$_{-0.054}^{+0.051}$	&	$<$2.0	\\		
F14378-3651	& -1.76$\pm$0.06	& 0.72$\pm$0.03	&	2.573$_{-0.034}^{+0.033}$	&	2.15$_{-0.15}^{+0.30}$	\\		
F15130-1958	& -	& -	&	2.740$_{-0.182}^{+0.172}$	&	-	\\		
F15462-0450	& -	& -	&	2.389$_{-0.110}^{+0.098}$	&	-	\\		
F16156+0146NW	& -1.33$\pm$0.09	& 0.32$\pm$0.06	&	2.078$_{-0.295}^{+0.196}$	&	3.45$_{-0.20}^{+0.20}$	\\		
F19254-7245S	& - 	& -	&	2.614$_{-0.072}^{+0.068}$	&	-	\\		

\hline			
\end{tabular}
\label{tab:densn}
\end{table*}
\end{center}

\begin{figure}
\centering
\includegraphics[scale=0.38, trim=0.1cm 4.6cm 0.1cm 3.6cm]{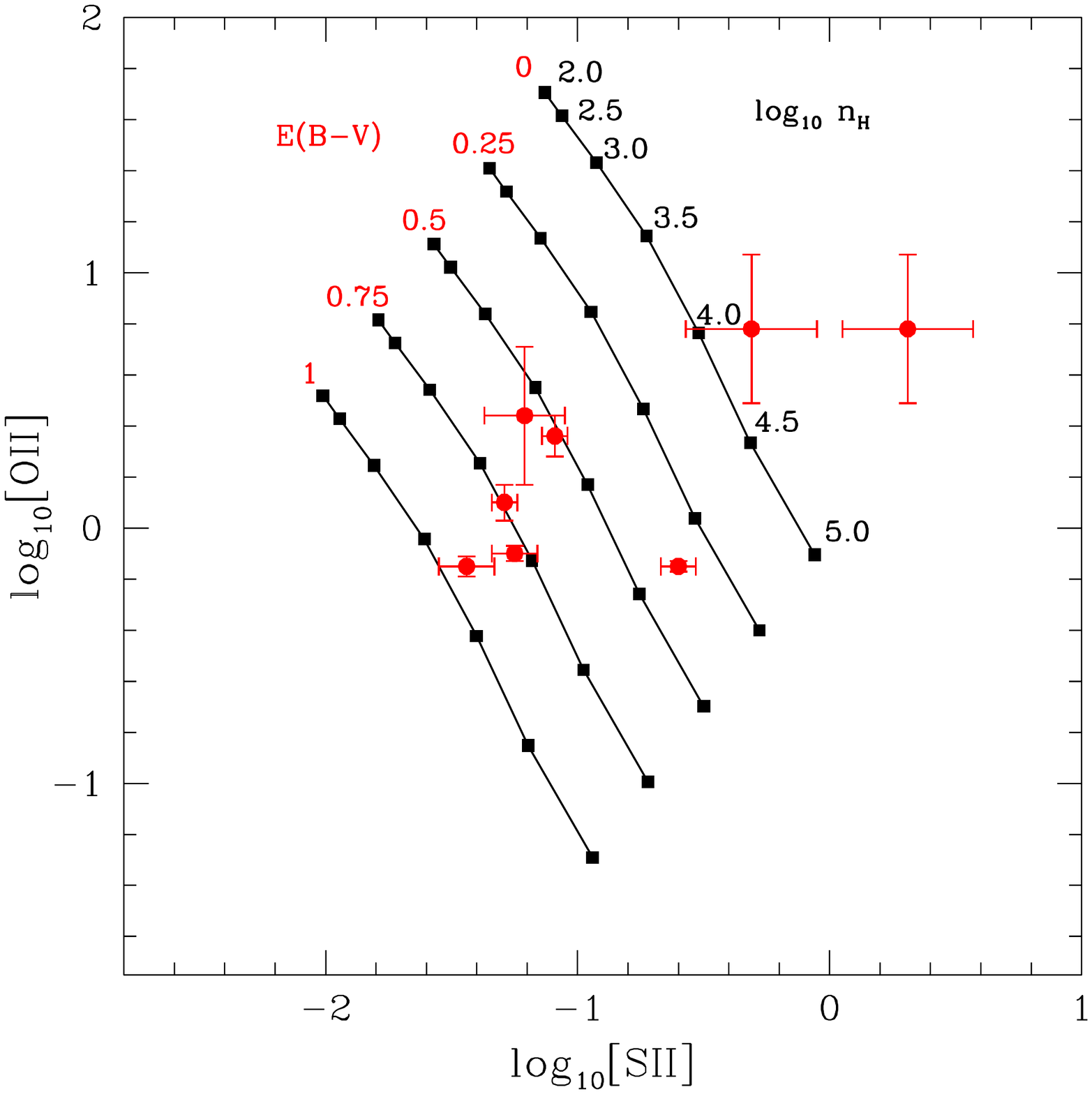}
\caption{Determining the densities and reddenings from the fluxes of the broad and intermediate trans-auroral [OII] and [SII] emission-line components. Log$_{10}$ (3726+3729)/(7319+7331) is plotted against log$_{10}$ [SII](4068+4076)/(6717+6731). The overplotted gridlines are described in the caption of Figure \ref{fig:trans}. The red points indicate the trans-auroral ratios measured for the  broad plus intermediate components of the emission-line profiles for the ULIRGs.}
\label{fig:transb}
\end{figure}

In Figure \ref{fig:transb} we plot log(TR([OII])) vs. log(TR([SII])) for the broad/intermediate components only, with the resulting
density and temperature estimates presented in Table \ref{tab:densb}. Taking into account the error bars, we find electron densities in the range 10$^{3.05}$ $<$ n$_{e}$ $<$ 10$^{4.65}$ cm$^{-3}$, with a median density of log$_{10}$ n$_e$=3.70$\pm$0.16. This is similar to the densities found for the total emission-line fluxes. Interestingly, when comparing the densities indicated by the narrow and broad components for each object separately, the broad/intermediate components indicate higher electron densities when compared to the narrow components. For the dust reddening we find a range of 0 $<$ E(B-V) $<$ 1.03, with an median value of E(B-V)=0.55$\pm$0.12. This is consistent with the reddening values found for the total profiles.

Note that for one object, F14378-3651, the broad/intermediate components for TR([OII]) and TR([SII]) were not detected. Therefore, when using n$_e$ in further calculations for the outflow in F14378-3651, we use median electron density measured for the broad kinematic components in the sample as a whole (i.e. log$_{10}$ n$_e$=3.70$\pm$0.16).  


\begin{center}
\begin{table*}
\centering
\caption{The log$_{10}$ n$_e$ and E(B-V) values determined using the broad and intermediate components of the emission-line profiles for the ULIRG sample.}
\begin{tabular}{lcccc}
\hline		
Object	&	log$_{10}$[SII] & log$_{10}$[OII] &	Log$_{10}$ n$_e$ (cm$^{-3}$)	&	E(B-V)	\\
\hline			
F12072-0444	& -1.44$\pm$0.11 & -0.15$\pm$0.04 &	3.75$_{-0.10}^{+0.10}$	&	0.93$_{-0.10}^{+0.10}$	\\
F13305-1739	& -0.31$\pm$0.26 & 0.78$\pm$0.29 &	4.15$_{-0.20}^{+0.15}$	&	---	\\
F13443+0802SE	& -1.25$\pm$0.09 & -0.10$\pm$0.03 &	3.65$_{-0.10}^{+0.10}$	&	0.73$_{-0.05}^{+0.05}$	\\
F13451+1232W	& -0.60$\pm$0.07 & -0.15$\pm$0.02 &	4.55$_{-0.10}^{+0.10}$	&	0.38$_{-0.10}^{+0.10}$	\\
F15130-1958	& -1.21$\pm$0.16 & 0.44$\pm$0.27 &	3.45$_{-0.40}^{+0.90}$	&	0.58$_{-0.20}^{+0.15}$	\\
F15462-0450$^{a}$	& 0.31$\pm$0.26 & 0.78$\pm$0.29 &	4.75$_{-0.25}^{+0.25}$	&	---	\\
F16156+0146NW	& -1.09$\pm$0.05 & 0.36$\pm$0.08 &	3.65$_{-0.10}^{+0.10}$	&	0.53$_{-0.05}^{+0.05}$	\\
F19254-7245S	& -1.29$\pm$0.05 & 0.10$\pm$0.07 &	3.65$_{-0.10}^{+0.10}$	&	0.73$_{-0.05}^{+0.05}$	\\
\hline			
\end{tabular}
\begin{tablenotes}
\item[a]$^a$ F15462-0405 is an outlier on Figure \ref{fig:transb}. The density estimated here is consistent with the
projection of the [OII] trans-auroral ratio and associated 1$\sigma$ error onto the zero reddening line.  
\end{tablenotes}
\label{tab:densb}
\end{table*}
\end{center}

\subsubsection{Reddening comparison}

\begin{figure}
\centering
\includegraphics[scale=0.38, trim=0.1cm 4.6cm 0.1cm 3.6cm]{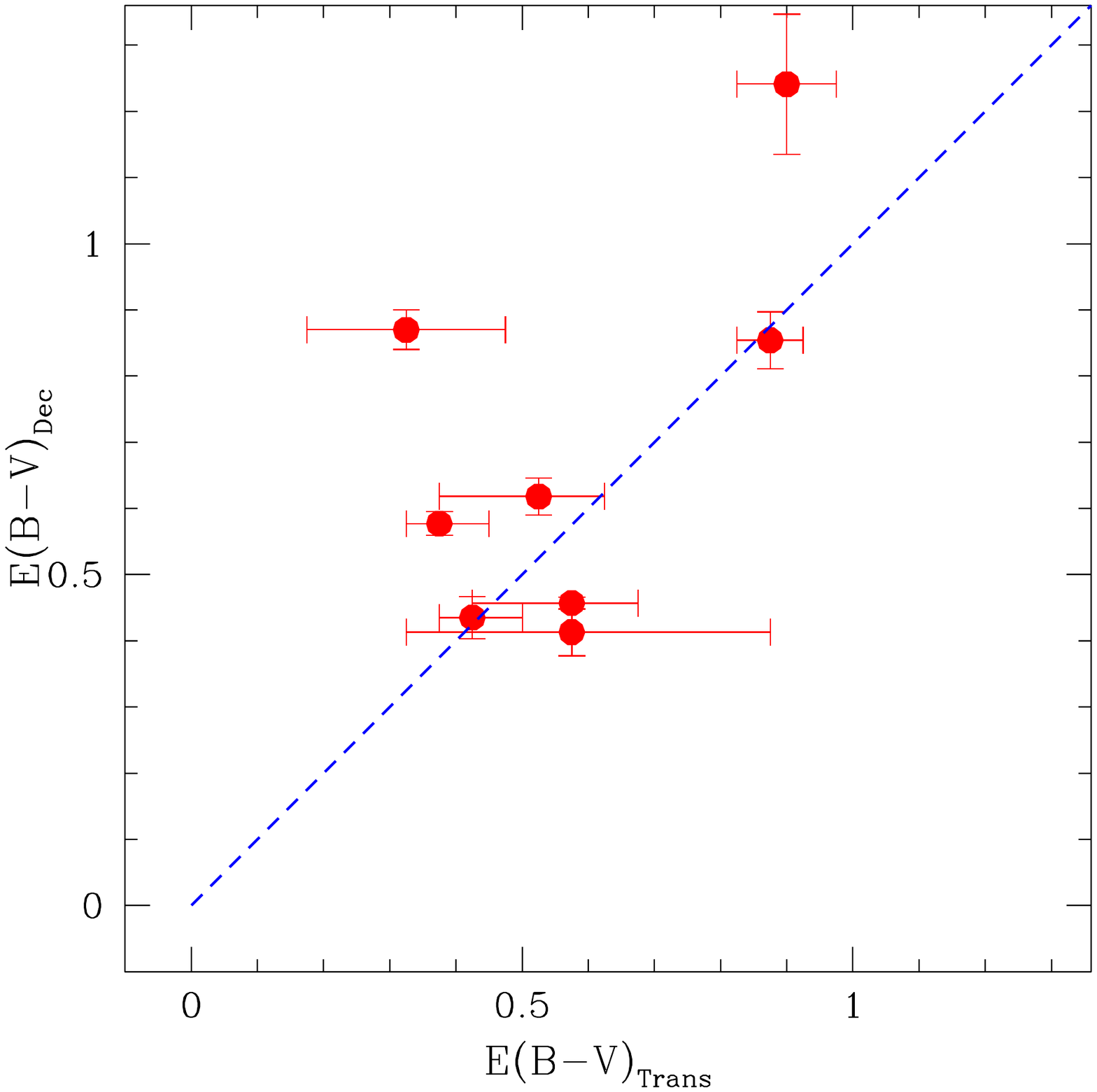}
\caption{E(B-V)$_{Dec}$ versus E(B-V)$_{Trans}$ for the total fluxes of the ULIRG sample. The blue dashed line indicates the one-to-one relationship.}
\label{fig:ext}
\end{figure}

Given that we have determined the reddening of the emission regions of the ULIRGs using two different methods, we are in a position to investigate the robustness of the different methods by directly comparing them. In Figure \ref{fig:ext} we plot the median E(B-V) as determined from the hydrogen line ratios (E(B-V)$_{DEC}$) in $\S$\ref{sect:red}, against the central E(B-V) as determined in $\S$\ref{sect:dens} (E(B-V)$_{Trans}$), for the total emission-line fluxes for both the hydrogen and trans-auroral emission-lines. 
All of the objects lie within three error bars of the one-to-one line (blue dashed line).

\section{Importance of the AGN-driven outflows in ULIRGs}

\subsection{Mass outflow rates and kinetic powers} \label{sect:outf}

In order to understand the role that AGN-driven outflows play in the evolution of galaxies, it is important to accurately quantify the mass outflow rates and the kinematic powers of the outflows. This is especially important given that there is a wide range in the AGN-driven outflow powers required by the models --- from 0.5\%  \citep{hopkins10} up to 10\% (\citealt{fabian99}; \citealt{springel05}) of the AGN bolometric luminosity. 

To determine the key outflow properties, we start by calculating the total gas mass in the outflow ($M$):

\begin{equation}
M = \frac{L(H\beta) m_{p}}{\alpha_{H\beta}^{eff} h \nu_{H\beta} n_{e}} .
\label{eqn:masstot2}
\end{equation}

\noindent where $L({H\beta})$ is the the H$\beta$ luminosity,  m$_p$ is the mass of the proton, $\alpha_{H\beta}^{eff}$ is effective Case B recombination coefficient of H$\beta$, $h$ is Planck's constant, $\nu_{H\beta}$ is the frequency of H$\beta$, and $n_{e}$ is the electron density of the outflowing gas.

We calculate the dynamical time scale of the outflow ($\tau$) using the average velocity of the outflowing gas ($\nu_{out}$) and the characteristic radius of the outflow ($r$):

\begin{equation}
\tau = r / v_{out} .
\label{eqn:time}
\end{equation}

The mass outflow rate ($\dot{M}$) is then the total mass of the outflowing gas divided by the time scale of the outflow: 

\begin{equation}
\dot{M} =\frac{L(H\beta )m_p v_{out}}{\alpha_{H\beta}^{eff}h\nu_{H\beta}n_e r} .
\label{eqn:mass}
\end{equation}

To determine the mass outflow rates, we use the reddening corrected (see $\S$\ref{sect:red}) H$\beta$ luminosity ($L_{H\beta}$) for the outflowing component, the flux weighted $v_{out}$ values (see $\S$\ref{sect:kine}), the broad component $n_e$ values determined from the trans-auroral diagnostics (see $\S$\ref{sect:dens}), and the radii of the outflows (R$_{[OIII]}$) as estimated using the 2D Xshooter spectra (see $\S$\ref{tab:spatial}) to approximate $r$ or, where available, estimates based on HST ACS/STIS data (see $\S$\ref{tab:spatialcomp}). 

The kinetic powers of the outflows ($\dot{E}$) are then determined using $\dot{M}$ in the following equation:

\begin{equation} 
\dot{E} =\frac{\dot{M}}{2}\left (v_{out}^{2} + 3\sigma^2 \right ) ,
\label{eqn:pow}
\end{equation}

\noindent where $\sigma$ is the velocity dispersion ($\sigma \sim FWHM/2.355$), which has been calculated using the flux weighted FWHM of the outflowing gas (see $\S$\ref{sect:kine}). We then determine what fraction of the AGN power ($F_{kin}$) in the outflows by dividing $\dot{E}$ by $L_{BOL}$, with the latter estimated using the bolometric corrections presented in $\S$\ref{sect:lbol}. Note that we use the \citet{heckman04} correction for the majority of the sample. However, for F14378-3651 and F19254-7245S we use the \citet{lamastra09} correction, which we have argued is more appropriate for these objects (see $\S$\ref{sect:lbol}). 

Note that in this method we do not correct the velocities for projection effects, and assume that $v_{out}$ represents the true outflow velocity, while $\sigma$ represents the
turbulence in the outflowing gas. Given the strong dependence of the outflow kinetic power on $v_{out}$ (i.e. $\dot{E} \propto v_{out}^3$ for small $\sigma$), these conservative assumptions are likely to lead to lower limits on
the true outflow masses and kinetic powers. In $\S$\ref{sect:outmax} below we consider an alternative approach to calculating the outflow velocities that is less conservative.

The derived $\dot{M}$, $\dot{E}$, and $F_{kin}$ values are presented in Table \ref{tab:out2d}, where the upper and lower limits reflect the range in the estimated $n_e$ of the outflowing gas using the trans-auroral diagnostics. Note that the electron density used in the calculations for F14378-3651 is the median value of the broad component density for the ULIRG sample as a whole, because the broad components were not detected for all the trans-auroral blends in F14378-3651. 

\begin{center}
\begin{table*}
\centering
\caption{The key properties of the ionized, outflowing gas in the ULIRG sample. `$\dot{M}$' gives the range in the mass outflow rate of the gas in M$_{\sun}$ yr$^{-1}$. `$\dot{E}$' gives the range in the power of the outflowing gas in erg s$^{-1}$. `$F_{kin}$' gives the power of the outflowing gas as a fraction of the bolometric luminosity of the ULIRG as a percentage.}
\begin{tabular}{lccc}
\hline
Object	&	$\dot{M}$ (M$_{\sun}$ yr$^{-1}$)	&	$\dot{E}$ (erg s$^{-1}$)	&	$F_{kin}$ (\%)	\\
\hline							
F12072-0444	&	1.9$_{-0.6}^{+0.8}$	&	(3.9$_{-1.4}^{+2.0}$)$\times10^{41}$	&	(1.7$_{-0.5}^{+0.6}$)$\times10^{-2}$	\\
F13305-1739	&	$>$1.5	&	$>$5.1$\times10^{41}$	& $>$2.6$\times10^{-3}$		\\
F13443+0802SE	&	0.0064$_{-0.0033}^{+0.0059}$	&	(5.4$_{-2.9}^{+5.2}$)$\times10^{38}$	&	(4.3$_{-2.6}^{+4.9}$)$\times10^{-5}$	\\
F13451+1232W	&	2.8$_{-1.0}^{+1.5}$	&	(1.6$_{-0.7}^{+1.1}$)$\times10^{42}$	&	(3.3$_{-1.5}^{+2.4}$)$\times10^{-2}$	\\
F14378-3651	&	$>$0.039	&	$>$1.5$\times10^{40}$	&	$>$2.3$\times10^{-3}$	\\
F15130-1958	&	2.5$_{-2.3}^{+6.4}$	&	(1.1$_{-1.0}^{+3.0}$)$\times10^{42}$	&	0.15$_{-0.14}^{+0.42}$	\\
F15462-0450	&	$>$0.020	&	$>$6.2$\times10^{39}$	&	$>$4.2$\times10^{-4}$	\\
F16156+0146NW	&	0.93$_{-0.47}^{+0.87}$	&	(2.1$_{-1.2}^{+2.4}$)$\times10^{41}$	&	(8.3$_{-5.0}^{+9.8}$) $\times10^{-3}$	\\
F19254-7245S	&	$>$1.4	&	$>$5.2$\times10^{41}$	&	$>$3.2$\times10^{-2}$	\\
\hline
\end{tabular}
\label{tab:out2d}
\end{table*}
\end{center}

In the cases of the objects for which R$_{[OIII]}$ is resolved we find mass outflow rates in the range 0.006 $<$ $\dot{M}$ $<$ 3 M$_{\sun}$ yr$^{-1}$ and kinetic powers in the range 5.4$\times$10$^{38}$ $<$ $\dot{E}$ $<$ 1.6$\times$10$^{42}$ erg s$^{-1}$.  This range in $\dot{E}$ does not overlap with those calculated by \citet{harrison12}, \citet{liu13} and \citet{mcelroy15}. However, it is consistent with the ranges reported in \citet{harrison14}, and \citet{montse16} (see Table \ref{tab:parcomp}). In general, we find lower $\dot{E}$ values than in the previous studies.

In terms of the comparison with L$_{BOL}$, we find fractions in the range 4$\times$10$^{-5}$ $<$ $F_{kin}$ $<$ 0.2\%. This shows a spread of over 4 orders of magnitude. However, we note that a substantial part of this spread is likely to be due to the uncertainties in the $L_{BOL}$ estimates. When compared to theoretical expectations, most of the outflow power fractions appear to fall below even the lowest model estimates (e.g. $F_{kin}\sim$0.5\%; \citealt{hopkins10}). 

Finally, we find that the lower limits on $\dot{M}$, $\dot{E}$ and $F_{kin}$ for the four objects in
which we do not resolve the outflow regions -- F13305-1739, F14378-3651, F15462-0450 and F19254-7245S -- are consistent with the ranges found for the objects in which the outflows are resolved.

\subsection{Estimates based in maximal outflow velocities} \label{sect:outmax}

The estimates made for $\dot{M}$, $\dot{E}$ and $F_{kin}$ in the last section are likely to represent
conservative lower limits because they do not take into account the projection effects when
estimating the outflow velocities.

Less conservative, but perhaps more realistic, estimates can be obtained by assuming that the broadening of the emission-lines is entirely due to different projections of the velocity vector to the line of sight of an expanding
outflowing bubble of gas, rather than due to turbulence. In this case, the maximum velocities measured in the wings of the line profiles represent the parts of the outflow moving straight towards or away from us, and hence the true, de-projected outflow velocities. To determine these maximum velocities, we use non-parametric measures. First we use the [OIII] model presented in Table \ref{tab:gfits} to construct a function that describes the velocity profile of the outflowing [OIII]$\lambda$5007 emission components (i.e. the sum of the broad and intermediate Gaussian components). From this we determine the velocities measured relative to the galaxy rest frame that encompass 5\% (for most objects) or 95\% (for F19254-7245S)  of the total line flux (v$_{5}$ or v$_{95}$). Note we use the 5\% velocity rather than a velocity than encompasses a smaller fraction of the flux  to avoid any inaccuracy caused by the continuum. subtraction. We present the maximum velocities in Table \ref{tab:outc}. Note that, under this assumption, we remove the turbulence term ($\sigma$) from the equation for the kinetic power, because we are assuming that the line broadening is entirely due
to integrating different projections of the velocity vector across the (spatially unresolved) outflow. 

\begin{center}
\begin{table*}
\centering
\caption{The key properties of the ionized, outflowing gas in the ULIRG sample using the maximum possible outflow velocity. `$\dot{M}$' gives the mass out flow rate in M$_{\sun}$ yr$^{-1}$. `$\dot{E}$' gives the outflow power in erg s$^{-1}$. `$F_{kin}$' gives the fraction of the outflow power with respect to the bolometric luminosity of the ULIRG as a percentage.}
\begin{tabular}{lccc}
\hline
Object	&	$\dot{M}$ (M$_{\sun}$ yr$^{-1}$)	&	$\dot{E}$ (erg s$^{-1}$)	&	$F_{kin}$ (\%)	\\
\hline							
F12072-0444	&	$5.1_{-1.5}^{+2.1}$	&	$(2.3_{-0.7}^{+0.9})\times10^{42}$	&	$(9.5_{-2.1}^{+2.7})\times10^{-2}$	\\
F13305-1739	&	$>11$	&	$>4.9\times10^{42}$	&	$>2.5\times10^{-2}$	\\
F13443+0802SE	&	$0.069_{-0.029}^{+0.48}$	&	$(6.4_{-2.7}^{+4.4})\times10^{39}$	&	$(5.0_{-1.8}^{+2.8})\times10^{-4}$	\\
F13451+1232W	&	$10.5_{-2.3}^{+2.8}$	&	$(2.2_{-0.5}^{+0.6})\times10^{43}$	&	$0.45_{-0.12}^{+0.15}$	\\
F14378-3651	&	$>$0.082	&	5.7$\times10^{40}$	&	8.4$\times10^{-3}$	\\
F15130-1958	&	$6.1_{-5.6}^{+15.2}$	&	$(5.9_{-5.4}^{+14.6})\times10^{42}$	&	$0.80_{-0.74}^{+2.17}$	\\
F15462-0450	&	$>0.10$	&	$>1.2\times10^{41}$	&	$>0.008$	\\
F16156+0146NW	&	$6.6_{-2.8}^{+4.9}$	&	$(5.5_{-2.4}^{+4.1})\times10^{41}$	&	0.21$_{-0.09}^{+0.19}$	\\
F19254-7245S	&	$>46$	&	$>2.5\times10^{43}$	&	$>1.5$	\\
\hline
\label{tab:eout}
\end{tabular}
\end{table*}
\end{center}

Table \ref{tab:eout} presents the $\dot{M}$, $\dot{E}$ and $F_{kin}$ values for the ULIRGs obtained using the maximum outflow velocities. Considering the values for cases where the outflows are spatially resolved, we find that the values are higher than those estimated in $\S$\ref{sect:outf}: 0.07 $<$ $\dot{M}$ $<$ 11 M$_{\sun}$ yr$^{-1}$; 6.3$\times$10$^{39}$ $<$ $\dot{E}$ $<$ 2.3$\times$10$^{43}$ erg s$^{-1}$; 4$\times10^{-4}$ $<$ $F_{kin}$ $<$ 1\% for the objects in which the outflows
are resolved.
The latter estimates are more consistent, at least within their uncertainties, with the `multi-staged' outflow model proposed by \citet{hopkins10}. It is only in the case of F19254-7245S, which has an upper limit on the radius, and hence lower limits on  
$\dot{M}$, $\dot{E}$ and $F_{kin}$, that there evidence for a more massive and powerful warm outflow\footnote{For this object, however, the evidence for the existence of an outflow is perhaps less convincing than in the other objects in the sample, because although its emission-lines are broad, the are redshifted rather than blueshifted.}. 

\begin{center}
\begin{table*}
\centering
\caption{Comparison between the outflow properties deduced in various recent studies of emission-line outflows 
in samples of luminous AGN.}
\begin{tabular}{lllcccl}
\hline
Study &Sample &$N_{obj}$ &$R_{out}$ &$\dot{M}$ &log$_{10}$$\dot{E}$ &Method \\
& & &(kpc) &(M$_{\odot}$ yr$^{-1}$) &(erg s$^{-1}$) & \\
\hline
This paper &ULIRGs, $z<0.15$, & 9 & 0.05--1.2 &0.004--$3$ 
& 38.7--42.2 &Conservative $v_{out}$ \\
           & with optical AGN  & &''\hspace{0.5cm}'' &0.07--11 
& 39.8--42.9 & Less conservative $v_{out}$ \\
\\
Harrison et al. & Type 2 quasars, & 10 & 2--10 & - & 43.9--45.5 &Energy conserving bubble \\
 (2012) & ULIRGs, 1.4$<$z$<$3.0  & & & & & \\
\\
Liu et al. &Type 2 quasars,  & 11 & $\sim$15 & 11--56 & 43.3--44.3 & Similar to this paper \\
 (2013) &z$\sim$0.5 & &''\hspace{0.5cm}'' &3,300--20,000 & 44.6--45.5 &Clouds embedded in a wind \\
\\
Harrison et al. & Luminous type 2  & 16 & 1.5--4.3 & 3--70 & 41.5--43.5 & Similar to this paper \\
 (2014) & AGN, z$<$0.2  & & 6 & 9,000--23,000 & 43.2--45.7 & Energy conserving bubble \\
\\
McElroy et al. & Type 2 quasars,  & 17 & 1.6--8.3 &167--3,333 & 42.7--44.3 &  Similar to  this paper \\
(2015) &z$<$0.11  &  &  &  &  & \\
\\
Villar Mart\'{i}n  & Luminous type 2  & 15 & 0.12--2.1  & 0.1--114 & 37.5--44.1 & Similar to this paper\\ 
et al. (2016) &AGN, z$<$0.6  &  &  &  &  & \\
\\
Sun et al. &Luminous type 2  & 12 & 2.5--33.5 & - & 41.5--43.8 & Similar to this paper \\ 
(2017) &AGN, z$<$0.2  & & 1.5--8.2 & - & 43.8--45.8 & Energy conserving bubble  \\
\hline
\label{tab:parcomp}
\end{tabular}
\end{table*}
\end{center}

In general, however, even with this less conservative assumption about the true outflow velocities, the mass outflow rates and outflow kinetic powers remain significantly lower than those determined in some recent studies \citep[][see Table \ref{tab:parcomp} for a comparison]{liu13,harrison14,mcelroy15,sun17}, but are consistent with those deduced by \citet{montse16}. Concentrating first on estimates of $\dot{M}$ and $\dot{E}$ obtained using methods similar to that of this paper, the various studies make different assumptions about how the de-projected outflow velocities and radii are determined from the data. However, the largest single factor that causes the apparent discrepancies in the calculated $\dot{M}$ and $\dot{E}$ values is the different assumptions made about the electron densities: \citet{liu13}, \citet{harrison14}, \citet{mcelroy15} \& \citet{sun17} all assume relatively low densities ($n_e \sim$ 100 cm$^{-3}$) when compared to this work and that of \citet{montse16}. 

Considering the previous studies that used the alternative method of assuming an energy-conserving bubble expanding into a uniform medium, the $\dot{E}$ values are considerably higher
than those obtained with the method used in this paper. However, the latter values are likely to substantially over-estimate the true kinetic powers of the outflows, because
the assumption of a {\it uniform} medium of relatively high density (typically $n_e \sim 0.5$~cm$^{-3}$)\footnote{This density is comparable to, or higher than,  the central densities of the hot ISM ($T > 10^7$~K) in the centres of rich clusters of galaxies; although the cooler phases of the ISM will have a higher densities, their
volume filling factors are expected to be much lower.} leads to unrealistically high total outflow gas masses (e.g. $M_{gas} \sim 5\times10^{10}$~M$_{\odot}$ for $n_e = 0.5$~cm$^{-2}$ and $R_{out} = 10$~kpc: $\sim$10$\times$ the total gas mass of the Milky Way).

It is also notable that the masses and kinetic powers we estimate for the warm, AGN-driven outflows are a factor 10 -- 1000$\times$ lower than those deduced for
the neutral and molecular outflows detected on similar spatial scales in the near-nuclear regions of some ULIRGs (\citealt{rupke05}; \citealt{cicone14}, \citealt{gonz17},\citealt{veilleux17}). The warm outflows would only achieve parity in terms of their $\dot{M}$, $\dot{E}$ and $F_{kin}$ values with the neutral and molecular outflows if the electron densities were orders of magnitude lower than we have assumed -- more in line with the density estimates deduced for the narrow kinematic components using the [SII]6731/6717 ratio (see Table \ref{tab:densn}).

It has been argued that, given the high critical densities of the [OII]$\lambda\lambda$7319,7330 and [SII]$\lambda\lambda$4068.4076 lines, the trans-auroral diagnostics may be biased towards higher density clumps embedded in a lower density medium, with the latter containing most of the mass in the outflows \citep{sun17}. However, we note that any high density clumps would nonetheless be expected to radiate strongly in the hydrogen recombination lines that we have used to estimate the gas masses. Moreover, the electron densities that we measure from the trans-auroral line ratios remain below the critical density of the [OIII]$\lambda$5007 line ($n_{crit} = 7\times10^5$~cm$^{-3}$) used to determine the emission-line kinematics and outflow radii. Therefore, depending on the ionization level, we might also expect the high density clumps  to radiate significant [OIII]$\lambda$5007 emission. Indeed, the electron densities we derive in this paper are consistent with those recently deduced for the kpc-scale outflows responsible for the blue-shifted high-ionisation broad absorption line (BAL) or narrow absorption line (NAL) systems detected in some type 1 AGN (see $\S$\ref{sect:abs}). While we cannot entirely rule the presence of a lower density outflow component that has a high mass but contributes relatively little to the emission-line fluxes, there is currently no clear observational evidence for such a component in observations of the near-nuclear regions of ULIRGs.

\section{Comparison with absorption-line outflows} \label{sect:abs}

Whereas in this paper we have investigated the warm, AGN-driven outflows using emission line observations -- a typical approach for type 2 or type 1 AGN observed at optical wavelengths -- an alternative technique is to use observations of outflows detected as blue-shifted broad- or narrow-line absorption line systems in the UV resonance lines of type 1 AGN \citep{weymann81,crenshaw03}. However, there has been relatively little work done
on the relationship between kinematically-disturbed absorption- and emission-line components: are these phenomena entirely distinct, or do they represent the same outflows observed in different ways, perhaps depending on the inclination of the systems relative to the line of sight?

In this context, it is interesting that the relatively high densities ($3\times10^3 < n_e < 6\times10^4$ cm$^{-3}$) and compact scales
($0.05 < R_{[OIII} < 1.2$ kpc) that we determine for the emission-line outflows in ULIRGs 
are similar to those deduced for the outflows detected as blueshifted broad- or narrow-line absorption systems in the UV spectra
of a type 1 AGN (\citealt{dekool01}; \citealt{moe09}; \citealt{dunn10}; \citealt{borguet13}; \citealt{aravall}; \citealt{chamberlain15a}; \citealt{chamberlain15b}). Therefore it is important
to consider whether in the ULIRG emission-line outflows would be detected as BAL or NAL systems if they were intercepted by a line of sight to their (mainly hidden) type 1 AGN nuclei.

With  measured  velocity shifts  of -100 $>$ $\Delta$V $>$ -2000 km s$^{-1}$ and line widths of $600 < FWHM < 2000$ km s$^{-1}$ (see Table \ref{tab:gfits}), the kinematics of the emission-line outflows in ULIRGs are typically less extreme than those observed in BAL
quasars ($v_{max} > 5000$~km s$^{-1}$ and $FWHM > 2000$ km s$^{-1}$; \citealt{weymann81}), and  it is doubtful that this difference is entirely due to projection effects. On the other hand, while the velocity shifts 
of the emission-line outflows in ULIRGs cover a similar range to NAL systems, the line widths are generally larger than observed in such systems ($FWHM < 400$ km s$^{-1}$; \citealt{crenshaw03}). However, it is important to
recognise that in individual NAL systems we only see the outflow components that happen to intercept the line of sight (LOS) to the type 1 AGN, and that there may be many outflowing clouds with different velocities that do not intercept the LOS. Therefore it is plausible that, if we could integrate all the possible LOS to the AGN in such systems, the
absorption lines would appear as broad as the outflowing emission-line components in the ULIRGs. In this case, the ULIRG emission-line outflows may be linked to the NAL outflow phenomenon. 

\section{Conclusions}

By taking advantage of the wide spectral coverage and high spectral resolution of Xshooter on the VLT we have, for the first time, accurately quantified the mass outflow rates and kinetic powers of the warm, AGN-driven outflows in some of the most rapidly evolving galaxies in the local universe. Our analysis of a representative sample nearby ULIRGs 
from the QUADROS project reveals the following.

\begin{itemize}

\item [-] The radial extents of the warm outflows in all the objects in our sample are small ($R_{[OIII]}$ $<$ 1.2kpc) when compared to galaxy scales.

\item [-] If we make the conservative assumption that the flux-weighted mean velocity shift of the broad and intermediate kinematic components represents the true outflow velocity, the estimated mass outflow rates and kinetic powers are relatively modest ($6\times10^{-3} < \dot{M} < 3$ M$_{\sun}$ yr$^{-1}$; 
$4\times10^{-5} < \dot{E}/L_{BOL} < 0.2$\%). This is lower than the rates estimated for other samples of AGN (e.g., \citealt{liu13}; \citealt{harrison14}), and  required
in some models (5-10\%; \citealt{fabian99}; \citealt{dimatteo05}), casting doubt on whether the warm outflows alone are capable of disrupting star formation in rapidly evolving galaxies.

\item [-] If instead we assume that the far wings of the emission-lines represent the true, de-projected outflow velocities, the mass outflow rates, and particularly the kinetic powers of the outflows, are higher ($7\times10^{-2} < \dot{M} < 11$ M$_{\sun}$ yr$^{-1}$; $5\times10^{-4} < \dot{E}/L_{BOL} < 1$\%). However, the kinetic powers still fall well short of what is required in many galaxy evolution models \citep[e.g.][]{fabian99, dimatteo05}.

\end{itemize}

Therefore, on the sub-kpc scales of the nuclear starbursts in ULIRGs, the warm outflows are likely to have a substantial
impact, potentially quenching the star formation. However, there remains a lack of observational evidence that the AGN-induced outflows are genuinely galaxy-wide and influence the star formation histories of the galaxies on larger scales.

In future papers on the QUADROS project, we will extend our spectroscopic work on outflows to a sample of 9 ULIRGs observed with the WHT telescope and consider whether the outflow properties of the QUADROS sample as a whole correlate with those of the AGN nuclei \citep{spence17}; we will also investigate the radial extents of the outflows in more detail using HST observations \citep{tadhunter17}.

\section*{Acknowledgements}

MR \& CT acknowledge support from STFC. CRA acknowledges the Ram\'{o}n y Cajal Program of the Spanish Ministry of Economy and Competitiveness through project RYC-2014-15779 and the Spanish Plan Nacional de Astronom\'{i}a y Astroﬁs\'{i}ca under grant AYA2016-76682-C3-2-P. FS gratefully acknowledges support from the European Research Council under the European Union’s Seventh Framework Programme (FP/2007-2013) ERC Advanced Grant RADIOLIFE-320745 (PI R.Morganti). We thank the referee for useful comments and suggestions which improved this work. Based on observations collected at the European Organisation for Astronomical Research in the Southern Hemisphere under ESO programme 091.B-0256(A), and on observations taken with the NASA/ESA Hubble Space Telescope, obtained at the Space Telescope Science Institute (STScI), which is operated by AURA, Inc. for NASA under contract NAS5-26555. The authors acknowledge the data analysis facilities provided by the Starlink Project, which was run by CCLRC on behalf of PPARC. This publication makes use of data products from the Two Micron All Sky Survey, which is a joint project of the University of Massachusetts and the Infrared Processing and Analysis Center/California Institute of Technology, funded by the National Aeronautics and Space Administration and the National Science Foundation. This research has made use of the NASA/IPAC Extragalactic Database (NED) which is operated by the Jet Propulsion Laboratory, California Institute of Technology, under contract with the National Aeronautics and Space Administration.

\appendix

\section{Flux Calibration} \label{sect:fluxcal}

\begin{figure*}
\centering
\includegraphics[scale=0.5]{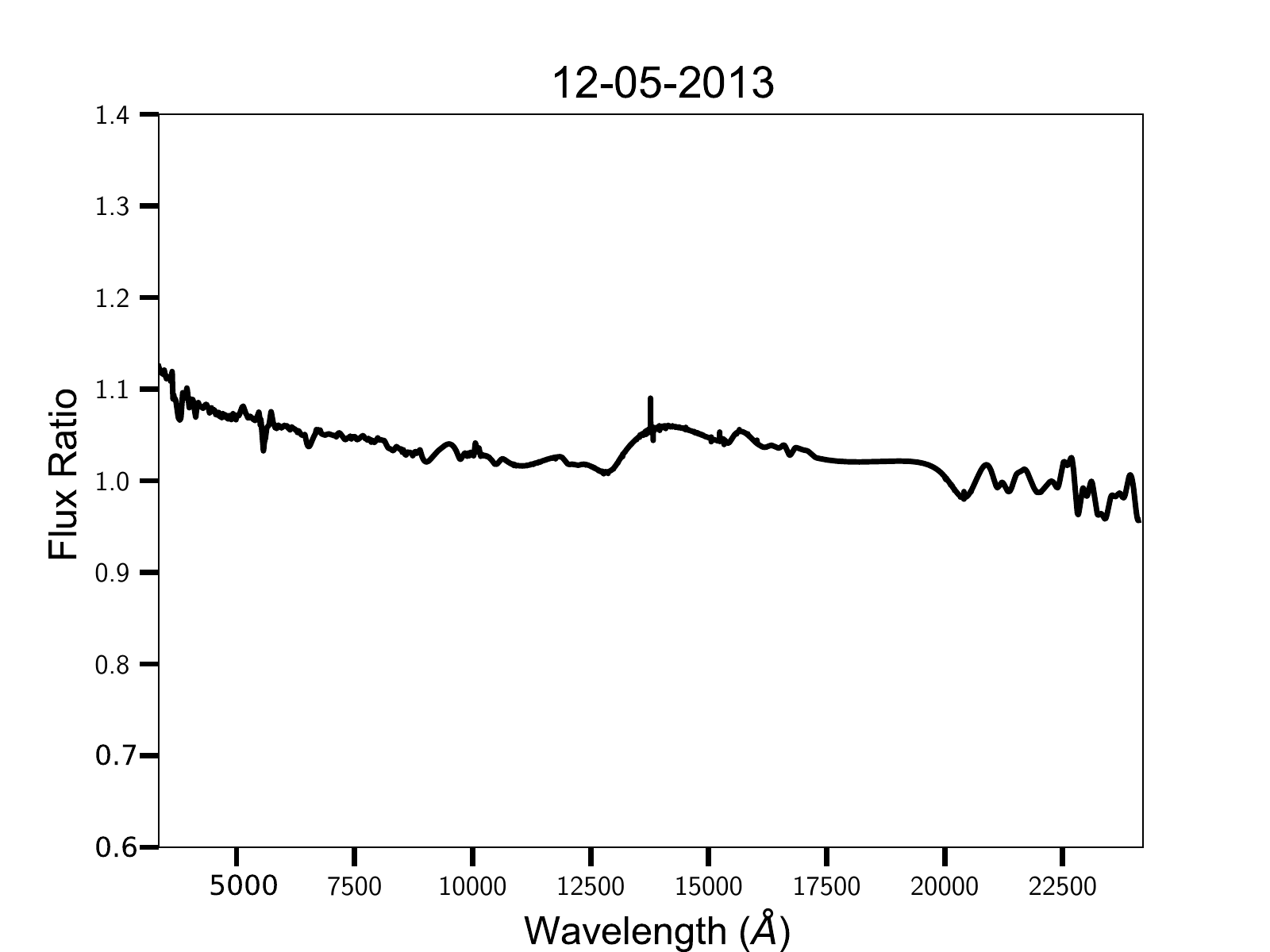}
\includegraphics[scale=0.5]{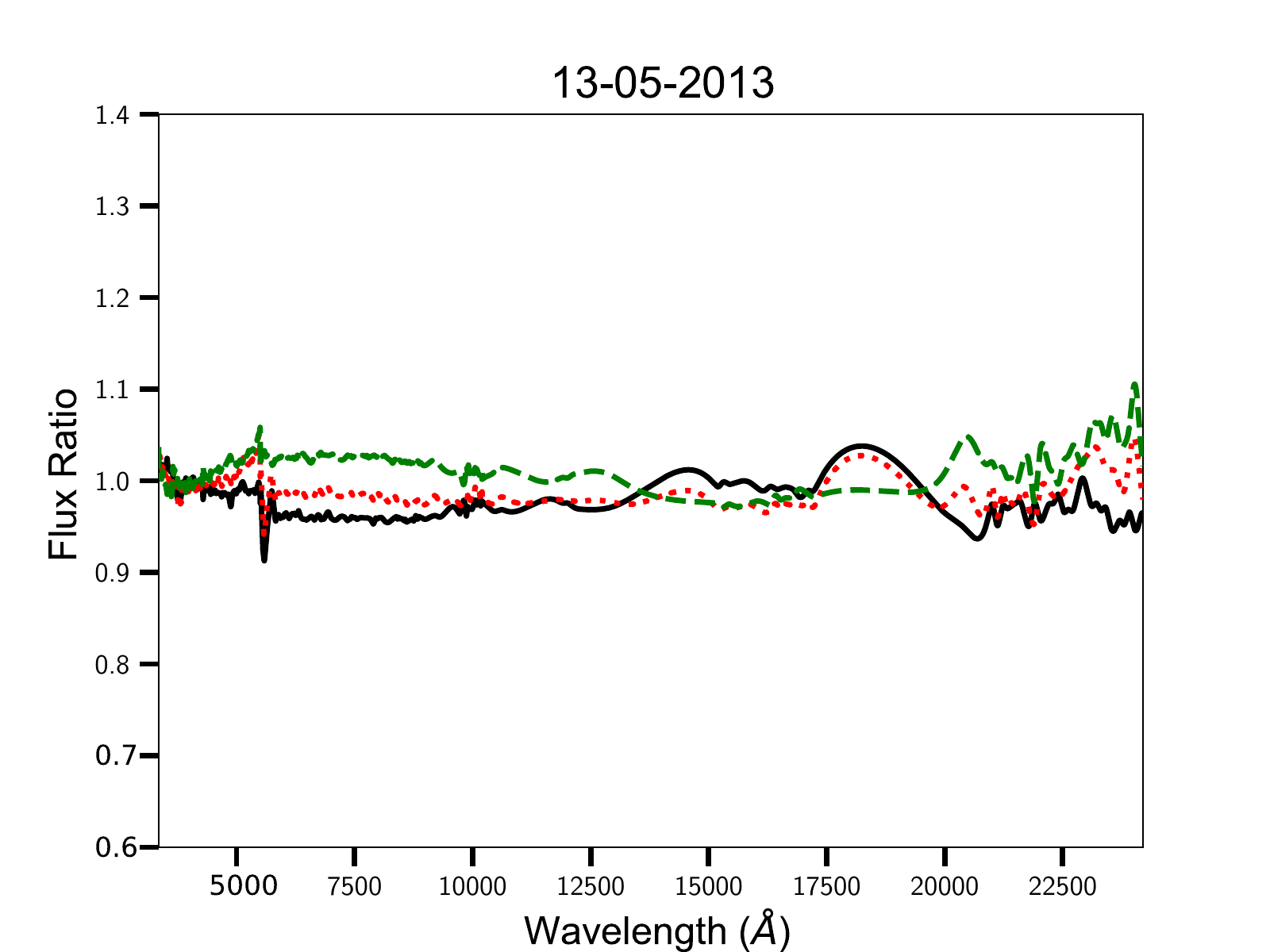}
\caption{The flux ratios obtained by dividing the spectra of
individual objects that have been flux calibrated using different standard stars. The black line shows the flux ratio EG274/LT3218, red line represents LTT987/LT3218 and green line LTT987/EG274.}
\label{fig:fluxcal}
\end{figure*}

In order to flux calibrate the Xshooter spectra, we observed 3 standard stars on the nights of 12-05-2013 and 13-05-2013: LT3218, EG274 and LTT987. For the additional objects observed on 19-05-2013 (F15130-1958 and F19254-7245S) we used the master response curve provided by ESO. To assess the flux calibration accuracy we reduced the ULIRGs F12072-0444 and F13305-1739 using all the flux standard stars observed for their respective nights, and then divided the extracted, flux calibrated spectra derived from the different flux standards to assess
the level of variation. The results are plotted on Figure \ref{fig:fluxcal} for both nights, where the black line shows the flux ratio EG274/LT3218, red line represents LTT987/LT3218 and green line LTT987/EG274. 

For the night of 12/05/13 there was an issue with the standard LTT987, since the flux ratios involving this star were systematically displaced by $>$10\%. In addition, there appear to be no data for this standard in the NIR arm. Therefore we did not include LTT897 when reducing the data for this night. Considering the full spectral range encompassed by Xshooter, we find a maximum variation of 8\% relative to the mean ratio. Similarly, for the night of 13/05/13 we find a maximum variation from the mean ratio of 6\%.

\section{Notes on individual objects} \label{sect:specdes}

In this section we provide more details of the general properties and Xshooter spectra of individual objects in our sample.

\vglue 0.3cm\noindent
{\bf F12072-0444.}  Although in the optical and near-IR ground-based images of \citet{kim02}  F12072-0444 appears as a single nucleus system, the higher resolution H-band HST image of \citet{veilleux06} reveals a 
double nucleus with a separation of $\sim$1 arcsec. For the Xshooter observations presented in this
paper, the slit was centred on the single nucleus visible in the optical acquisition images. The detection of a warm ionised outflow in this system was previously described by
\citet{javier13}, and evidence that  it contains a molecular outflow is provided by Herschel spectroscopic
observations  which show P-Cygni OH profiles at far-IR wavelengths, with the blueshifted OH absorption-line wings extending to a maximum velocity of -967~km~s$^{-1}$ \citep{spoon13,veilleux13}. However, no evidence has been found for
a neutral outflow in this source based on optical observations of the NaID absorption lines \citep{rupke05}.

The optical to near-IR SED of F12072-0444 is relatively flat, with only a small drop in the flux at the shortest wavelengths ($<$4000\AA\ ), and weak absorption lines. The spectrum shows a rich variety of emission-lines associated with the AGN activity. In addition to typical AGN recombination and emission-lines from both neutral and ionized gas species\footnote{These include emission from HI, HeI, HeII, [NI], [NII], [OI], [OII], [OIII], [NeIII],  [PII], [SII] and [SIII].}, detect OIII Bowen resonance fluorescence emission-lines in the near-UV, along with coronal emission-lines\footnote{Ionization potential $>$ 54.4 eV: the HeII edge.} from [NeV], [FeVII], [FeXI] and [SiX] at both optical and near-IR wavelengths, and near-IR molecular hydrogen lines including H$_2$ 1--0 S(1), H$_2$ 1--0 S(2), H$_2$ 1--0 S(3), H$_2$ 1--0 S(4) and H$_2$ 1--0 S(5).

The best fitting model to the [OIII]$\lambda\lambda$4959,5007 lines comprises three components: a narrow component with $\Delta$v=-156$\pm$40 km s$^{-1}$ and FWHM=346$\pm$25 km s$^{-1}$, an intermediate component with $\Delta$v=-413$\pm$44 km s$^{-1}$ and FWHM=604$\pm$26 km s$^{-1}$, and a broad component with $\Delta$v=-652$\pm$59 km s$^{-1}$ and FWHM=1160$\pm$100 km s$^{-1}$. Note that all velocity shifts in this section are given with respect to the host galaxy rest frame. Notably, {\it all} the components of the [OIII] emission-lines are shifted relative to the host galaxy rest frame for this object. While the [OIII] kinematic model provided a good fit to the hydrogen recombination lines, it is less successful at fitting the trans-auroral blends. Instead, we fitted the latter blends by constraining the narrow components using the [OIII] model, and the broad components using a fit to the [OII]$\lambda\lambda$3726,3729 blend.  

\vglue 0.3cm\noindent
{\bf F13305-1739.} This object appears as a single nucleus system in the ground-based images of \citet{kim02}. Its emission-line outflows were first discussed by \citet{javier13}, but the system shows no evidence for neutral or molecular outflows based on observations of the optical NaID and far-IR OH absorption lines respectively \citep{rupke05, veilleux13}.

As pointed out by \citet{javier13}, based on its [OIII] emission-line luminosity, F13305-1739 is the most luminous object in our entire sample (see Table \ref{tab:bol}); according to the criterion of \citet{zakamska03} it qualifies as a type 2 quasar even before its [OIII] luminosity is corrected for dust extinction. It also has one of the richest spectra in the sample in terms of its detected spectral features, and its optical to near-IR continuum SED is relatively blue, with a strong Balmer break and Balmer absorption lines indicating the presence of a 
young stellar population (YSP). The relatively unreddened nature of its YSP is consistent with its  very blue 2MASS J-K$_s$ colour: 1.13$\pm$0.11. This colour is bluer than both `typical' type 1 AGN (J-K$_s$ $\sim$ 1.70) and 2Jy Radio galaxies (J-K$_s$ $\sim$ 1.40; see \citealt{rose13b} for full discussion). The emission-line spectrum of F13305-1739 shows dozens of strong lines. As well as emission-lines typically associated with AGN spectra, we have detected strong coronal emission-lines from [NeV] and [FeVII], as well as near-IR molecular hydrogen lines including H$_2$ 1--0 S(1), H$_2$ 1--0 S(2), H$_2$ 1--0 S(3), H$_2$ 1--0 S(4) and H$_2$ 1--0 S(5). 

The best fitting model to the [OIII]$\lambda\lambda$4959,5007 lines comprises five components, three of which are narrow (FWHM $<$ 500 km s$^{-1}$). The first narrow component has $\Delta$v=+195$\pm$40 km s$^{-1}$ and FWHM=80$\pm$9 km s$^{-1}$, the second narrow component has $\Delta$v=-94$\pm$34 km s$^{-1}$ and FWHM=188$\pm$14 km s$^{-1}$ and the third narrow component has $\Delta$v=+370$\pm$35 km s$^{-1}$ and FWHM=432$\pm$23 km s$^{-1}$. In addition to the narrow components, there is a broad component with $\Delta$v=-115$\pm$34 km s$^{-1}$ and FWHM=1140$\pm$90 km s$^{-1}$, and a very broad component with $\Delta$v=-279$\pm$40 km s$^{-1}$ and FWHM=2150$\pm$210 km s$^{-1}$. While our [OIII] model provides an adequate fit to the hydrogen recombination lines, it was less successful in fitting the trans-auroral blends. Instead, we fitted the latter blends by constraining the narrow components using the [OIII] model, and the broad components using a fit to the [OII]$\lambda\lambda$3726,3729 blend.  

\vglue 0.3cm\noindent
{\bf F13443+0802SE.} This is a triple system consisting of a close pair to the east (hereafter F13443+0802NE and F13443+0802SE) with a separation of $\sim$12kpc and a third component located at $\sim$37kpc to the south west of this pair (F13443+0802SW) \citep{kim02,tadhunter17}. Optical HST imaging of the southern component of the close pair (F13443+0802SE) shows strong [OIII] emission in a peculiar `Y-shaped' morphology \citep{tadhunter17}. The slit for the VLT/Xshooter observations was centred on this `Y-shaped' component in F13443+0802SE, which is unresolved
in our ground-based observations and represents the true AGN in the triple system.  F13443+0802SE shows no evidence
for a neutral outflow based on NaID optical absorption line measurements \citep{rupke05}.

The optical to near-IR continuum spectrum of F13443+0802SE shows a strong Balmer break and Balmer absorption lines characteristic of a young stellar population, and the near-IR continuum slope is consistent with the relatively blue 2MASS J-K$_s$ colour: 1.40$\pm$0.11. However, the emission-line spectrum of F13443+0802SE is less rich than in other objects in our Xshooter sample. There are moderate strength emission-lines including typical AGN lines, coronal emission-lines of [NeV] and [FeVII],  and near-IR molecular lines including H$_2$ 1--0 S(2), H$_2$ 1--0 S(3), H$_2$ 1--0 S(4) and H$_2$ 1--0 S(5). 

F13443+0802SE is  the least kinematically-disturbed ULIRG in our Xshooter sample. The best-fitting [OIII]$\lambda\lambda$4959,5007  model comprises just three components: a narrow component with $\Delta$v=+60$\pm$48 km s$^{-1}$ and FWHM=350$\pm$6 km s$^{-1}$, a second narrow component with $\Delta$v=-57$\pm$45 km s$^{-1}$ and FWHM=73$\pm$14 km s$^{-1}$ and an intermediate component with $\Delta$v=-50$\pm$46 km s$^{-1}$ and FWHM=698$\pm$42 km s$^{-1}$. This model successfully fitted both the recombination lines and trans-auroral blends in F13443+0802SE.  

\vglue 0.3cm\noindent
{\bf F13451+1232W.} Optical images of this system show a close double nucleus system with separation of 2.1 arcsec \citep{kim02,tadhunter17}. The slit for our Xshooter observations was centred on the more westerly of the two nuclei (F13451+1232W), which also contains the bright AGN nucleus. Like F13305-1739, this object would qualify as a type 2 quasar based on its [OIII] emission-line luminosity according to the criteria of \citet{zakamska03} without correcting for dust extinction. Along with clear
evidence for a warm emission-line outflow at optical wavelengths \citep{holt03,holt11}, F13451+1232W shows evidence for neutral outflows detected in the radio HI~21cm line and the optical NaID line \citep{morganti05,rupke05}, as well as molecular outflows detected in CO emission-lines \citep{dasyra12,morganti13}. We further note that F13451+1232W
is one of the few ULIRGs in the QUADROS sample that is classified as a radio-loud AGN ($P_{1.4GHz} = 2\times10^{26}$ W Hz$^{-1}$).

The optical spectrum of F13451+1232W has been studied in detail by \citet{holt03}, \citet{tadhunter05}, \citet{holt11} and \citet{javier13}. Its continuum spectrum is dominated by stellar populations of old/intermediate age \citep{tadhunter05}, but in the nuclear spectrum the stellar absorption features are relatively weak, with only the MgIB multiplet being securely detected. emission-lines including typical AGN lines, coronal emission from [NeV], [FeVII], [SiVI] and [SiX], and near-IR molecular hydrogen lines including H$_2$ 1--0 S(1), H$_2$ 1--0 S(2), H$_2$ 1--0 S(3), H$_2$ 1--0 S(4) and H$_2$ 1--0 S(5) have been detected.

F13451+1232W is one of the most kinematically disturbed of all the objects in our Xshooter sample. The best-fitting [OIII]$\lambda\lambda$4959,5007 model consists of four components: a narrow component with $\Delta$v=+69$\pm$46 km s$^{-1}$ and FWHM=319$\pm$6 km s$^{-1}$, a broad component with $\Delta$v=-311$\pm$60 km s$^{-1}$ and FWHM=1160$\pm$150 km s$^{-1}$, a second broad component with $\Delta$v=-1841$\pm$98 km s$^{-1}$ and FWHM=1900$\pm$150 km s$^{-1}$, and a very broad component with $\Delta$v=-262$\pm$40 km s$^{-1}$ and FWHM=3100$\pm$200. This model successfully fitted the recombination lines, however it did not provide an adequate fit to the trans-auroral blends. Instead, we fitted the latter blends by constraining the narrow components using the [OIII] model, and the broad components using a fit to the [OII]$\lambda\lambda$3726,3729 blend. When comparing the kinematic components of the [OIII] model to those in previous work, we find that 3/4 of the components are consistent with those reported in \citet{holt03}, \citet{holt11} and \citet{javier13}. The very broad component was not reported in the latter papers and is therefore reported here for the first time.

\vglue 0.3cm\noindent
{\bf F14378-3651.} This system shows a single compact nucleus in HST images \citep{westmoquette12}. An emission-line outflow was previously detected in H$\alpha$ in the nuclear regions of F14378-3651 by \citep{westmoquette12}, and the detection of a P-Cygni profile to the far-IR OH features detected by Herschel provides clear evidence for a molecular outflow \citep{sturm11,veilleux13,gonz17} with blueshifted velocities up to -800 -- -1000~km~s$^{-1}$. In this paper we report the first evidence for an outflow in this system in the higher ionisation [OIII] emission-lines.

Although the detection of a strong Balmer break and Balmer+Paschen absorption lines suggest that the optical to near-IR continuum of F14378-3651 is dominated by a young stellar population (YSP), the red shape of optical SED provides evidence that this YSP is strongly reddened, consistent with the relatively high degree of emission-line reddening deduced for this object (see $\S$\ref{sect:red} \& \ref{sect:dens}). The strong HI absorption prevents the detection of higher order (e.g. H$\gamma$) recombination lines in the spectrum. While other typical AGN emission-lines are detected, no coronal lines have been detected. We also report the detection of warm molecular lines including  H$_2$ 1--0 S(1), H$_2$ 1--0 S(2) and H$_2$ 1--0 S(3).


The best fitting model to the [OIII]$\lambda\lambda$4959,5007 lines comprises three components: a narrow component with $\Delta$v=+3$\pm$62 km s$^{-1}$ and FWHM=236$\pm$5 km s$^{-1}$, a second narrow component with $\Delta$v=-729$\pm$64 km s$^{-1}$ and FWHM=177$\pm$13 km s$^{-1}$, and a broad component with $\Delta$v=-650$\pm$85 km s$^{-1}$ and FWHM=1240$\pm$190 km s$^{-1}$. Notably, the second narrow component is only detected in the [OIII] emission-lines. Overall, with the exception of the second narrow component, the [OIII] model successfully fits both the optical recombination lines and trans-auroral blends in F14378-3651. However, no broad outflow components were detected for the P$\alpha$ and Br$\gamma$ lines. 

\vglue 0.3cm\noindent
{\bf F15130-1958.} This object shows a single, compact nucleus in optical ground-based and HST images \citep{kim02,tadhunter17}. Its optical to near-IR SED is unusual: whereas the blue-green end of the spectrum is relatively flat and shows a clear Balmer break and Balmer absorption lines characteristic of young stellar populations, the spectrum rises steeply from the long wavelength end of the optical to the near-IR. This steep rise at longer wavelengths is consistent with the red 2MASS J-K$_s$ colour measured for the integrated light of the galaxy (J-K$_s$ $=$ 2.17$\pm$0.14) and the unusually red R-K' colour measured in a 4~kpc aperture centred on the
nucleus ((R-K')$_{4} =$ 5.8; Kim et al. 2002). While the shape of the longer-wavelength end of the SED might be consistent with the presence of a moderately obscured type 1 AGN, we do not detect a BLR component to Pa$\alpha$ that would be expected in such a case. 

As well as emission-lines typically associated with AGN spectra, we have detected coronal emission-lines from [NeV] and [FeVII], as well as near-IR molecular hydrogen lines, including H$_2$ 1--0 S(1), H$_2$ 1--0 S(2) and H$_2$ 1--0 S(3). Note that while \citet{javier09} report the detection of Wolf-Rayet features in their optical WHT/ISIS spectrum for F15130-1958, we do not detect these features.

The best-fitting [OIII]$\lambda\lambda$4959,5007 model consists of just two components: an intermediate component with $\Delta$v=-516$\pm$38 km s$^{-1}$ and FWHM=828$\pm$38 km s$^{-1}$, and a broad component with $\Delta$v=-1186$\pm$77 km s$^{-1}$ and FWHM=1250$\pm$190 km s$^{-1}$. It is notable that no narrow component has been detected and that all the kinematic components of the [OIII] emission-lines are substantially shifted relative to the host galaxy rest frame, similar to the case of PKS1549-79 \citep{tadhunter01}. We also note that \citet{javier13} report an additional intermediate component to the [OIII] lines that is not detected in the spectrum presented in this paper.

\vglue 0.3cm\noindent
{\bf F15642-0450.} The optical spectrum of this single-nucleus system is dominated by type 1 AGN emission, as indicated by the presence of a strong BLR (FWHM=4100$\pm$90 km s$^{-1}$) components to the hydrogen lines and broad FeII emission-lines. No stellar features have been detected in the spectrum. Evidence for a molecular outflow 
in this source is provided by the presence of broad, blueshifted wings to the OH absorption lines
detected in far-IR wavelengths Herschel spectra, with these wings extending to a -927 km s$^{-1}$ \citep{veilleux13,spoon13}. 

As well as BLR and FeII emission, we detect emission-lines typically associated with AGN spectra, as well as OIII Bowen resonance fluorescence emission-lines, coronal lines including [NeV], [CaV], [FeX] and [FeXI], and near-IR molecular hydrogen lines including H$_2$ 1--0 S(1), H$_2$ 1--0 S(2), H$_2$ 1--0 S(3) and H$_2$ 1--0 S(4).

The best fitting model to the [OIII]$\lambda\lambda$4959,5007 lines comprises three components: a narrow component which we use to define the rest frame of the system with FWHM=186$\pm$9 km s$^{-1}$, a second narrow component with $\Delta$v=-742$\pm$10 km s$^{-1}$ and FWHM=411$\pm$17 km s$^{-1}$, and a broad component with $\Delta$v=-915$\pm$16 km s$^{-1}$ and FWHM=1460$\pm$20 km s$^{-1}$. Note that \citet{javier13} detect only one narrow component in their [OIII] emission-lines. For this object, the [OIII] model did not provide an adequate fit to either the hydrogen recombination lines or the trans-auroral blends.  In this case, all the recombination lines were fitted using the parameters required to fit the H$\beta$ line, while the trans-auroral lines were fitted by constraining the narrow components using the [OIII] model, and the broad components using a fit to the [OII]$\lambda\lambda$3726,3729 blend.   

\vglue 0.3cm\noindent
{\bf F16156+0146NW.} In optical ground-based and HST images this object shows two nuclei separated by $\sim$3.5 arcseconds. For our Xshooter observations, the slit was centred on the NW nucleus that contains the optical AGN
(F16156+0146NW). As well as emission-lines typically associated with AGN spectra, we have detected  coronal lines from [NeV], [FeVII] and [SiVI], as well as near-IR molecular lines including H$_2$ 1--0 S(1), H$_2$ 1--0 S(2), H$_2$ 1--0 S(3), H$_2$ 1--0 S(4) and H$_2$ 1--0 S(5).

The best-fitting [OIII]$\lambda\lambda$4959,5007 model consists of three components: a narrow component with $\Delta$v=+36$\pm$78 km s$^{-1}$ and FWHM=343$\pm$12 km s$^{-1}$, an intermediate component with $\Delta$v=-181$\pm$79 km s$^{-1}$ and FWHM=910$\pm$110 km s$^{-1}$, and a broad component with $\Delta$v=-381$\pm$82 km s$^{-1}$ and FWHM=1780$\pm$210 km s$^{-1}$. A similar [OIII] model was derived by \citet{javier13}, and while it provided an adequate fit to the hydrogen recombination lines, it was less successful at fitting the trans-auroral blends. Instead, we fitted the latter blends by constraining the narrow components using the [OIII] model, and the broad components using a fit to the [OII]$\lambda\lambda$3726,3729 blend. 

Note that we detect an additional blueshifted narrow component in the P$\alpha$ and P$\beta$ recombination line profiles which is not detected in emission-lines at shorter wavelengths. This component appears to be highly reddened (E(B-V)=1.1$_{-0.5}^{+0.4}$; see $\S$\ref{sect:red}) suggesting that it is too highly extinguished to be detected at shorter wavelengths.

\vglue 0.3cm\noindent
{\bf F19254-7245S.} Also known as the Superantennae,  F19254-7245 is a well studied merger system with two optical nuclei separated by 8.5 arcsec  in the north-south direction \citep{melnick90,westmoquette12}. The slit used for our Xshooter spectrum of this source was centred on the southern nucleus (F19254-7245S), which also contains the optical AGN \citep{mirabel91}. This southern nucleus shows evidence for a molecular outflow in the form of blue wings to the far-IR OH absorption lines detected in Herschel spectra, that extend to a maximum blueshifted  velocity of -1126~km~s$^{-1}$ \citep{spoon13}. A warm outflow has also been detected in the southern nucleus in the form of extremely broad wings to several optical
recombination and forbidden lines that extend 1000s of km s$^{-1}$ from the line centres \citep{colina91,mirabel91,westmoquette12}.  

Although the optical to near-IR continuum SED of  F19254-7245S is clearly dominated by an underlying young stellar population, with a strong Balmer break, as well as stellar absorption features including those of HI and CaII, the
starlight is highly reddened, so the SED falls steeply from the optical to the near-UV. As for F14378-3651, this is consistent with the relatively high degree of emission-line reddening deduced for this object (see $\S$\ref{sect:red} \& \ref{sect:dens}). F19254-7245S has a rich emission-line spectrum that includes typical AGN lines, coronal emission from [NeV], [SiVI] and [SiX], and several near-IR molecular hydrogen lines lines (H$_2$ 1--0 S(0), H$_2$ 1--0 S(1), H$_2$ 1--0 S(2), H$_2$ 1--0 S(3), H$_2$ 1--0 S(4) and H$_2$ 1--0 S(5)).

The best-fitting [OIII]$\lambda\lambda$4959,5007 model consists of four components: a narrow component with $\Delta$v=+30$\pm$38 km s$^{-1}$ and FWHM=118$\pm$11 km s$^{-1}$, an intermediate component with $\Delta$v=+855$\pm$80 km s$^{-1}$ and FWHM=604$\pm$37 km s$^{-1}$, a second intermediate component with $\Delta$v=+120$\pm$36 km s$^{-1}$ and FWHM=656$\pm$72 km s$^{-1}$, and a broad component with $\Delta$v=+139$\pm$38 km s$^{-1}$ and FWHM=1890$\pm$130 km s$^{-1}$. F19254-7245S is the only object in our Xshooter sample for which outflowing gas component appears to be predominantly redshifted with respect to the rest frame of the host galaxy. The [OIII] model did not provide an adequate fit to either the recombination lines or trans-auroral blends. All the hydrogen recombination lines were successfully fit using the model that fits the H$\beta$ line, whereas for the fits to the trans-auroral lines we constrained the narrow components using the [OIII] model and  the broad components using the fit to the broad components of the [OII]$\lambda\lambda$3726,3729 blend.

\label{lastpage}

\end{document}